\documentclass[aps,preprint,prd,superscriptaddress,showpacs]{revtex4-1}
\usepackage{graphicx}
\usepackage{amsmath,amssymb}
\usepackage{subfigure}
\usepackage[T1]{fontenc} 
\usepackage[above,below]{placeins}
\usepackage{multirow}
\usepackage{longtable}
\usepackage{array}
\newcommand{\tabincell}[2]{\begin{tabular}{@{}#1@{}}#2\end{tabular}}
\allowdisplaybreaks[2]
\begin{document}
\title{Quantum entanglement in three accelerating qubits coupled to scalar fields}
\author{Yue Dai}
\affiliation{Center for Field Theory and Particle Physics, Department of Physics  \&  State Key Laboratory of Surface Physics,   Fudan University,\\ Shanghai 200433, China}
\author{Zhejun Shen}
\affiliation{Center for Field Theory and Particle Physics, Department of Physics  \&  State Key Laboratory of Surface Physics,   Fudan University,\\ Shanghai 200433, China}
\author{Yu Shi }  \email{yushi@fudan.edu.cn}
\affiliation{Center for Field Theory and Particle Physics, Department of Physics  \&  State Key Laboratory of Surface Physics,   Fudan University,\\ Shanghai 200433, China}
\affiliation{Collaborative Innovation Center of Advanced Microstructures, Fudan University, \\Shanghai 200433, China}

\begin{abstract}
We consider quantum entanglement of three accelerating qubits, each of which is locally coupled with a real scalar field, without causal influence among the qubits or among the fields. The initial states are assumed to be the GHZ and W states, which are the two representative three-partite entangled states. For each initial state, we study how various kinds of entanglement depend on the accelerations of the three  qubits. All kinds of entanglement  eventually suddenly die if at least two of three qubits have large enough accelerations. This result implies the eventual sudden death of all kinds of entanglement among three particles coupled with scalar fields when they are sufficiently close to the horizon of a black hole.
\end{abstract}


published as Phys. Rev. D {\bf 94}, 025012 (2016). 

\maketitle
\section{Introduction}

How quantum states are affected by gravity or  acceleration  is a subject of longstanding interest~\cite{mukhanov}. In the presence of a black hole, the physical vacuum is that in the Kruskal-Szekers coordinates, which are nonsingular and cover the whole Schwarzschild spacetime, while a remote  observer can observe particles. Known as Hawking radiation, it underlies the paradox of information loss~\cite{hawking}.
Analogously, an accelerating object coupled with a field detects a thermal bath of particles of this field, even though this  field is in the Minkowski vacuum~\cite{davies,fulling,dewitt,unruh}: this is known as the Unruh effect.

For a composite quantum system, the characteristic quantum feature is quantum entanglement.  An interesting question is how quantum  entanglement among objects coupled with fields is affected by the Unruh effect. Various investigations were made on two entangled detectors, one or both of which accelerate~\cite{matsas1,doukas1,doukas2,huyu,hu0,dai}.

Naturally, one may wonder about the situation of three entangled field-coupled qubits. This is interesting and nontrivial because there are various different types of entanglement among three qubits $A$, $B$, and $C$. There is bipartite entanglement between two qubits, and there is  bipartite entanglement between one qubit and the remaining two qubits as one  party. Most interestingly, there is tripartite entanglement among all three qubits, which cannot be reduced to any combination  of all kinds of  bipartite entanglement~\cite{bennett}. This is profound and important in understanding many-body correlations.  In quantum information theory, recent years witnessed much development in quantifying entanglement in terms of some measures. A convenient one is the so-called negativity, which is the sum of the negative eigenvalues of the partial transpose of the  density matrix, ranging from $0$ to $1/2$~\cite{vidal}. It can be used to quantify various kinds of bipartite entanglement in a three-qubit state.  Twice the negativity that quantifies the bipartite entanglement between a qubit and the remaining two qubits is called a one-tangle, ranging from $0$ to $1$. Twice the negativity quantifies the bipartite entanglement between two qubits is called a two-tangle,  ranging from $0$ to $1$. Using all the one-tangles and two-tangles, one can define a measure of tripartite entanglement called a three-tangle, ranging from $0$ to $1$~\cite{ou}.

In the present paper, by using these negativity-based entanglement measures, we make  detailed investigations on how various types of entanglement in three field-coupled qubits vary with their accelerations, which have implications on particles near the horizon of a black hole. It is well known that there are two inequivalent types of tripartite entanglement~\cite{duer}, typified, respectively, by the GHZ state
\begin{equation}
|\mathrm{GHZ}\rangle = \frac{1}{\sqrt{2}}(|000\rangle + |111\rangle).
\end{equation}
and the W state
\begin{equation}
|\mathrm{W}\rangle=\frac{1}{\sqrt{3}}(|001\rangle + |010\rangle+|100\rangle).
\end{equation}
So both the GHZ and W states are considered in this paper.

\section{Method}

We consider three qubits $A$, $B$, and $C$ far away from each other. For simplicity, it is assumed that each qubit $q$ ($q=A,B,C$)  is  coupled only locally  with the field  $\Phi_q$ around it, as described by  the Unruh-Wald model~\cite{unruhwald}. The Hamiltonian of each qubit $q$ itself is $H_q=\Omega_q Q_q^\dagger Q_q$, where the creation operator $Q_q^\dagger$ and annihilation operator $Q_q$ are defined by $Q_q | 0 \rangle_q = Q_q^\dagger  | 1 \rangle_q = 0$, $Q_q^\dagger  | 0 \rangle_q =  | 1 \rangle_q$, and $Q_q | 1 \rangle_q = | 0\rangle_q$, and $\Omega_q$ is  the energy difference between the two eigenstates. Its coupling with the field is described by  $H_{I_q}(t_q) = \epsilon_q  ( t_q  ) \int_{\Sigma_q}  {\Phi_q  (  \mathbf{x}  ) [ {\psi_q  ( \mathbf{ x}  )Q_q  + {\psi_q^*} ( \mathbf{x} ){Q_q^\dag }}  ]\sqrt { - g} {d^3}x}$, where $\mathbf{x}$ and $t_q$ are spacetime coordinates in the comoving frame of the qubit, the integral is over the spacelike Cauchy surface $\Sigma_q$ at the given time $t_q$, $\epsilon_q  ( t_q  )$ is the coupling constant with a finite duration, and $\psi_q (\mathbf{ x} )$ is a smooth nonvanishing function within a small region around the qubit.   The total Hamiltonian of the system is thus
\begin{equation}
\sum_{q=A,B,C} (H_q+H_{\Phi_q}+ H_{I_q}),
\end{equation}
where $H_{\Phi_q}$ is the Klein-Gordon Hamiltonian for $\Phi_q$.

It is assumed that the qubits are  far away from each other such that during the  interaction times, there is no physical coupling or influence between the  fields around  different qubits or between a qubit and the field around another qubit.  We make this assumption to avoid the issue of a global time slice and  those complications caused by the different accelerations of the qubits. Consequently, one can describe the quantum state of the qubits by considering  each qubit in its comoving reference frame. The interesting case where all qubits are coupled with the same field will be explored in a future work.

Therefore, in our consideration, after a time duration longer than the interacting times $T_q \gg 1/\Omega_q$, the state of the whole system in the interaction picture is transformed  by  $$U_{A}\otimes U_{B}\otimes U_{C},$$
where $U_{q}$ is the unitary transformation acting on qubit  $q$ and the field $\Phi_q$ in its neighboring region. It can  be obtained that~\cite{dai} \begin{equation}
U_{q} \approx 1+iQ_{q} a^\dagger (\Gamma_q^*)  -  i  Q_{q}^\dagger a(\Gamma_q^*),
\end{equation}
where    $a(\Gamma_q^*)$  and $a^\dagger(\Gamma_q^*)$ are the  annihilation and    creation operators of  $\Gamma_q^*$, with
\begin{equation}
\Gamma_q (x) \equiv -2i \int [G_{Rq}(x;x')-G_{Aq}(x;x')]\epsilon_q(t')e^{i\Omega_q t}\psi_q^*(\mathbf{x}')\sqrt{-g'}
d^4x',
\end{equation}
where $G_{Rq}$ and $G_{Aq}$ are the retarded and advanced Green functions of the field $\Phi_q$~\cite{unruhwald}.

For each field $\Phi_q$, it has been argued that approximately the qubit $q$ is only coupled with the field mode $\Gamma_q^*$, with frequency $\Omega_q$, with the other modes   decoupled~\cite{unruhwald,dai}. Consider the Fock state $|n\rangle_{\Gamma^*_q}$  containing $n$ particles in the mode $\Gamma^*_{q}$, as observed in the Rindler wedge confining qubit $q$. We have~\cite{dai}
\begin{eqnarray}
U_{q} | 0 \rangle_q  | n \rangle_{\Gamma^*_q}&= & | 0 \rangle_q | n \rangle_{\Gamma^*_q}  - i\sqrt n \mu_q | 1 \rangle_q  | {n - 1} \rangle_{\Gamma^*_q}, \label{s1}  \\
U_{q} | 1 \rangle_q | n \rangle_{\Gamma^*_q} & =  & | 1 \rangle_q | n \rangle_{\Gamma^*_q}
+ i\sqrt {n + 1} \mu_q^* | 0 \rangle_q | {n + 1} \rangle_{\Gamma^*_q},   \label{s2}
\end{eqnarray}
where  $\mu_q  \equiv \langle \Gamma_q^*,\Gamma^*_q\rangle$. For an arbitrary mode $\chi$, $\langle \Gamma_q^*,\chi\rangle
=  \int \epsilon_q(t)  e^{i\Omega_q t}\psi_q^*(\mathbf{x}) \chi(t,\mathbf{x})\sqrt{-g} d^4x$~\cite{unruhwald}.

Suppose the initial state of the three qubits to be $|\Psi_i\rangle$.   Without causal connection either between the qubits or between the fields, each qubit independently detects a thermal bath of the Unruh particles  determined by its own acceleration. With each qubit in its own Rindler wedge,  the initial state of the whole system, as observed by the observers comoving with the qubits respectively, is described by the density matrix
\begin{equation} \rho_i = \rho_{\Gamma^*_A}\otimes \rho_{\Gamma^*_B} \otimes  \rho_{\Gamma^*_C}\otimes |\Psi_i\rangle\langle \Psi_i|, \end{equation}
where $\rho_{\Gamma^*_q}$ is the density matrix of the mode $\chi_{\Gamma^*_q}$ of  the field around qubit $q$, and the decoupled modes are neglected.

For a uniformly moving qubit $q$,
\begin{equation}
\rho_{\Gamma^*_q}=|0\rangle_{\Gamma^*_q}\langle 0|,
\end{equation}
because the uniformly moving qubit sees a Minkowski vacuum.

For an accelerating qubit,
\begin{equation}
\rho_{\Gamma^*_q}=\eta_q\sum_{n_q} e^{ - 2\pi n_q \Omega_q/a_q}|n_q\rangle_{\Gamma^*_q}\langle n_q|, \end{equation}
where $a_q$ is the acceleration of qubit $q$, and $n_q$ denotes the particle number of mode $\Omega_q$, $\eta_q \equiv \sqrt {1 - {e^{ - 2\pi \Omega_q/a_q}}}$.

The final state of the system with respect to the  comoving observers is
\begin{equation} \rho_f = U_{C}^\dagger  U_{B}^\dagger U_{A}^\dagger \rho_i U_{A} U_{B}U_{C}, \end{equation}
from which we obtain the reduced density matrix of the three qubits
\begin{equation}
\rho_{ABC} = Tr_{{\Gamma^*_A},{\Gamma^*_B},{\Gamma^*_C}} (\rho_f).
\end{equation}
Then we study the dependence of various types of entanglement on the accelerations.

The one-tangle between qubit $\alpha$ and the remaining qubits $\beta$ and $\gamma$ is
\begin{equation}
{\cal N}_{\alpha(\beta\gamma)} \equiv \|\rho_{ABC}^{T_\alpha}\|-1,
\end{equation}
where $T_\alpha$ represents a partial transpose with respect to $\alpha$, and $\|\rho\|\equiv \textrm{Tr}\sqrt{\rho\rho^\dagger}$ represents the trace norm of $\rho$.  The two-tangle between qubits $\alpha $ and $\beta$ is
\begin{equation}
{\cal N}_{\alpha\beta} \equiv \|\rho_{\alpha\beta}^{T_A}\|-1,
\end{equation}
where $\rho_{\alpha\beta}= \textrm{Tr}_\gamma \rho_{ABC}$ is the  reduced density matrix of $\alpha$ and $\beta$.
The three-tangle is
\begin{equation}
\pi \equiv \frac{1}{3}\sum_{\alpha=A,B,C}\pi_\alpha, \end{equation}  where
\begin{equation} \pi_\alpha \equiv {\cal N}_{\alpha(\beta\gamma )}^2-{\cal N}_{\alpha\beta}^2- {\cal N}_{\alpha\gamma}^2. \label{pia}
\end{equation}

Note that a monogamy relation
\begin{equation}  {\cal N}_{\alpha(\beta\gamma )}^2 \geq {\cal N}_{\alpha\beta}^2+ {\cal N}_{\alpha\gamma}^2 \label{mono}
\end{equation}
is always valid, and is the basis for the definition (\ref{pia}).

In the following, for the GHZ and W states, we study various cases of the accelerations of the three qubits. Note the permutation symmetry of each of these two states.

\section{GHZ state}

First  we consider the initial state to be the GHZ state,
\begin{equation}\label{ints}
|\Psi_i\rangle = |\mathrm{GHZ}\rangle.
\end{equation}

In the GHZ state, tracing over one qubit always yields a disentangled two-qubit state. On the other hand, the coupling between each qubit and the field around it does not increase the interqubit entanglement. Therefore, each two-tangle always remains zero,
\begin{equation}
{\cal N}_{AB} =  {\cal N}_{BC}  = {\cal N}_{AC} = 0.
\end{equation}

\subsection{$C$ accelerating}

Let us assume qubit $C$ accelerates while $A$ and $B$ move uniformly. In this case, the density matrix of the three qubits is obtained as

\begin{equation}
\begin{split}
\rho_{ABC} =&\eta_C^2\sum\limits_{{n_C}} {\frac{{{e^{ - 2\pi {n_C}{\Omega _C}/{a_C}}}}}{{{Z_{{n_C}}}}} } \left[ {\left( {1 + \left( {{n_C} + 1} \right)|\mu _A|^2|\mu _B|^2|\mu _C|^2} \right)\left| {000} \right\rangle \left\langle {000} \right|} \right.\\
& + \left| {000} \right\rangle \left\langle {111} \right|  + \left| {111} \right\rangle \left\langle {000} \right|
 + \left| {111} \right\rangle \left\langle {111} \right| + \left( {{n_C} + 1} \right)|\mu _B|^2|\mu _C|^2\left| {100} \right\rangle \left\langle {100} \right|\\
& + \left( {{n_C} + 1} \right)|\mu _A|^2|\mu _C|^2\left| {010} \right\rangle \left\langle {010} \right|
 + \left( {{n_C}|\mu _C|^2 + |\mu _A|^2|\mu _B|^2} \right)\left| {001} \right\rangle \left\langle {001} \right|\\
& + \left( {{n_C} + 1} \right)|\mu _C|^2\left| {110} \right\rangle \left\langle {110} \right| + |\mu _B|^2\left| {101} \right\rangle \left\langle {101} \right|\left. { + |\mu _A|^2\left| {011} \right\rangle \left\langle {011} \right|} \right],
\end{split}
\end{equation}
where
\begin{equation}
\begin{split}
{Z_{{n_C}}} =& 2+ |\mu _A|^2 + |\mu _B|^2  + \left( {2{n_C} + 1} \right)|\mu _C|^2 + |\mu _A|^2|\mu _B|^2\\&
+\left( {{n_C} + 1} \right)|\mu _A|^2|\mu _C|^2 + \left( {{n_C} + 1} \right)|\mu _B|^2|\mu _C|^2 + \left( {{n_C} + 1} \right)|\mu _A|^2|\mu _B|^2|\mu _C|^2.
\end{split}
\end{equation}

In the GHZ state, any qubit is maximally entangled with the other two qubits as a single party. Hence the one-tangles are
\begin{equation}
{\cal N}_{A(BC)}={\cal N}_{B(AC)} =  {\cal N}_{C(AB)}=1.
\end{equation}

\begin{figure}
\subfigure[]{
\label{g1}
\includegraphics[width=0.48\textwidth]{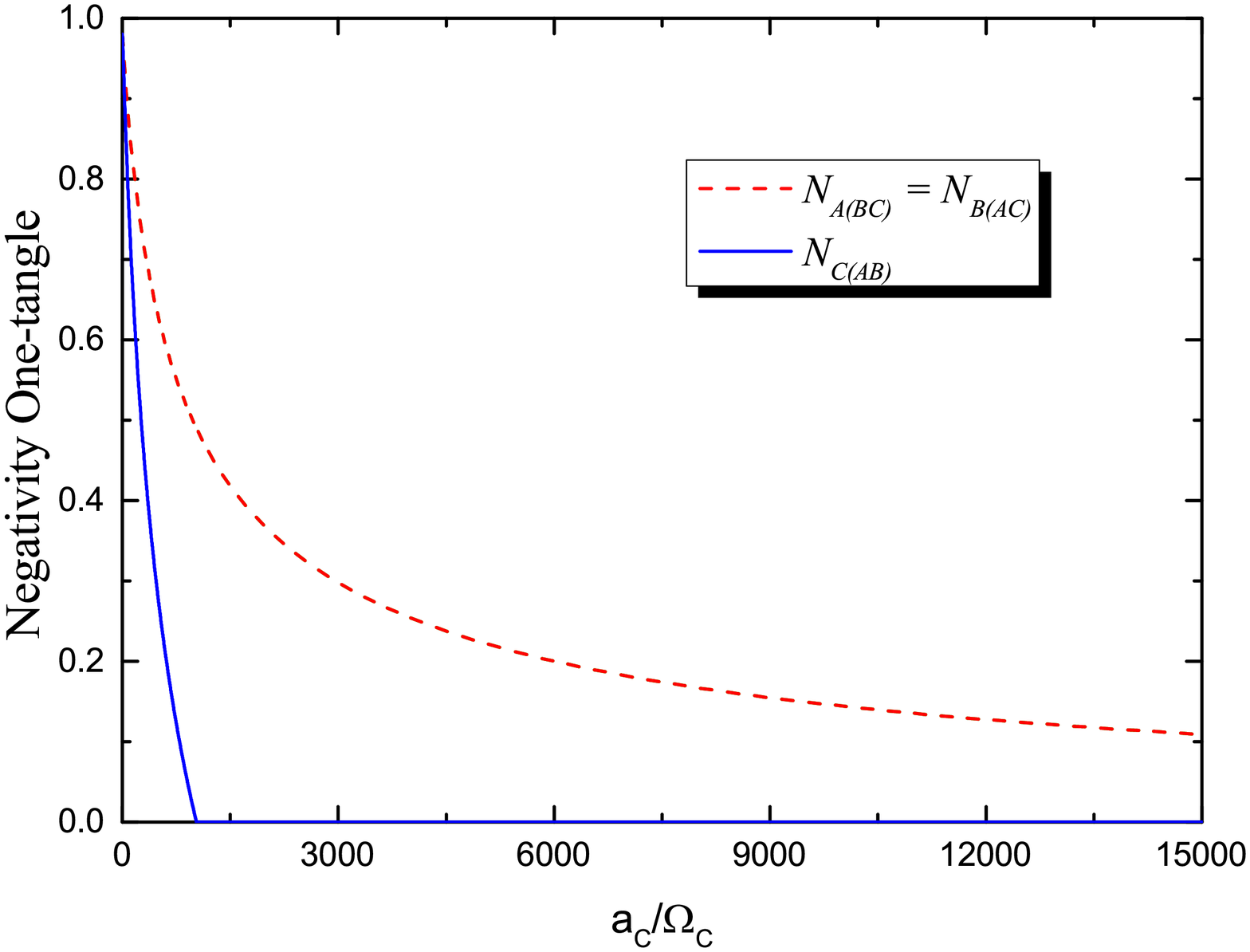}}
\subfigure[]{
\label{g2}
\includegraphics[width=0.48\textwidth]{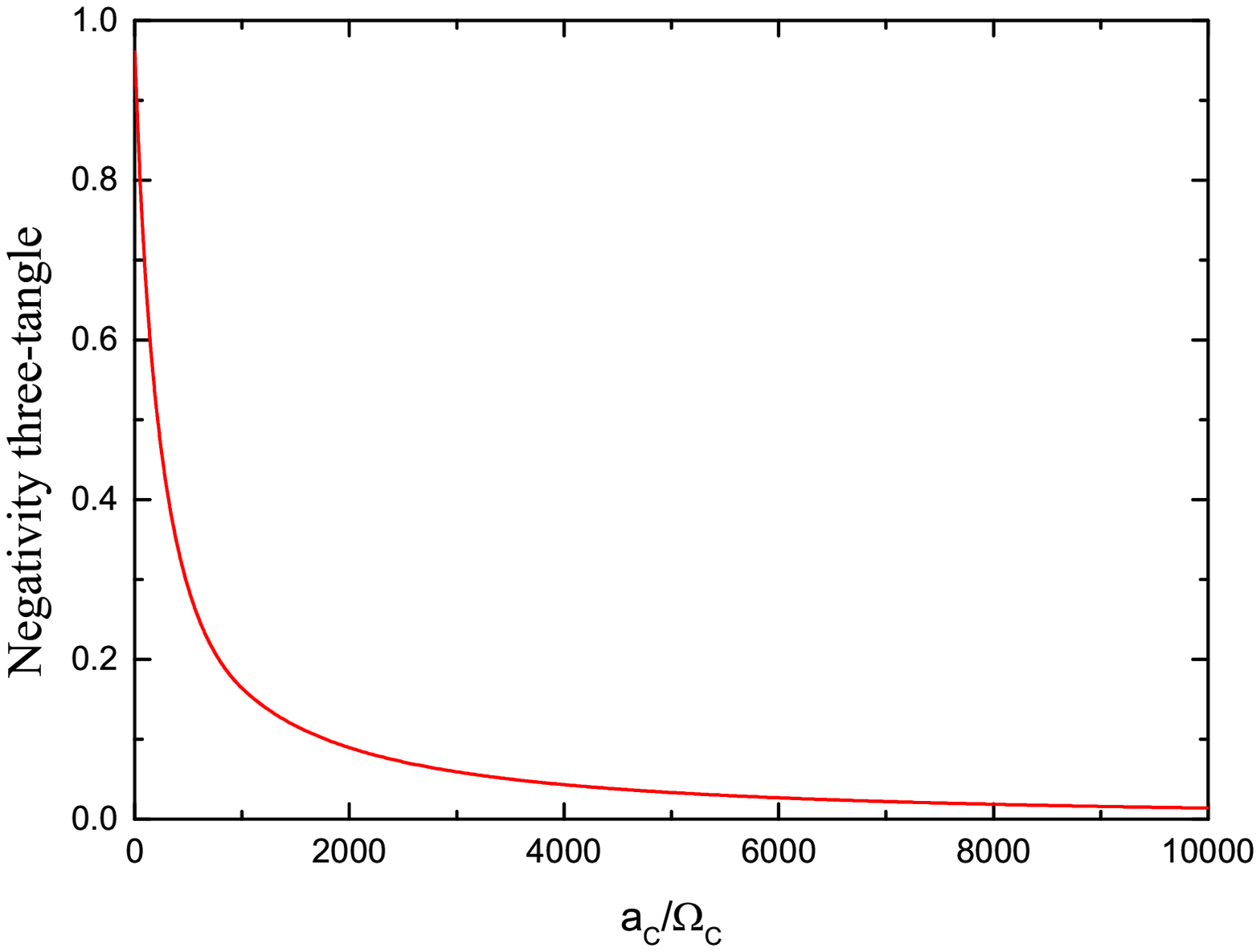}}
\caption{Dependence of (a) the one-tangles and (b) the three-tangle on the AFR of qubit $C$. Qubits $A$ and $B$ move uniformly. The qubits are in the GHZ state.}
\end{figure}

When $a_C\neq 0$, the entanglement decreases. ${\cal N}_{A(BC)}$ decreases with the increase of the acceleration-frequency ratio (AFR) $a_C/\Omega_C$ until its sudden death, as shown in Fig.~\ref{g1}. This is the phenomenon of entanglement sudden death~\cite{yu}.  The result here on one-tangles extends the previous result of bipartite entanglement~\cite{dai} from pure states to mixed states.

However, ${\cal N}_{A(BC)}={\cal N}_{B(AC)}$ approaches zero asymptotically, presumably because for these two one-tangles, $C$ is only one of the two qubits constituting a party, with the other qubit moving uniformly.

In the present case, since the two-tangles remain zero, the three-tangle turns out to be the average of the sum of the three one-tangles, as depicted in Fig.~\ref{g2}. Note that due to the sudden death of ${\cal N}_{A(BC)}$, there is a sudden change in the three-tangle, after which the three-tangle is just  $2{\cal N}_{A(BC)}/3=2{\cal N}_{B(AC)}/3$.

\subsection{$B$ and $C$ accelerating}

In the case where $B$ and $C$ accelerate while $A$ moves uniformly, we obtain
\begin{equation}
\begin{split}
{\rho _{ABC}} =& \eta_B^2\eta_C^2\sum\limits_{{n_B},{n_C}} {\frac{{{e^{ - 2\pi \left( {n_B{\Omega _B}/{a_B} + {n_C}{\Omega _C}/{a_C}} \right)}}}}{{{Z_{{n_B},{n_C}}}}}} \left[ {\left| {000} \right\rangle \left\langle {111} \right| + \left| {111} \right\rangle \left\langle {000} \right|} \right.\\
 &+ \left| {111} \right\rangle \left\langle {111} \right| + \left( {1 + \left( {{n_B} + 1} \right)\left( {{n_C} + 1} \right){{\left| {{\mu _A}} \right|}^2}{{\left| {{\mu _B}} \right|}^2}{{\left| {{\mu _C}} \right|}^2}} \right)\left| {000} \right\rangle \left\langle {000} \right|\\
 &+ \left( {{n_B} + 1} \right)\left( {{n_C} + 1} \right){\left| {{\mu _B}} \right|^2}{\left| {{\mu _C}} \right|^2}\left| {100} \right\rangle \left\langle {100} \right| + \left( {{n_B} + 1} \right){\left| {{\mu _B}} \right|^2}\left| {101} \right\rangle \left\langle {101} \right|\\
 &+ \left( {{n_C} + 1} \right){\left| {{\mu _C}} \right|^2}\left| {110} \right\rangle \left\langle {110} \right| + \left( {{n_B}{n_C}{{\left| {{\mu _B}} \right|}^2}{{\left| {{\mu _C}} \right|}^2} + {{\left| {{\mu _A}} \right|}^2}} \right)\left| {011} \right\rangle \left\langle {011} \right|\\
 &+ \left( {{n_B}{{\left| {{\mu _B}} \right|}^2} + \left( {{n_C} + 1} \right){{\left| \mu _A \right|}^2}{{\left| {{\mu _C}} \right|}^2}} \right)\left| {010} \right\rangle \left\langle {010} \right|\\
 &+ \left. {\left( {{n_C}{{\left| {{\mu _C}} \right|}^2} + \left( {{n_B} + 1} \right){{\left| \mu _A \right|}^2}{{\left| {{\mu _B}} \right|}^2}} \right)\left| {001} \right\rangle \left\langle {001} \right|} \right],
\end{split}
\end{equation}
where
\begin{equation}
\begin{split}
{Z_{{n_B},{n_C}}} =& 2 + |\mu _A|^2 + (2{n_B}+1)|\mu _B|^2 + (\frac{3}{2}{n_C}+1)|\mu _C|^2 \\
&+ \left( {{n_B} + 1} \right)|\mu _A|^2|\mu _B|^2 + \left[ n_B n_C + \left( {{n_B} + 1} \right)\left( {{n_C} + 1} \right)\right]|\mu _B|^2|\mu _C|^2
 \\
 &+\left( {{n_C} + 1} \right)|\mu _A|^2|\mu _C|^2 +  \left( {{n_B} + 1} \right)\left( {{n_C} + 1} \right)|\mu _A|^2|\mu _B|^2|\mu _C|^2.
\end{split}
\end{equation}

\begin{figure}

\centering
\subfigure[]{
\label{g3}
\includegraphics[width=0.45\textwidth]{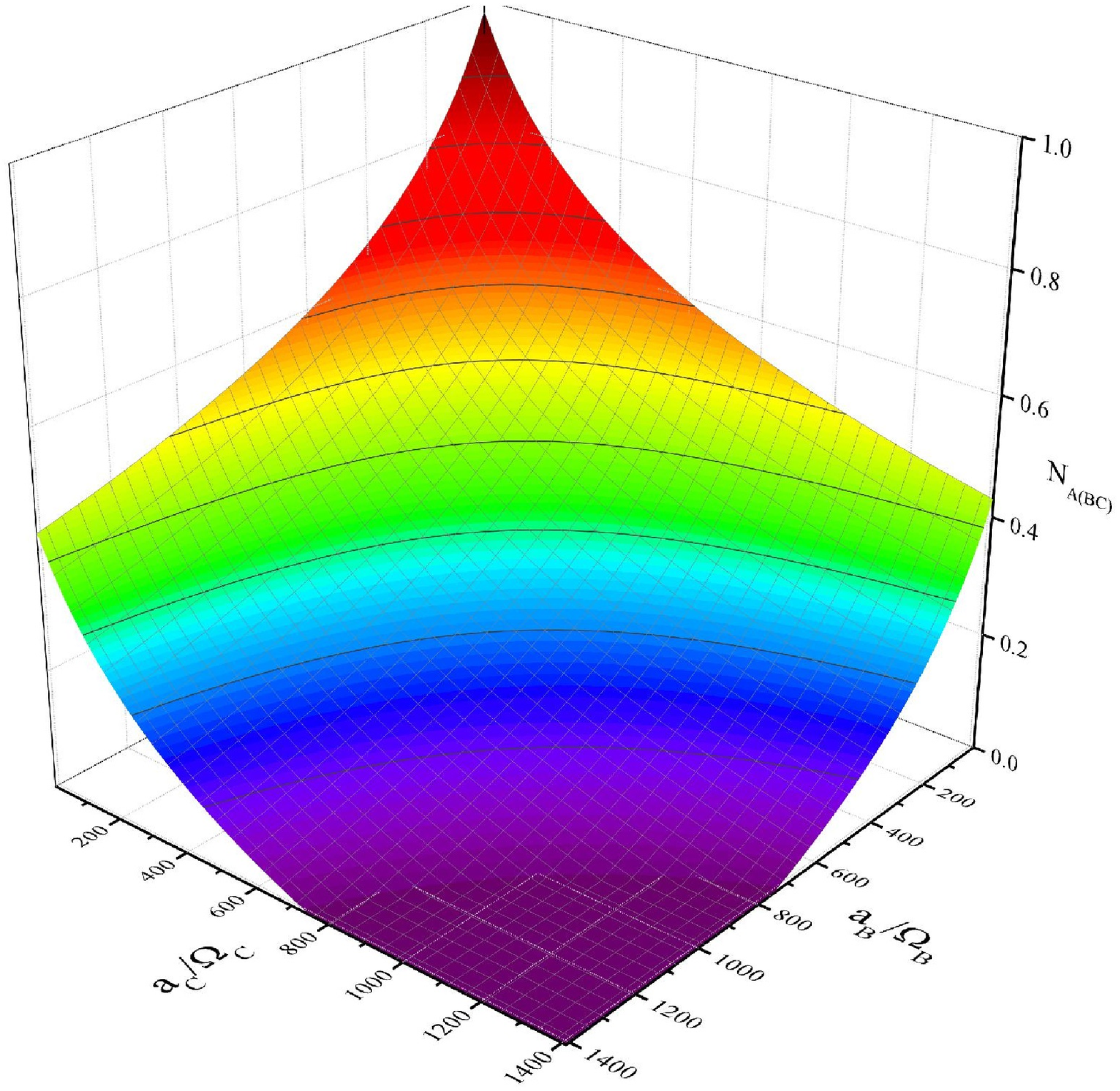}}\\
\subfigure[]{
\label{g4}
\includegraphics[width=0.45\textwidth]{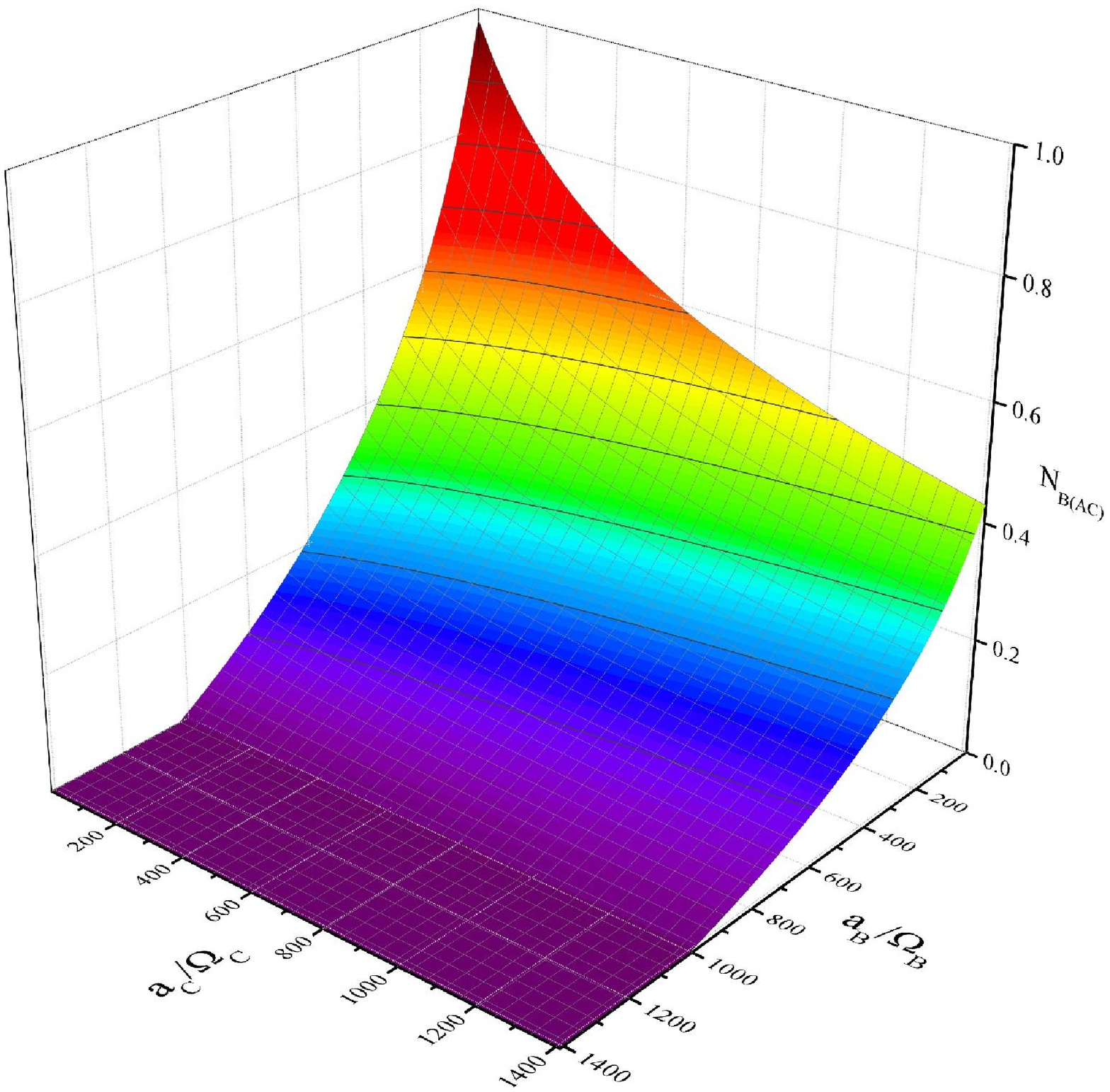}}
\subfigure[]{
\label{g5}
\includegraphics[width=0.45\textwidth]{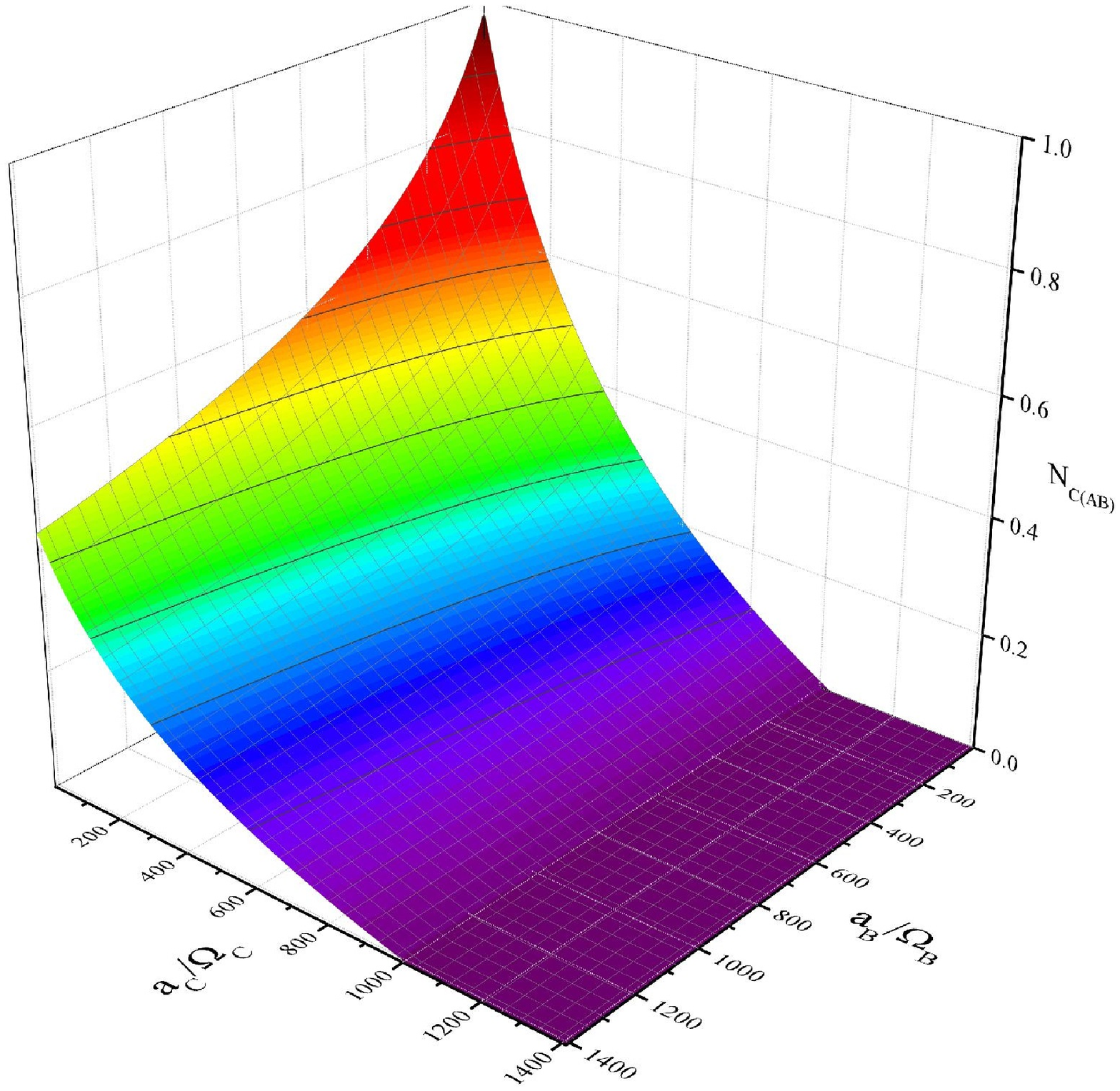}}
\caption{\label{g3to5}Dependence of the one-tangles (a) ${\cal N}_{A(BC)}$, (b) ${\cal N}_{B(AC)}$, and (c) ${\cal N}_{C(AB)}$ on the AFRs of qubits $B$ and $C$. Qubit $A$ moves uniformly. The qubits are in the GHZ state.}
\end{figure}

The dependence of ${\cal N}_{A(BC)}$ on $a_B/\Omega_B$ and $a_C/\Omega_C$ is symmetric, as shown in Fig.~\ref{g3}.  When one of the accelerations is zero, ${\cal N}_{A(BC)}$ decreases   towards zero asymptotically,  as discussed in the preceding subsection. When both are nonzero, ${\cal N}_{A(BC)}$ decreases quickly towards zero, reaching sudden death,  at finite values of the two AFRs.

As shown in Fig.~\ref{g4}, ${\cal N}_{B(AC)}$ strongly depends on $a_B/\Omega_B$ and suddenly dies at a finite value of $a_B/\Omega_B$, while it depends on $a_C/\Omega_C$ weakly, especially when $a_B/\Omega_B$ is so large that ${\cal N}_{B(AC)}$ is close to zero. This is because $B$ is one party by itself, while $C$ is only one of the two qubits constituting the other party. ${\cal N}_{C(AB)}$ can be obtained from ${\cal N}_{B(CA)}$ by exchanging $B$ and $C$, as shown in Fig.~\ref{g5}.

\begin{figure}

\centering
\subfigure[]{
\label{g6}
\includegraphics[width=0.48\textwidth]{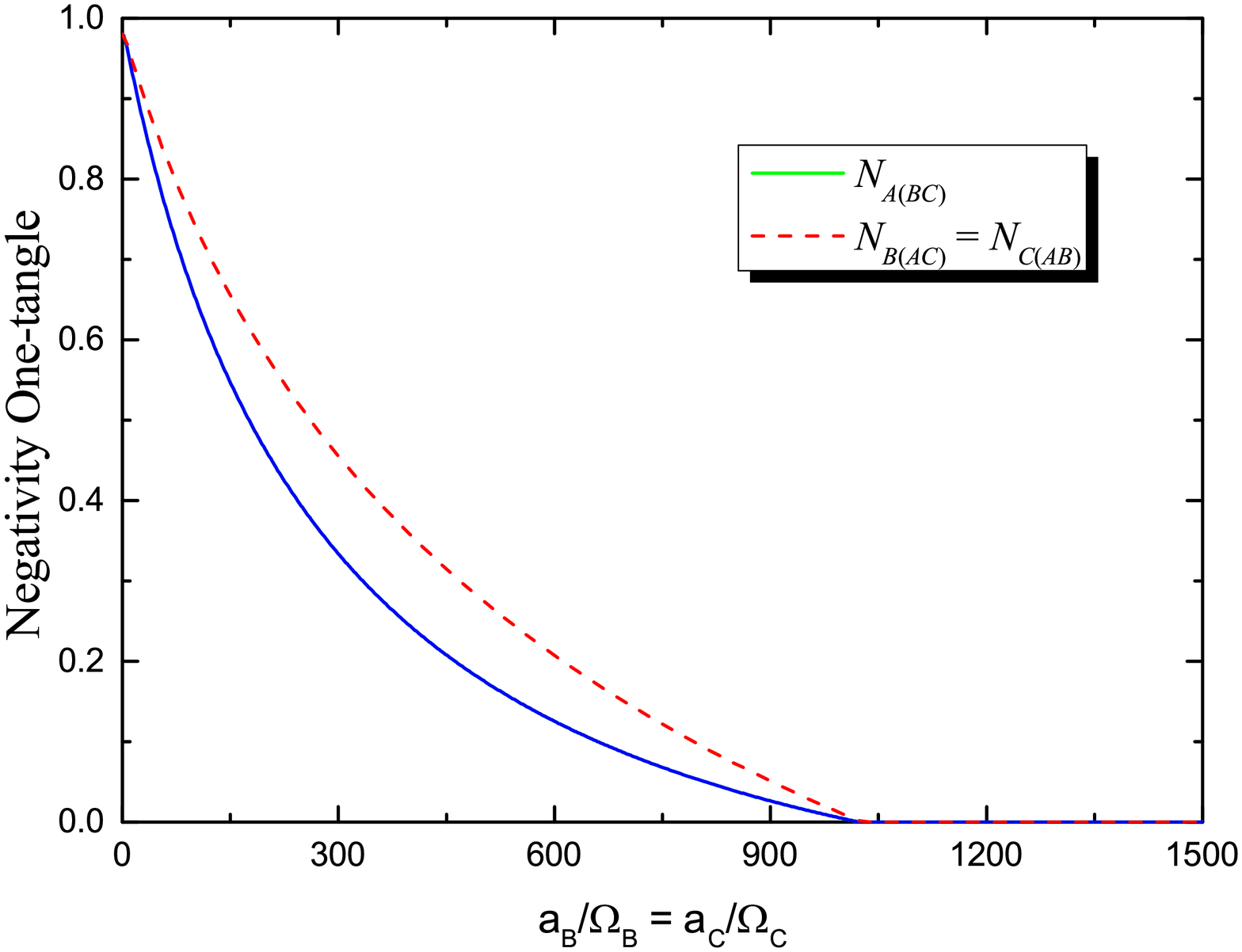}}
\subfigure[]{
\label{g7}
\includegraphics[width=0.48\textwidth]{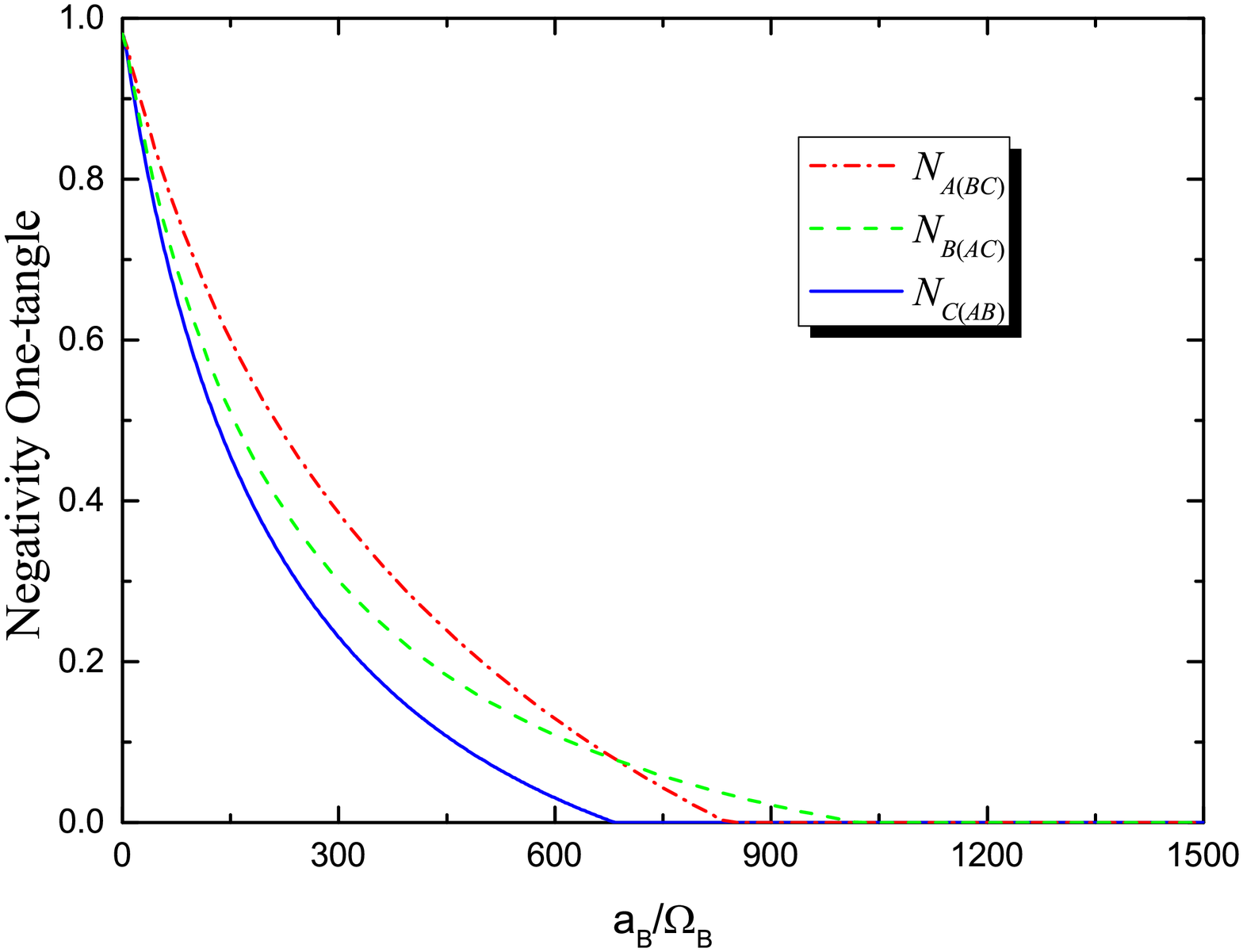}}
\subfigure[]{
\label{g8}
\includegraphics[width=0.48\textwidth]{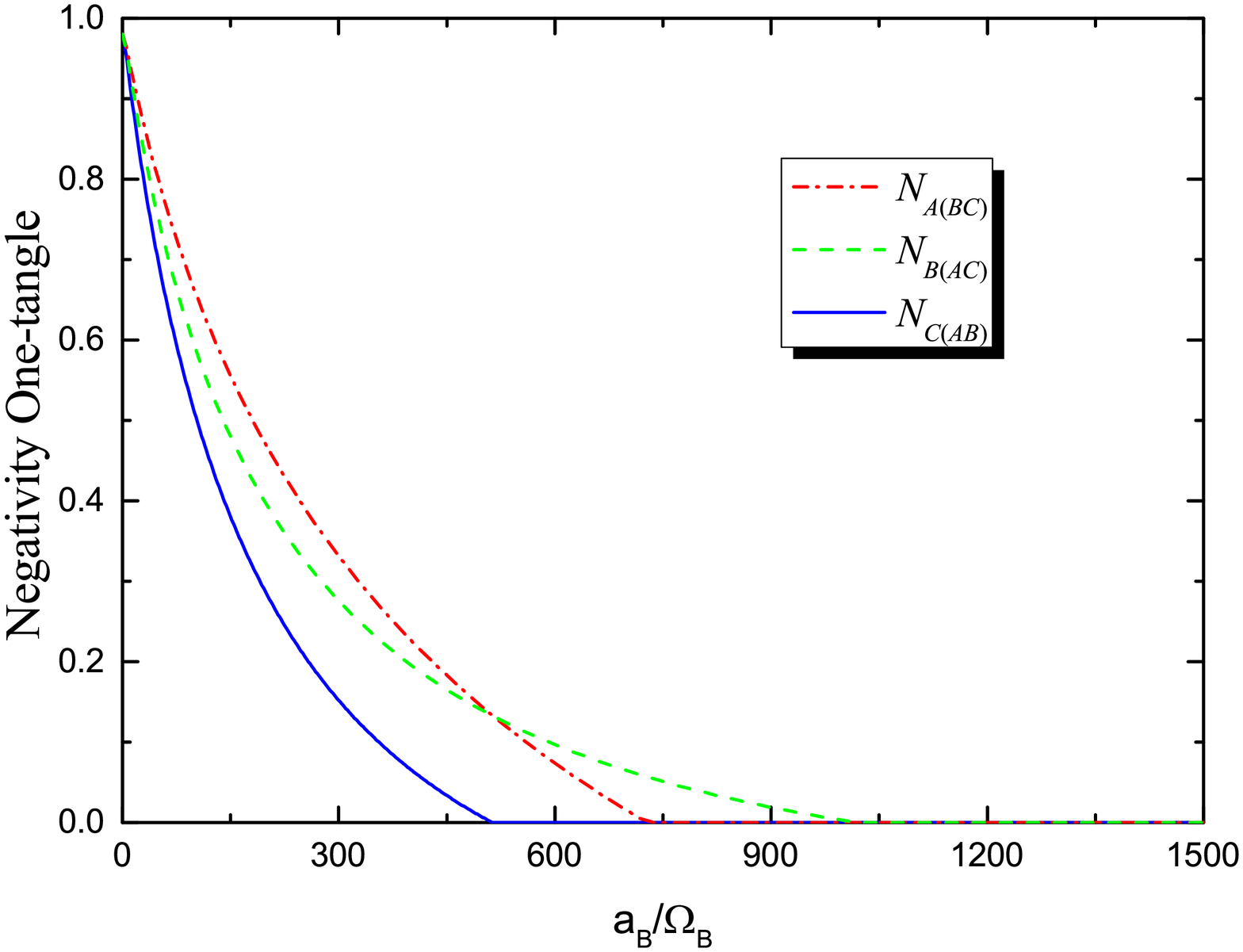}}
\subfigure[]{
\label{g13}
\includegraphics[width=0.48\textwidth]{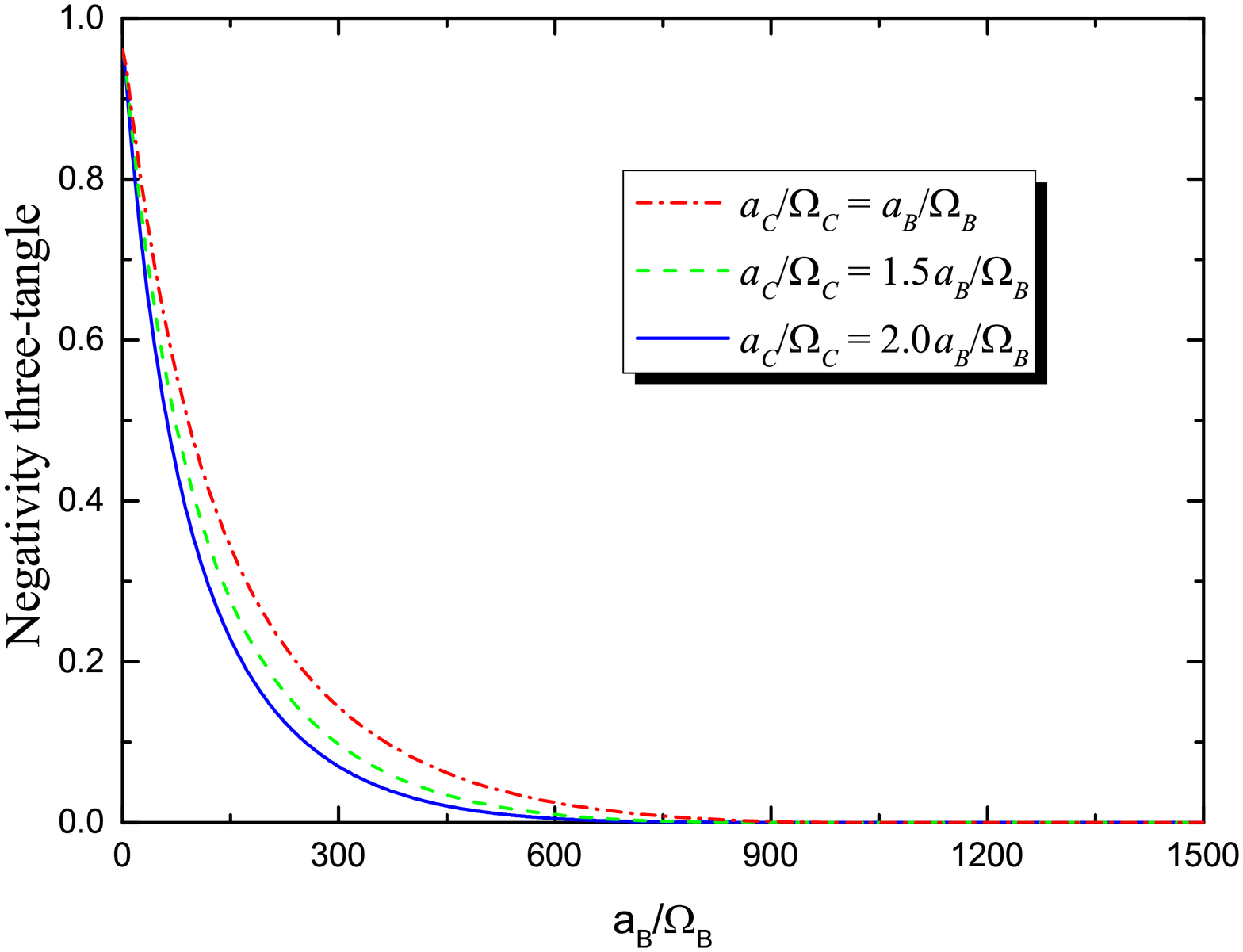}}
\caption{Dependence of the one-tangles on the AFRs of qubits $B$ and $C$ in the case that (a) $a_C/\Omega_C= a_B/\Omega_B$, (b) $a_C/\Omega_C= 1.5a_B/\Omega_B$, and (c) $a_C/\Omega_C= 2a_B/\Omega_B$. (d) The three-tangle in these cases. Qubit $A$ moves uniformly. The qubits are in the GHZ state.}
\end{figure}

We now look at some two-dimensional (2D) cross sections of the three-dimensional (3D) plots in Fig.~\ref{g3to5}. Figure~\ref{g6} is for $a_B/\Omega_B=a_C/\Omega_C$, and hence ${\cal N}_{B(AC)}={\cal N}_{C(AB)}$. Among the three one-tangles, ${\cal N}_{A(BC)}$ is the smallest, presumably because $B$ and $C$ constituting the party ($BC$) both accelerate.  The three one-tangles die at the same value of $a_B/\Omega_B=a_C/\Omega_C$.

Figure~\ref{g7} is for $a_C/\Omega_C= 1.5a_B/\Omega_B$, while Fig.~\ref{g8} is for $a_C/\Omega_C=2a_B/\Omega_B$. The values of $a_B/\Omega_B$ for the sudden death of ${\cal N}_{B(AC)}$ in these two cases are close to that in Fig.~\ref{g6}, as ${\cal N}_{B(AC)}$ is mainly determined by $a_B/\Omega_B$ when it is close to zero.  On the other hand, ${\cal N}_{C(AB)}$ dies first, as $a_C/\Omega_C$ is larger than $a_B/\Omega_B$ in these cross sections, while $a_A/\Omega_A=0$. When $a_B/\Omega_B$ is less than the value at which   ${\cal N}_{C(AB)}$ suddenly dies, ${\cal N}_{B(AC)} < {\cal N}_{A(BC)}$ as $a_B/\Omega_B > a_A/\Omega_a=0$. When ${\cal N}_{C(AB)}$ suddenly dies, ${\cal N}_{B(AC)} = {\cal N}_{A(BC)}$. When $a_B/\Omega_B$ is larger than the value for the death of ${\cal N}_{C(AB)}$, $ {\cal N}_{A(BC)} <  {\cal N}_{B(AC)}$, until ${\cal N}_{B(AC)}$  dies at a larger value of $a_B/\Omega_B$. The above-mentioned feature in the case of $a_B/\Omega_B=a_C/\Omega_C$ that the three one-tangles die at the same value of $a_B/\Omega_B$ is a special case, because when it is constrained that ${\cal N}_{C(AB)}={\cal N}_{B(AC)}$, the sudden death of ${\cal N}_{C(AB)}$ implies that of ${\cal N}_{B(AC)}$, as the two are equal. On the other hand, when ${\cal N}_{C(AB)}$ suddenly dies, there must be ${\cal N}_{A(BC)}= {\cal N}_{B(AC)}$, and hence the three have to suddenly die altogether; in other words, the only option for ${\cal N}_{A(BC)}$ to be between these two is that it dies also.

\begin{figure}

\centering
\subfigure[]{
\label{g9}
\includegraphics[width=0.48\textwidth]{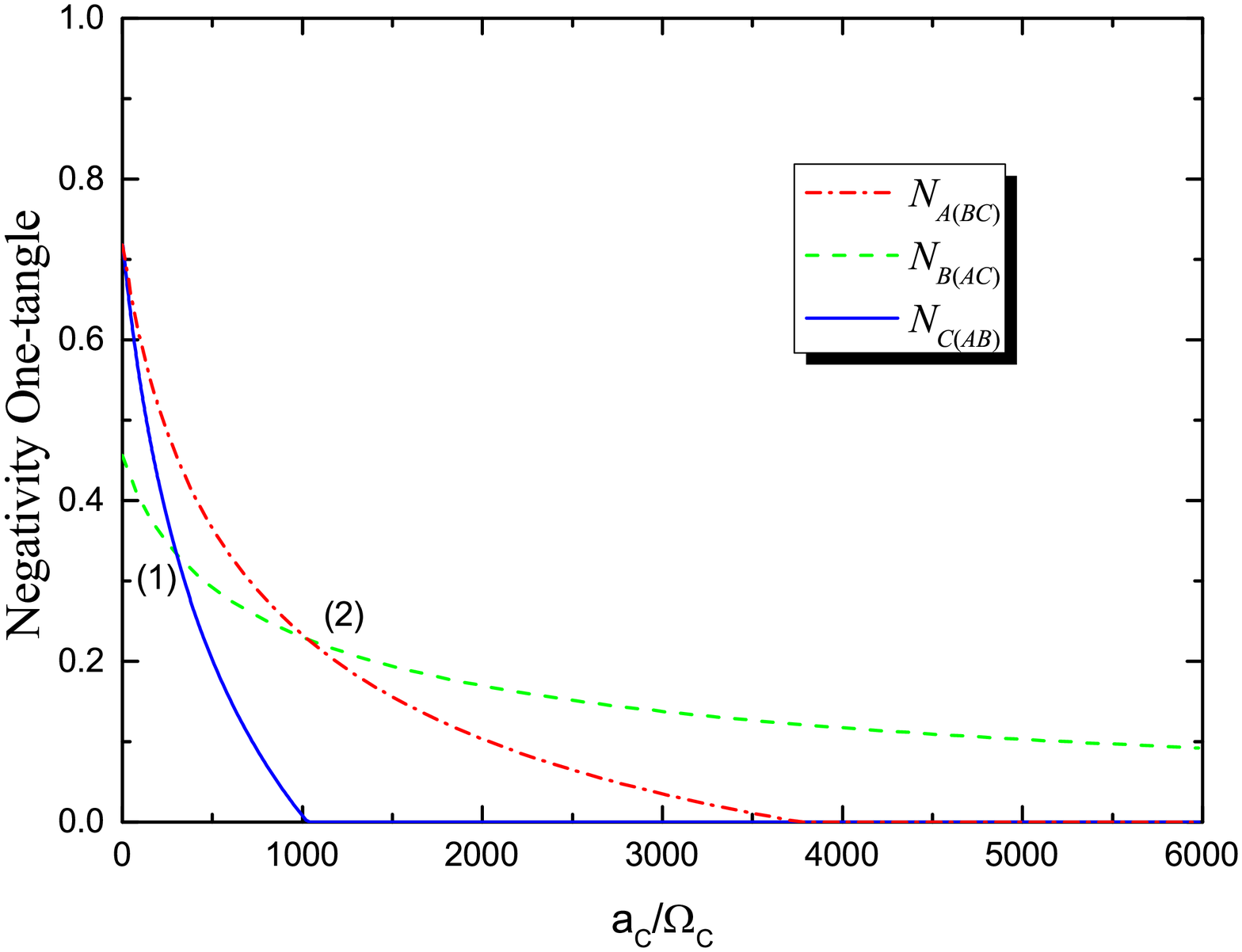}}
\subfigure[]{
\label{g10}
\includegraphics[width=0.48\textwidth]{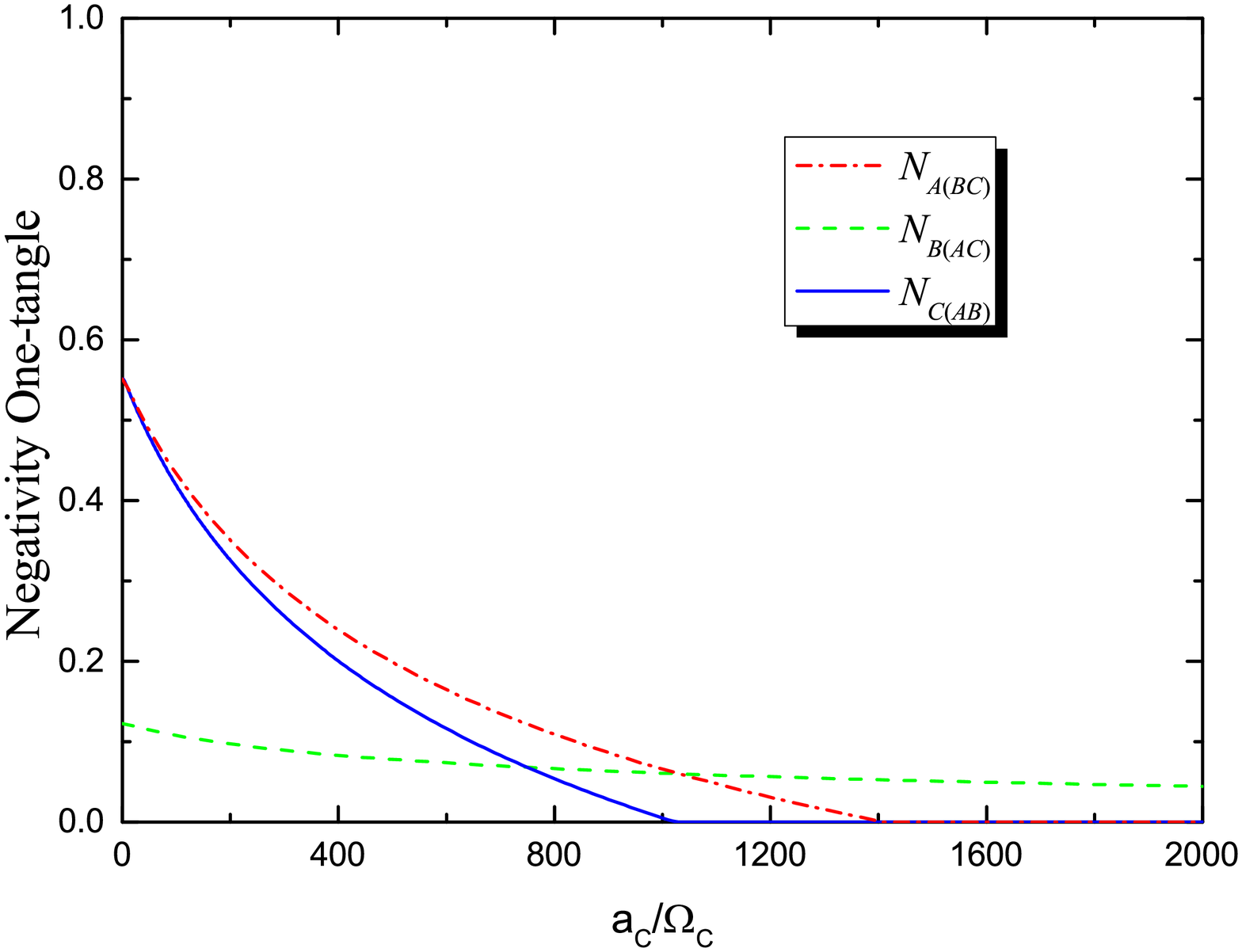}}
\subfigure[]{
\label{g11}
\includegraphics[width=0.48\textwidth]{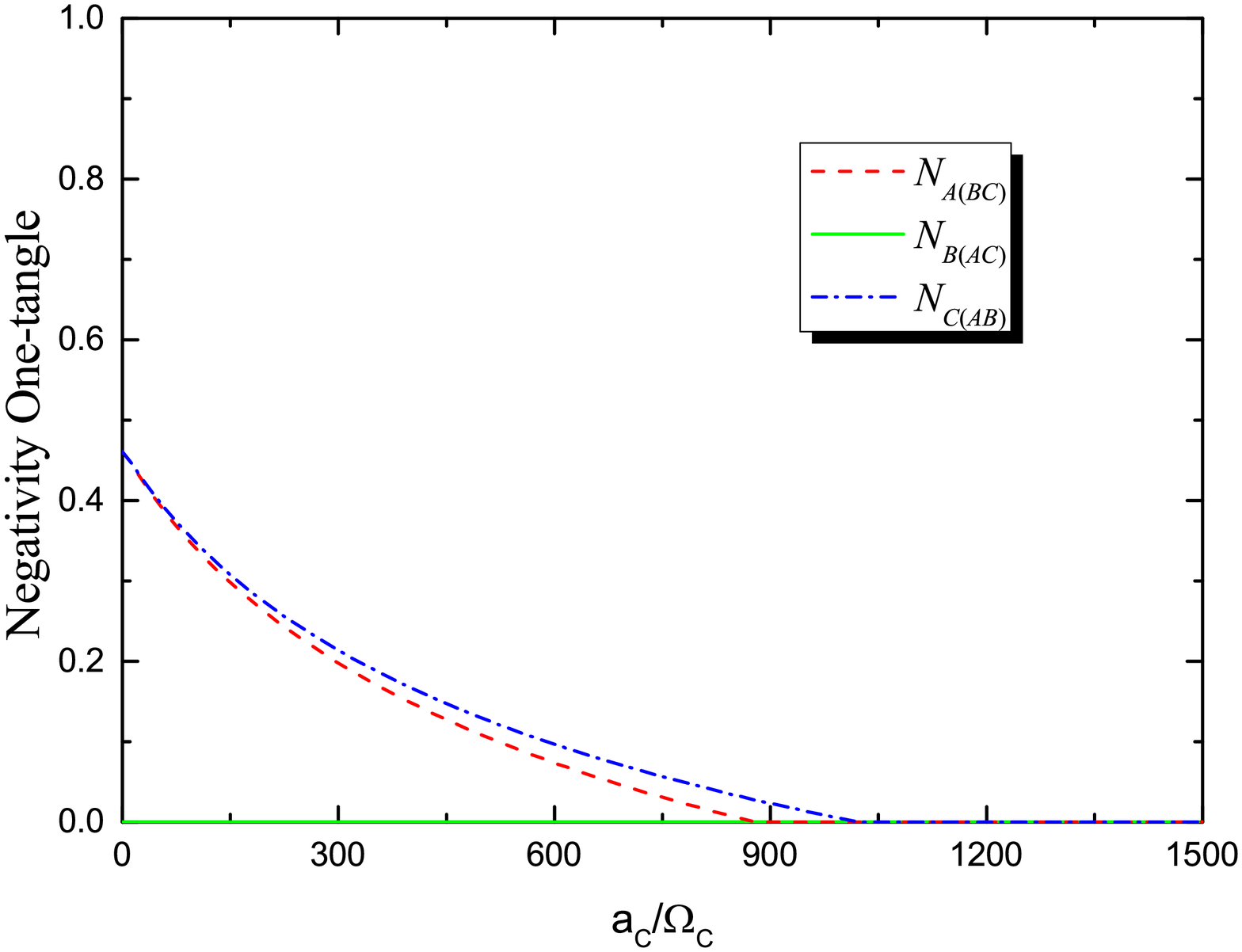}}
\subfigure[]{
\label{g14}
\includegraphics[width=0.48\textwidth]{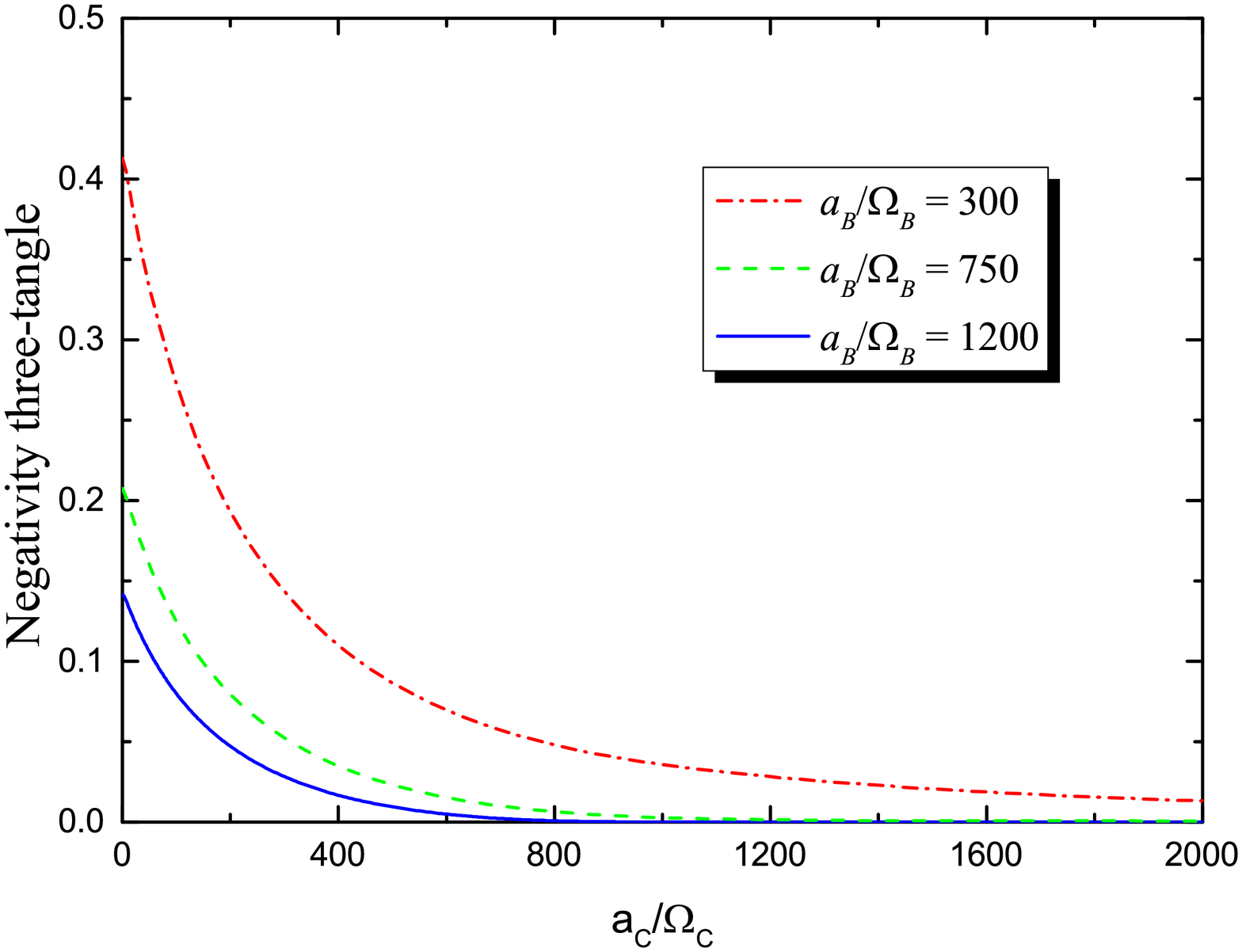}}
\caption{Dependence of the one-tangles on the AFR of qubit $C$ in
the case that (a) $a_B/\Omega_B = 300$, (b) $a_B/\Omega_B = 750$, and (c) $a_B/\Omega_B = 1200$. (d) The three-tangle in these cases. Qubit $A$ moves uniformly. The qubits are in the GHZ state.}
\end{figure}

We have also examined several cases of given values of $a_B/\Omega_B$, as shown in Figs.~\ref{g9}, \ref{g10}, and \ref{g11}, with $a_B/\Omega_B=300$, $750$, and $1200$ respectively. For $a_C/\Omega_C < a_B/\Omega_B$, ${\cal N}_{B(AC)}<{\cal N}_{C(AB)} < {\cal N}_{A(BC)}$. For  $a_C/\Omega_C > a_B/\Omega_B$ up to the death of ${\cal N}_{C(AB)}$,  ${\cal N}_{C(AB)} <   {\cal N}_{B(AC)}<{\cal N}_{A(BC)}$. After the death of ${\cal N}_{C(AB)}$, ${\cal N}_{A(BC)} < {\cal N}_{B(AC)} $.

\begin{figure}
\centering
\includegraphics[width=0.6\textwidth]{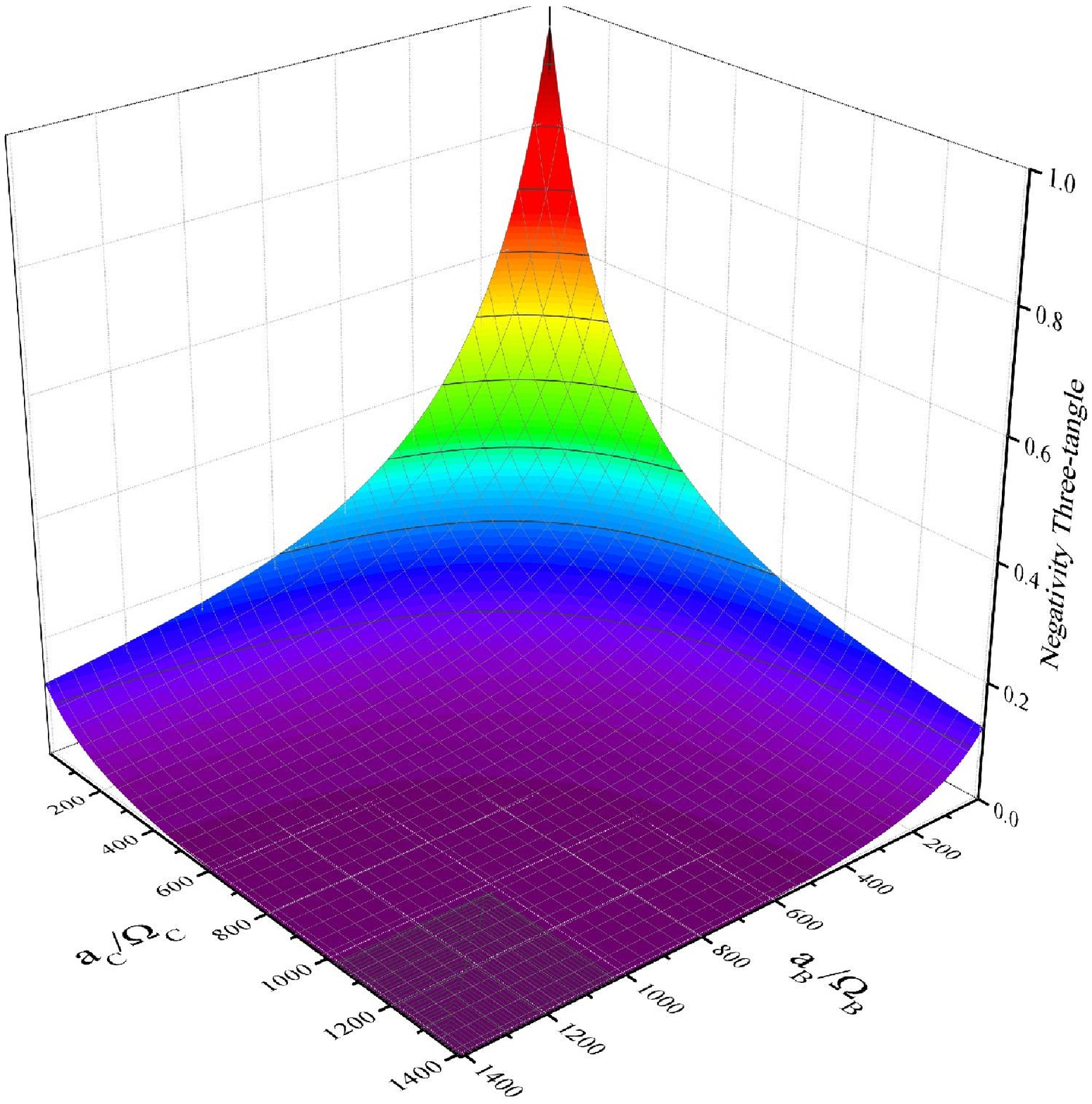}
\caption{\label{g12} Dependence of the three-tangle on the AFRs of qubits $B$ and $C$. Qubit $A$ moves uniformly. The qubits are in the GHZ state.}
\end{figure}

The three-tangle is shown in the 3D plot in Fig.~\ref{g12}. The condition of the three-tangle sudden death is that each of the two nonzero AFRs should be large enough. The reason is that the three-tangle is now the average of the squares of one-tangles. Hence it suddenly dies when all one-tangles suddenly die.  Some 2D cross sections of Fig.~\ref{g12} are shown in Fig.~\ref{g13} and Fig.~\ref{g14}. Note in the cases that $a_B/\Omega_B=300$ and $a_B/\Omega_B=750$ while $a_A=0$, as shown in Fig.~\ref{g14}, the three-tangle only approaches zero asymptotically with the increase of  $a_C/\Omega_C$, since the values of $a_B/\Omega_B$ are not large enough.

\subsection{$A$, $B$, and $C$ all accelerating}

In the case that all qubits accelerate, we have
\begin{eqnarray}
{\rho _{ABC}} &=& \frac{1}{2}\eta_A^2\eta_B^2\eta_C^2\sum\limits_{{n_A},{n_B},{n_C}} {\frac{{{e^{ - 2\pi \left( {{n_A}{\Omega_A}/{a_A} + {n_B}{\Omega_B}/{a_B} + {n_C}{\Omega_C}/{a_C}} \right)}}}}{{{Z_{{n_A},{n_B},{n_C}}}}}  } \nonumber\\
&&\times\left[ {\left( {1 + \left( {{n_A} + 1} \right)\left( {{n_B} + 1} \right)\left( {{n_C} + 1} \right)|\mu _A|^2|\mu_B|^2|\mu_C|^2} \right)\left| {000} \right\rangle \left\langle {000} \right|} \right.\nonumber\\
 &&+ \left( {1 + {n_A}{n_B}{n_C}|\mu_A|^2|\mu_B|^2|\mu_C|^2} \right)\left| {111} \right\rangle \left\langle {111} \right| + \left| {111} \right\rangle \left\langle {000} \right|\nonumber\\
 &&+ \left| {000} \right\rangle \left\langle {111} \right| + \left( {{n_A}|\mu_A|^2 + \left( {{n_B} + 1} \right)\left( {{n_C} + 1} \right)|\mu _B|^2|\mu_C|^2} \right)\left| {100} \right\rangle \left\langle {100} \right|\nonumber\\
 &&+ \left( {{n_B}|\mu_B|^2 + \left( {{n_A} + 1} \right)\left( {{n_C} + 1} \right)|\mu_A|^2|\mu_C|^2} \right)\left| {010} \right\rangle \left\langle {010} \right|\nonumber\\
 &&+ \left( {{n_C}|\mu_C|^2 + \left( {{n_A} + 1} \right)\left( {{n_B} + 1} \right)|\mu_A|^2|\mu_B|^2} \right)\left| {001} \right\rangle \left\langle {001} \right|\nonumber\\
 &&+ \left( \left( {{n_A} + 1} \right)|\mu_A|^2 + {{n_B}{n_C}|\mu_B|^2|\mu_C|^2 } \right)\left| {110} \right\rangle \left\langle {110} \right|\nonumber\\
 &&+ \left( \left( {{n_B} + 1} \right)|\mu_B|^2 + {{n_A}{n_C}|\mu_A|^2|\mu_C|^2 } \right)\left| {101} \right\rangle \left\langle {101} \right|\nonumber\\
&&+\left. { \left( \left( {{n_C} + 1} \right)|\mu_C|^2 + {{n_A}{n_B}|\mu_A|^2|\mu_B|^2 } \right)\left| {011} \right\rangle \left\langle {011} \right|} \right],\nonumber\\*
\,
\end{eqnarray}
where
\begin{equation}
\begin{split}
{Z_{{n_A},{n_B},{n_C}}} =& 2 + \left[{n_A}{n_B}{n_C} + (n_A+1)(n_B+1)(n_C+1)\right]|\mu _A|^2|\mu _B|^2|\mu_C|^2\\
&+ (2{n_A}+1)|\mu_A|^2 + \left[ {n_B}{n_C} + (n_B+1)(n_C+1)\right]|\mu _B|^2|\mu_C|^2\\
&+ (2{n_B}+1)|\mu_B|^2 + \left[ {n_A}{n_C}+(n_A+1)(n_C+1)\right]|\mu _A|^2|\mu_C|^2\\
&+ (2{n_C}+1)|\mu_C|^2 + \left[ {n_A}{n_B}+(n_A +1)(n_B+1)\right]|\mu _A|^2|\mu_B|^2, \\
\end{split}
\end{equation}
${\eta_q} = \sqrt {1 - {e^{ - 2\pi {\Omega_q}/{a_q}}}}$, $(q = A, B, C)$.

\begin{figure}

\centering
\subfigure[]{
\label{g17}
\includegraphics[width=0.48\textwidth]{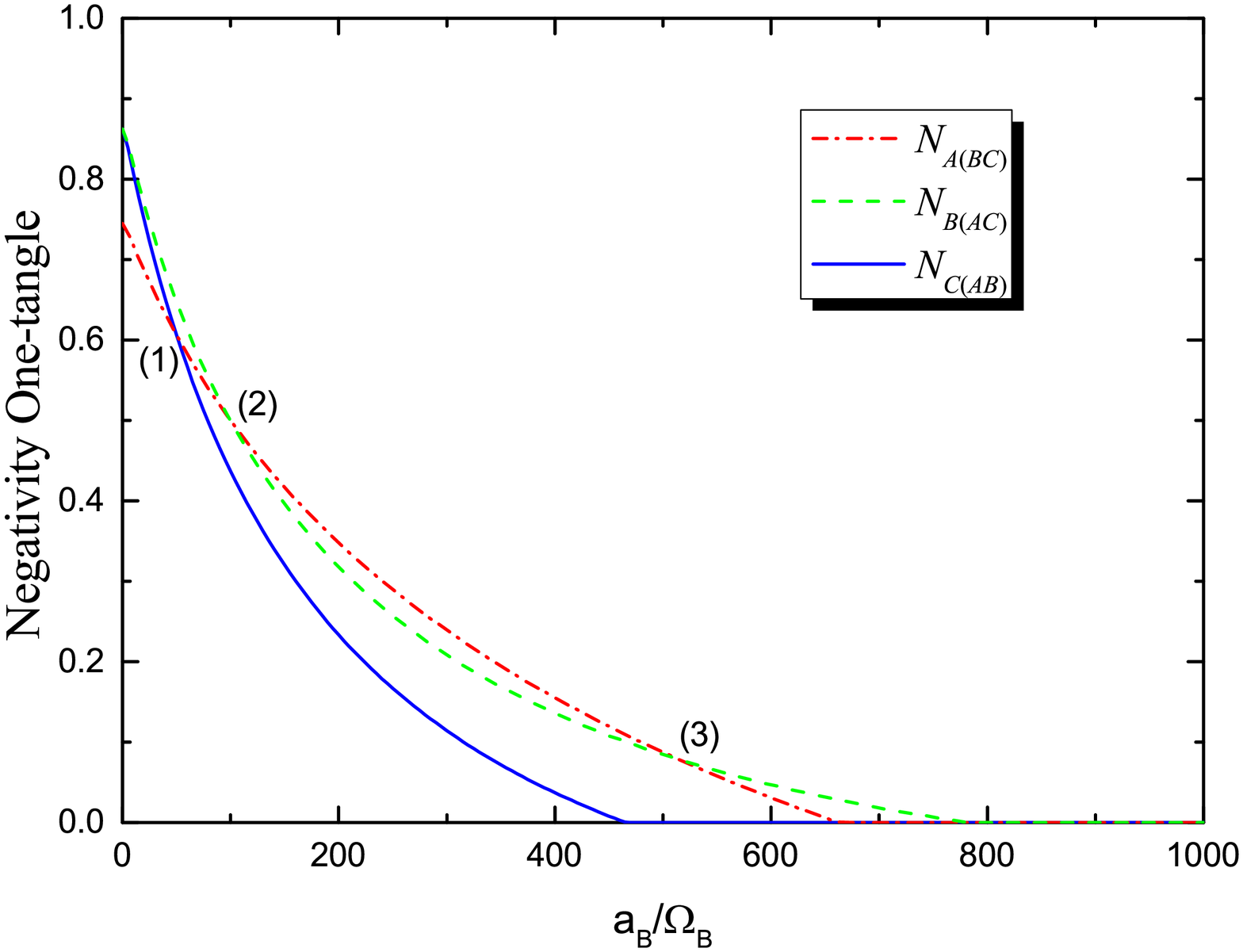}}
\subfigure[]{
\label{g18}
\includegraphics[width=0.48\textwidth]{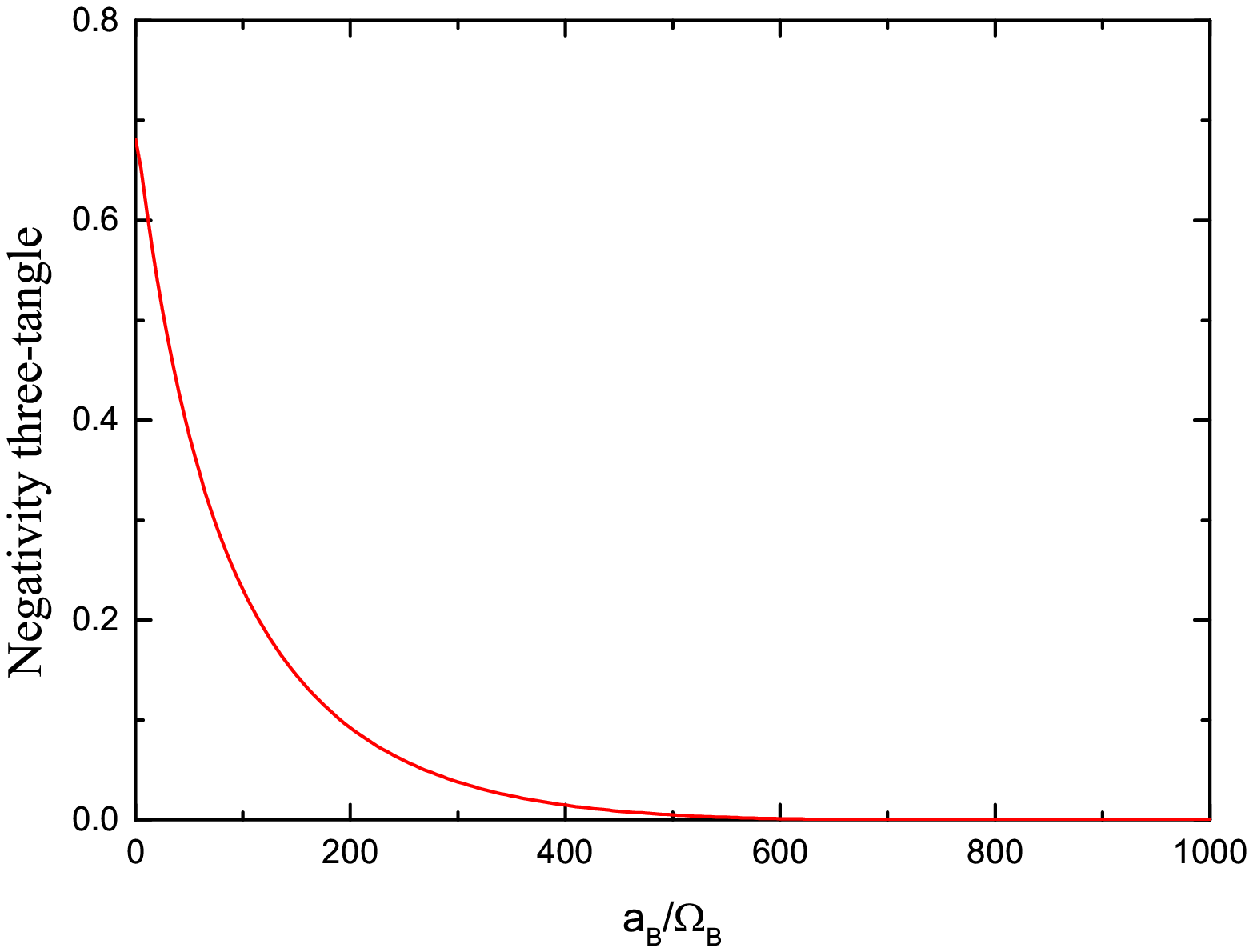}}
\caption{\label{g17to18}Dependence of (a) the one-tangles and (b) the three-tangle on the AFR of qubit $B$ in
the case that $a_A/\Omega_A=100$, $a_C/\Omega_C=2 a_B/\Omega_B$. The qubits are in the GHZ state.}
\end{figure}

Figure~\ref{g17to18} is for  $a_A/\Omega_A=100$ and  $a_C/\Omega_C=2 a_B/\Omega_B$, where  we show the three one-tangles. The plots have three intersections. Intersection (1) is at $a_A/\Omega_A = a_C/\Omega_C$ and thus ${\cal N}_{A(BC)} = {\cal N}_{C(AB)}$. Intersection (2) is at  $a_A/\Omega_A = a_B/\Omega_B$ and thus  ${\cal N}_{A(BC)} = {\cal N}_{B(AC)}$. Intersection (3) is another point where ${\cal N}_{A(BC)} = {\cal N}_{B(AC)}$. In the cases that qubit $A$ moves uniformly,  intersection (3) is at  the value of $a_B/\Omega_B$ where ${\cal N}_{C(AB)}$ suddenly dies,  as shown in Figs.~\ref{g7}, \ref{g8}, \ref{g9}, and \ref{g10}. Now the nonzero $a_A$ delays this intersection. The three-tangle suddenly dies after all three one-tangles become zero, as shown in Fig.~\ref{g18}.

\begin{figure}

\centering
\subfigure[]{
\label{g15}
\includegraphics[width=0.48\textwidth]{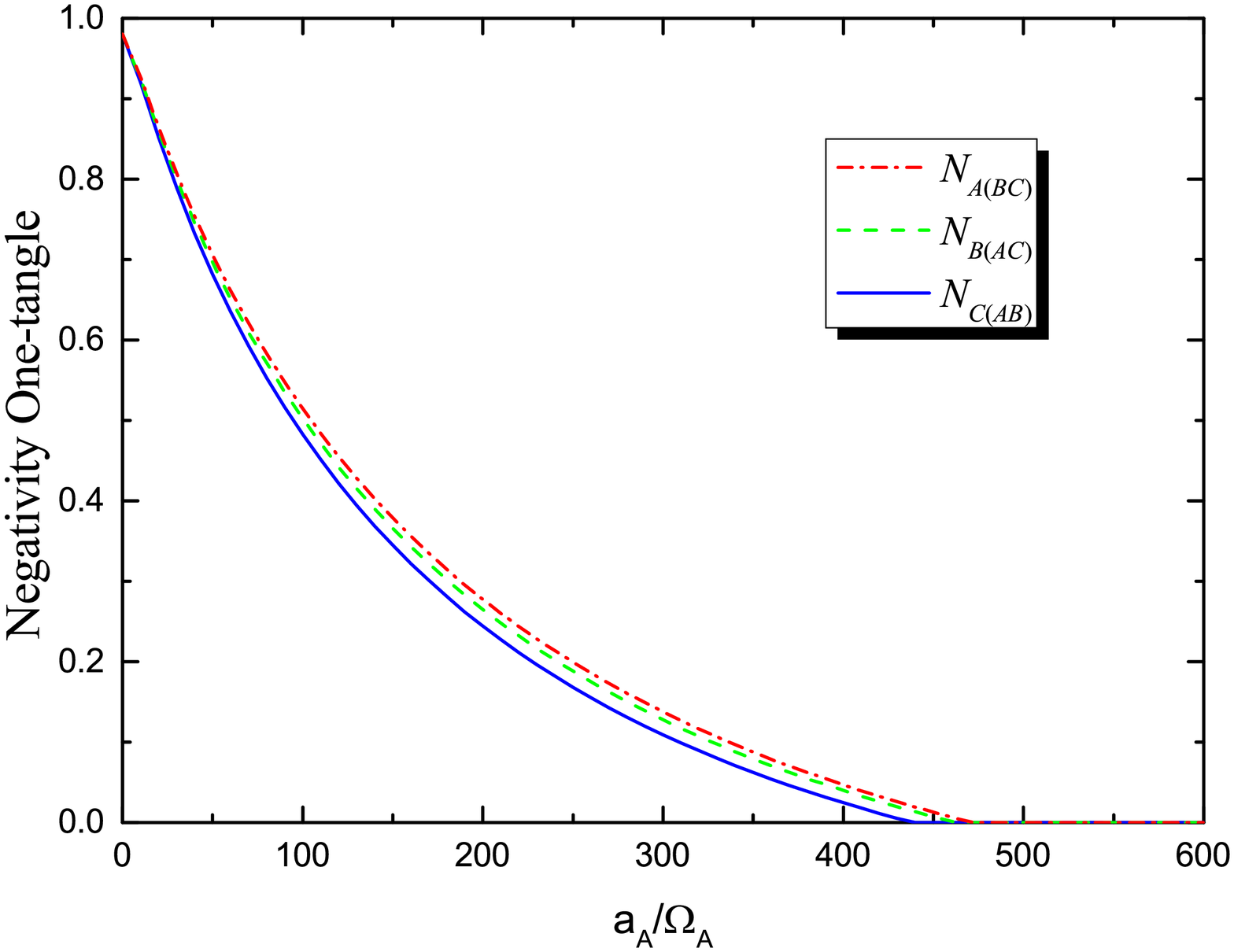}}
\subfigure[]{
\label{g16}
\includegraphics[width=0.48\textwidth]{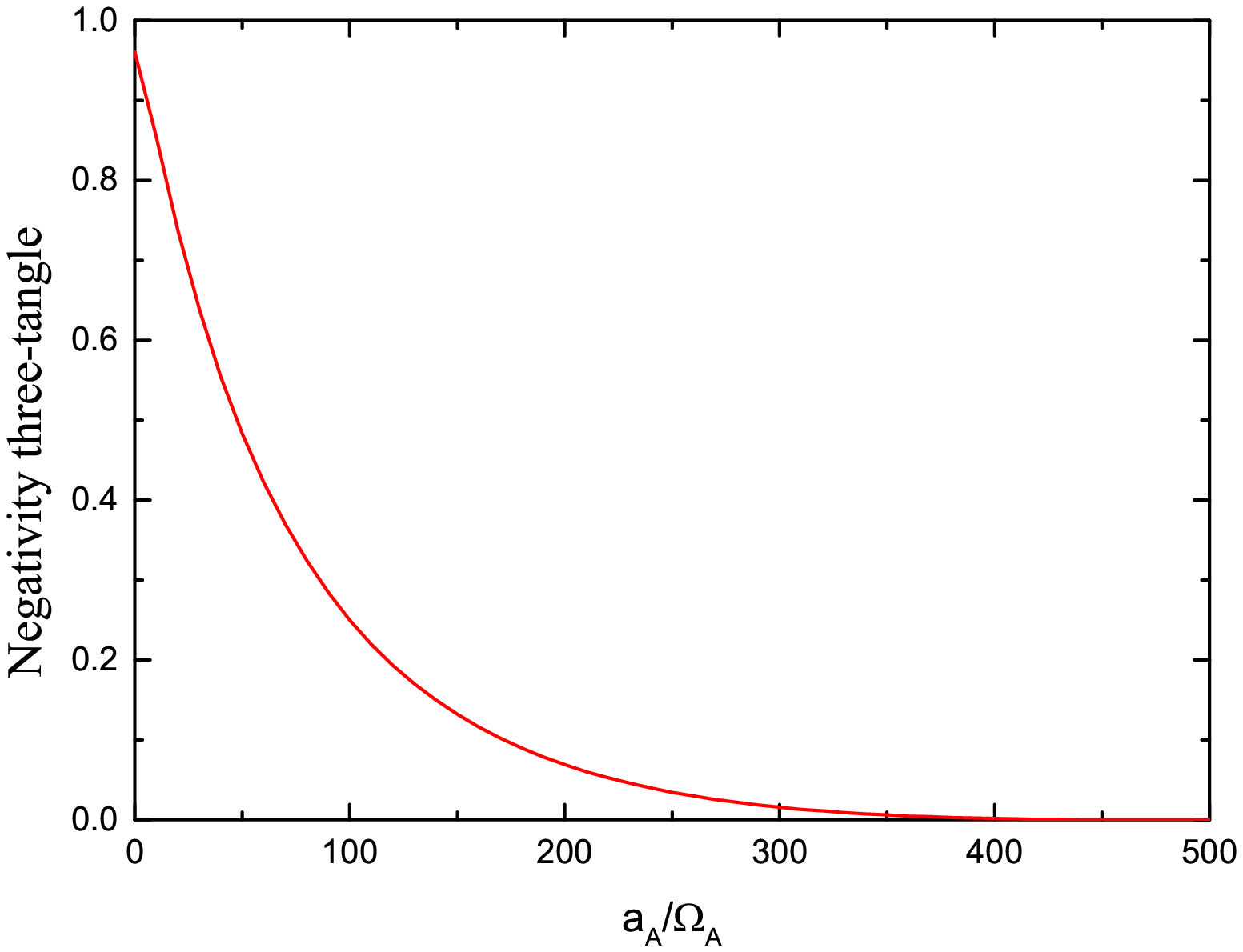}}
\caption{\label{15to16} Dependence of (a) the one-tangles and (b) the three-tangle on the AFR of qubit $A$ in
the case that $a_B/\Omega_B=1.2 a_A/\Omega_A$, $a_C/\Omega_C=1.5 a_A/\Omega_A$. The qubits are in the GHZ state.}
\end{figure}

In Fig.~\ref{15to16} we show the one-tangles and the three-tangle for $a_B/\Omega_B=1.2 a_A/\Omega_A$, $a_C/\Omega_C=1.5 a_A/\Omega_A$.
The three one-tangles suddenly die successively, and afterwards the three-tangle becomes zero. After ${\cal N}_{C(AB)}$ suddenly dies, ${\cal N}_{A(BC)}$ and ${\cal N}_{B(AC)}$ do not intersect. This is because $a_A/\Omega_A$ is so large that the would-be intersection is shifted to some value of  $a_A/\Omega_A$ larger than the value of sudden death of each of them.

\section{W state}

Now we consider the initial state to be the W state,
\begin{equation}
|\Psi_i\rangle =|\mathrm{W}\rangle.
\end{equation}
In the W state, the one-tangles are ${\cal N}_{A(BC)} = {\cal N}_{B(AC)} = {\cal N}_{C(AB)} = 2\sqrt{2}/3$. The two-tangles are nonzero now, because for the W state, tracing out one qubit does not yield a separable state. One obtains ${\cal N}_{AB} = {\cal N}_{BA} = {\cal N}_{AC} = {\cal N}_{CA} = {\cal N}_{BC} = {\cal N}_{CB} =(\sqrt{5}-1)/3 \approx 0.412$. The three-tangle is less than 1, because the two-tangles  are nonzero. One obtains $\pi =4(\sqrt{5}-1)/9  \approx 0.549$.

\subsection{$C$ accelerating}

First we consider the case that only qubit $C$ accelerates, while $A$ and $B$ move uniformly. We obtain

\begin{equation}
\begin{split}
{\rho _{ABC}} =& \eta_C^2\sum\limits_{{n_C}} {\frac{{{e^{ - 2\pi {n_C}{\Omega _C}/{a_C}}}}}{{{Z_{{n_C}}}}}} \left[ {|001\rangle \langle 010| + |001\rangle \langle 100|} \right. + |100\rangle \langle 001| + |100\rangle \langle 010|\\
 &+ |100\rangle \langle 100| + |010\rangle \langle 010| + \left( {|\mu _A|^2 + |\mu _B|^2 + \left( {{n_C} + 1} \right)|\mu _C|^2} \right)|000\rangle \langle 000|\\
 &+ |010\rangle \langle 001| + {n_C}|\mu _C|^2\left( {|101\rangle \langle 101| + |011\rangle \langle 101| + |101\rangle \langle 011| + |011\rangle \langle 011|} \right)\\
 &+\left.{|010\rangle \langle 100| + \left( {1 + {n_C}|\mu _A|^2|\mu _C|^2 + {n_C}|\mu _B|^2|\mu _C|^2} \right)|001\rangle \langle 001|} \right],
\end{split}
\end{equation}
where
\begin{equation}
{Z_{{n_C}}} = 3+|\mu_A|^2 + |\mu_B|^2 +(3{n_C}+1)|\mu_C|^2 + {n_C}|\mu_A|^2|\mu_C|^2 + {n_C}|\mu_B|^2|\mu_C|^2.
\end{equation}

\begin{figure}

\centering
\subfigure[]{
\label{w1}
\includegraphics[width=0.48\textwidth]{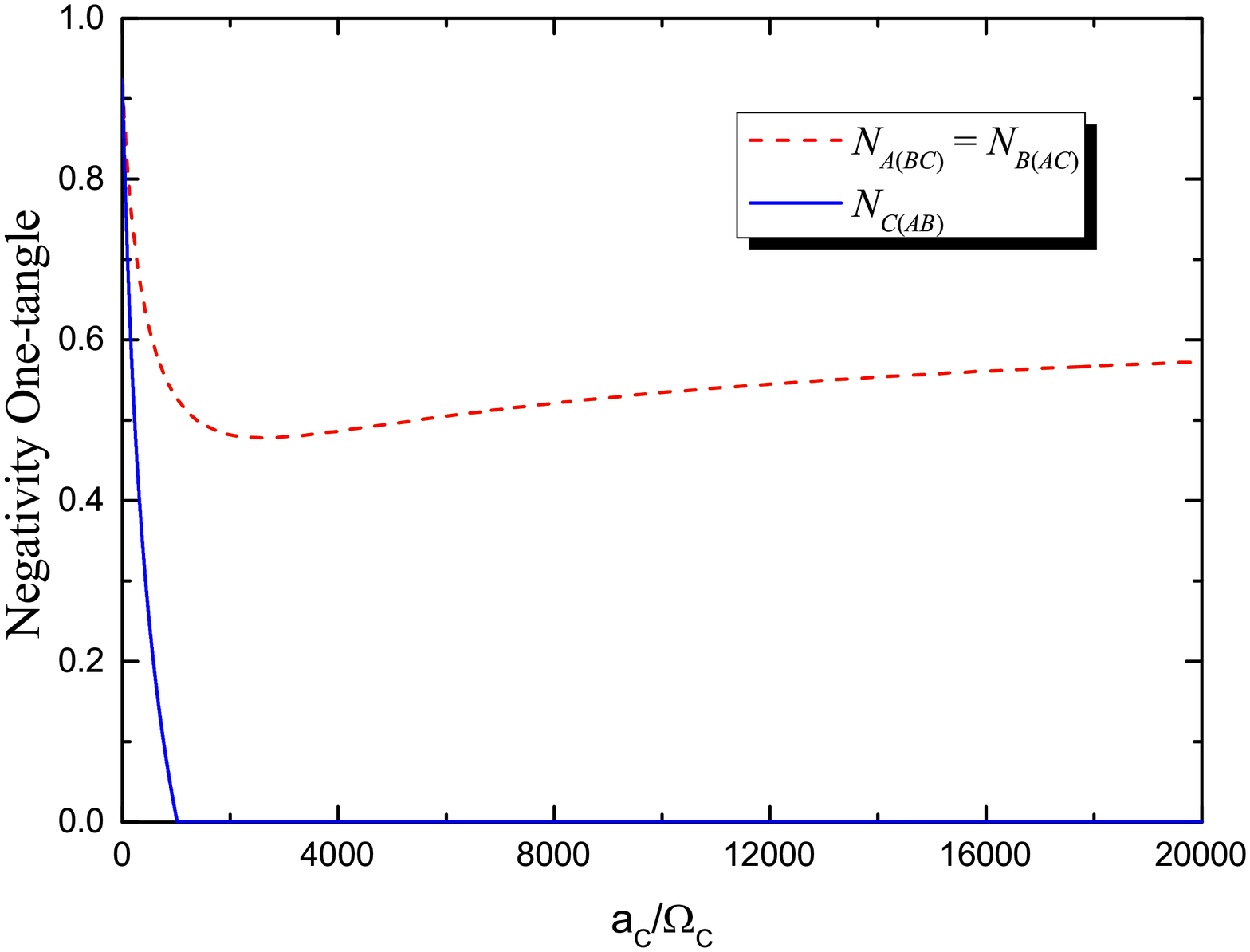}}
\subfigure[]{
\label{w2}
\includegraphics[width=0.48\textwidth]{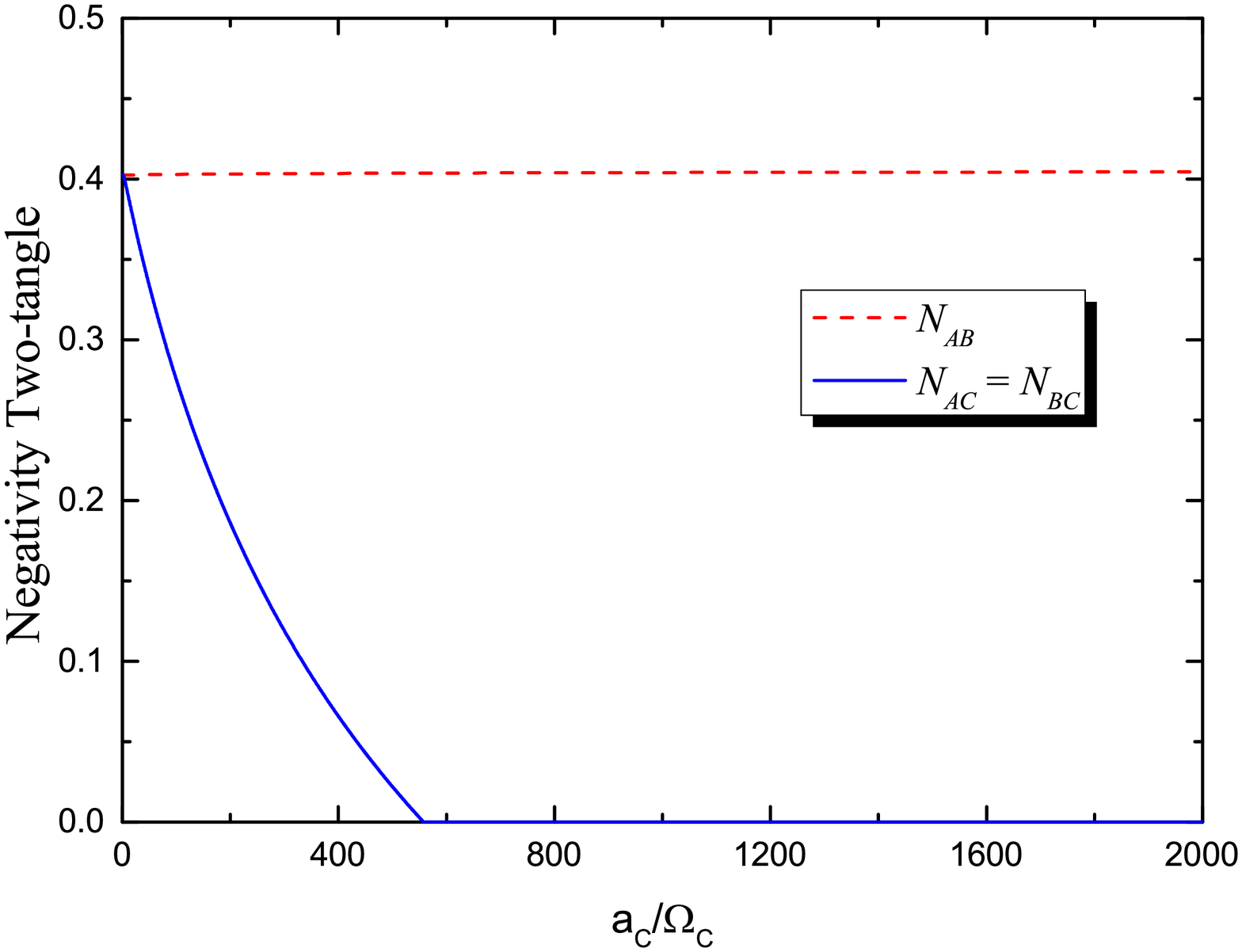}}
\subfigure[]{
\label{w3}
\includegraphics[width=0.48\textwidth]{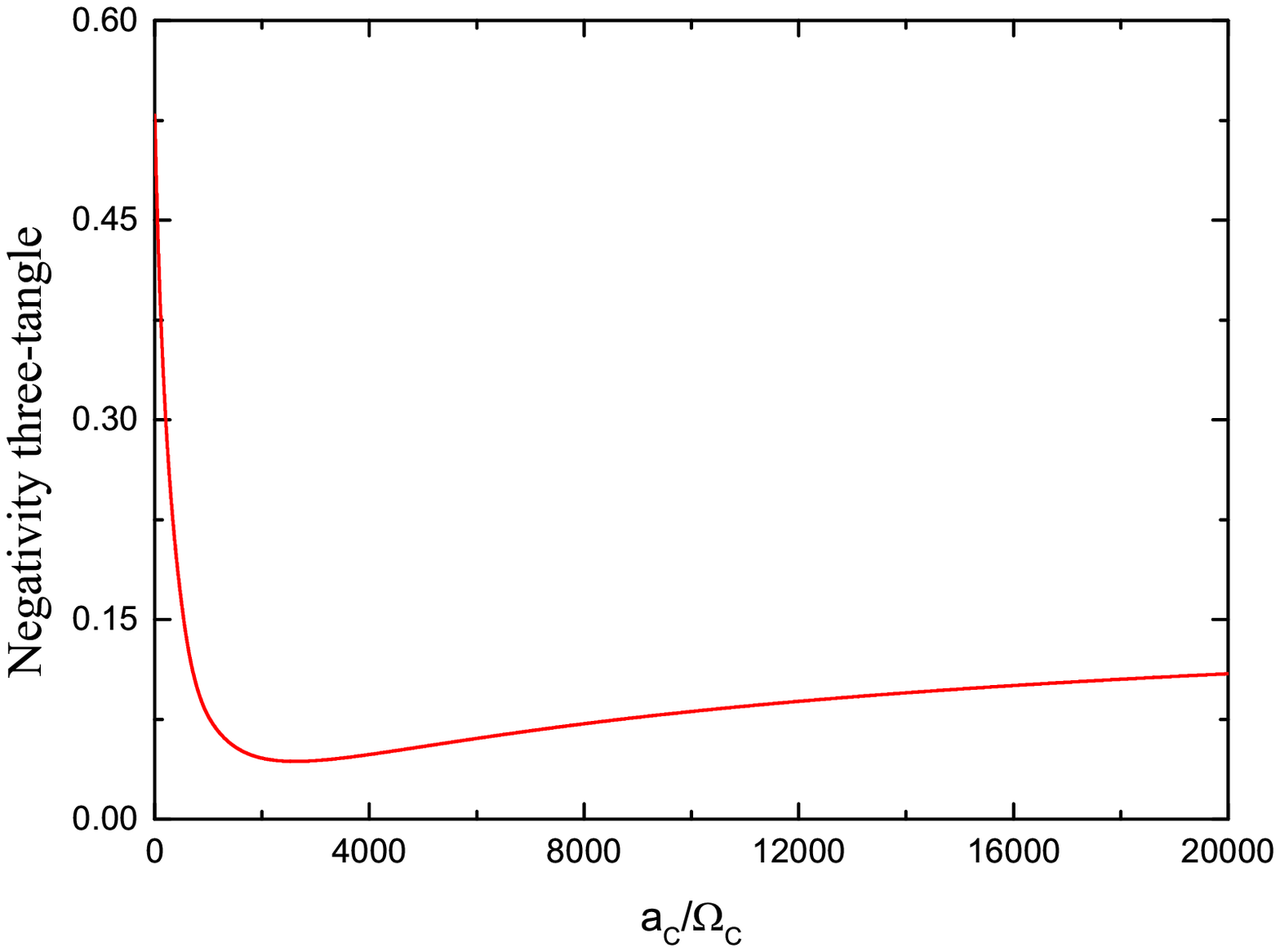}}
\caption{Dependence of (a) the one-tangles, (b) the two-tangles, and (c) the three-tangle on the AFR of qubit $C$. Qubits $A$ and $B$ move uniformly. The qubits are in the W state.}
\end{figure}

The entanglement among the qubits decreases when $C$ accelerates. As shown in Fig.~\ref{w1}, the one-tangle ${\cal N}_{C(AB)}$ suddenly dies at a certain value of $a_C/\Omega_C$. This is similar to the GHZ state. However, differing from the GHZ state, with $a_A/\Omega_A = a_B/\Omega_B = 0$,  ${\cal N}_{A(BC)} = {\cal N}_{B(CA)}$ has a minimum at a certain value of $a_C/\Omega_C$, but then increases towards a nonzero asymptotical value.

As shown in Fig.~\ref{w2}, with the increase of $a_C/\Omega_C$, ${\cal N}_{AC}= {\cal N}_{BC}$ decreases and suddenly dies at a certain value, while ${\cal N}_{AB}$ remains constant since it has nothing to do with $C$.

As shown in Fig.~\ref{w3}, with the increase of $a_C/\Omega_C$, the three-tangle first decreases towards a minimum, and then increases towards a nonzero asymptotic value. This can be inferred from the  features of the one-tangles and two-tangles, according to the definition of the three-tangle.

In the limit of $a_C/\Omega_C \to \infty$,
\begin{equation}
\rho_{ABC}\to\frac{1}{3}\left[ |000\rangle\langle000|+|011\rangle\langle011|+|101\rangle\langle101|+|101\rangle\langle011|+|011\rangle\langle101|\right],
\end{equation}
so the asymptotic values of the one-tangles are
\begin{eqnarray}
\lim\limits_{a_C\to\infty}\mathcal{N}_{A(BC)} &=& \lim\limits_{a_C\to\infty}\mathcal{N}_{B(AC)} = \frac{2}{3},\\
\lim\limits_{a_C\to\infty}\mathcal{N}_{C(AB)} &=& 0.
\end{eqnarray}
The two-tangles are the following: $\mathcal{N}_{AB}=(\sqrt{5}-1)/3$, which is a constant independent of $a_C/\Omega_C$; $\mathcal{N}_{AC}=\mathcal{N}_{BC}=0$ after its sudden death. Consequently, the three-tangle asymptotically approaches $4(\sqrt{5}-1)/27 \approx 0.183$.

\subsection{$B$ and $C$ accelerating }
In the case that $B$ and $C$ accelerate while $A$ moves uniformly, we have
\begin{eqnarray}
{\rho _{ABC}} &=& \eta_B^2\eta_C^2\sum\limits_{{n_B},{n_C}} {\frac{{{e^{ - 2\pi \left( {{n_B}{\Omega _B}/{a_B} + {n_C}{\Omega _C}/{a_C}} \right)}}}}{{{Z_{{n_B},{n_C}}}}}} \left[ {|001\rangle \langle 010| + |001\rangle \langle 100|} \right. + |100\rangle \langle 001|\nonumber\\
 &&+ |100\rangle \langle 010| + |010\rangle \langle 001| + |010\rangle \langle 100| + |100\rangle \langle 100| + n_B n_C|\mu_B|^2|\mu_C|^2|111\rangle \langle 111|\nonumber\\
 &&+ {n_B}|\mu_B|^2|011\rangle \langle 110| + {n_C}|\mu_C|^2|101\rangle \langle 101| + {n_B}|\mu_B|^2|110\rangle \langle 110|\nonumber\\
 &&+ {n_C}|\mu_C|^2|011\rangle \langle 101| + \left( {|\mu_A|^2 + \left( {n_B + 1} \right)|\mu_B|^2 + \left( {n_C + 1} \right)|\mu_C|^2} \right)|000\rangle \langle 000|\nonumber\\
 &&+ \left( {1 + {n_C}|\mu_A|^2|\mu_C|^2 + \left( {n_B + 1} \right)n_C|\mu_B|^2|\mu_C|^2} \right)|001\rangle \langle 001|\nonumber\\
 &&+ n_C|\mu_C|^2|101\rangle \langle 011| + \left( {1 + n_B|\mu _A|^2|\mu _B|^2 + n_B\left( {n_C + 1} \right)|\mu _B|^2|\mu_C|^2} \right)|010\rangle \langle 010|\nonumber\\
 &&+ n_B|\mu_B|^2|110\rangle \langle 011| + \left. {\left( {{n_B}{n_C}|\mu_A|^2|\mu _B|^2|\mu _C|^2 + n_B|\mu_B|^2 + n_C|\mu _C|^2} \right)|011\rangle \langle 011|} \right],\nonumber\\*
\,
\end{eqnarray}
where
\begin{equation}
\begin{split}
{Z_{{n_B},{n_C}}} =& 1 + \frac{1}{3}|\mu _A|^2 + \left(n_B + \frac{1}{3}\right)|{\mu _B}|^2 + \left(n_C + \frac{1}{3}\right)|{\mu _C}|^2\\
&+ \frac{1}{3}{n_B}|\mu_A|^2|\mu_B|^2 + \frac{1}{3}{n_C}|{\mu_A}|^2|\mu_C|^2+ \left({n_B}{n_C}+\frac{1}{3}n_B+\frac{1}{3}n_C\right)|\mu_B|^2|{\mu_C}|^2\\
&+ \frac{1}{3}\left( n_A + 1 \right){n_B}{n_C}|\mu _A|^2|\mu _B|^2|\mu _C|^2.
\end{split}
\end{equation}

\begin{figure}

\centering
\subfigure[]{
\label{w4}
\includegraphics[width=0.35\textwidth]{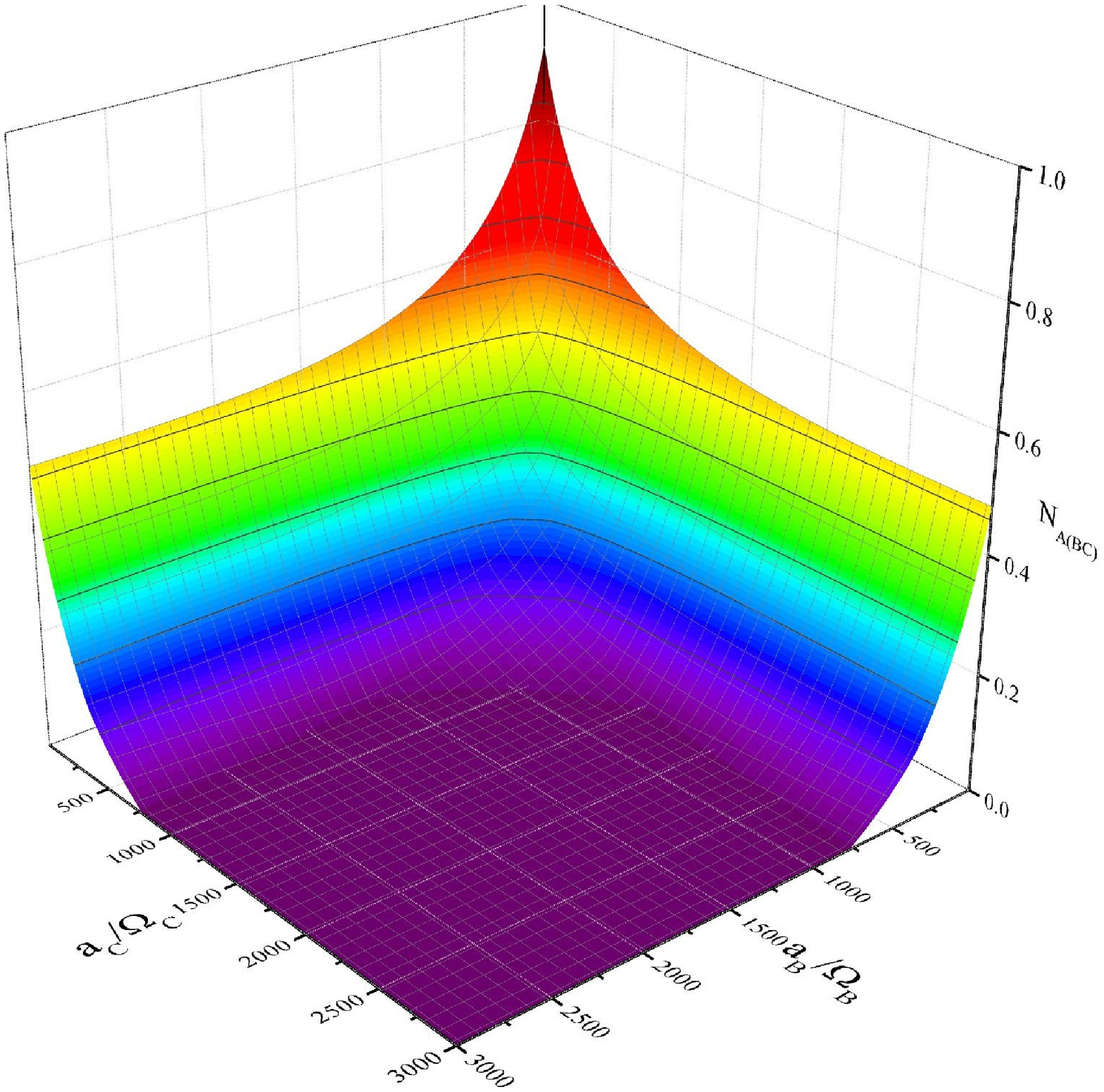}}
\subfigure[]{
\label{w5}
\includegraphics[width=0.35\textwidth]{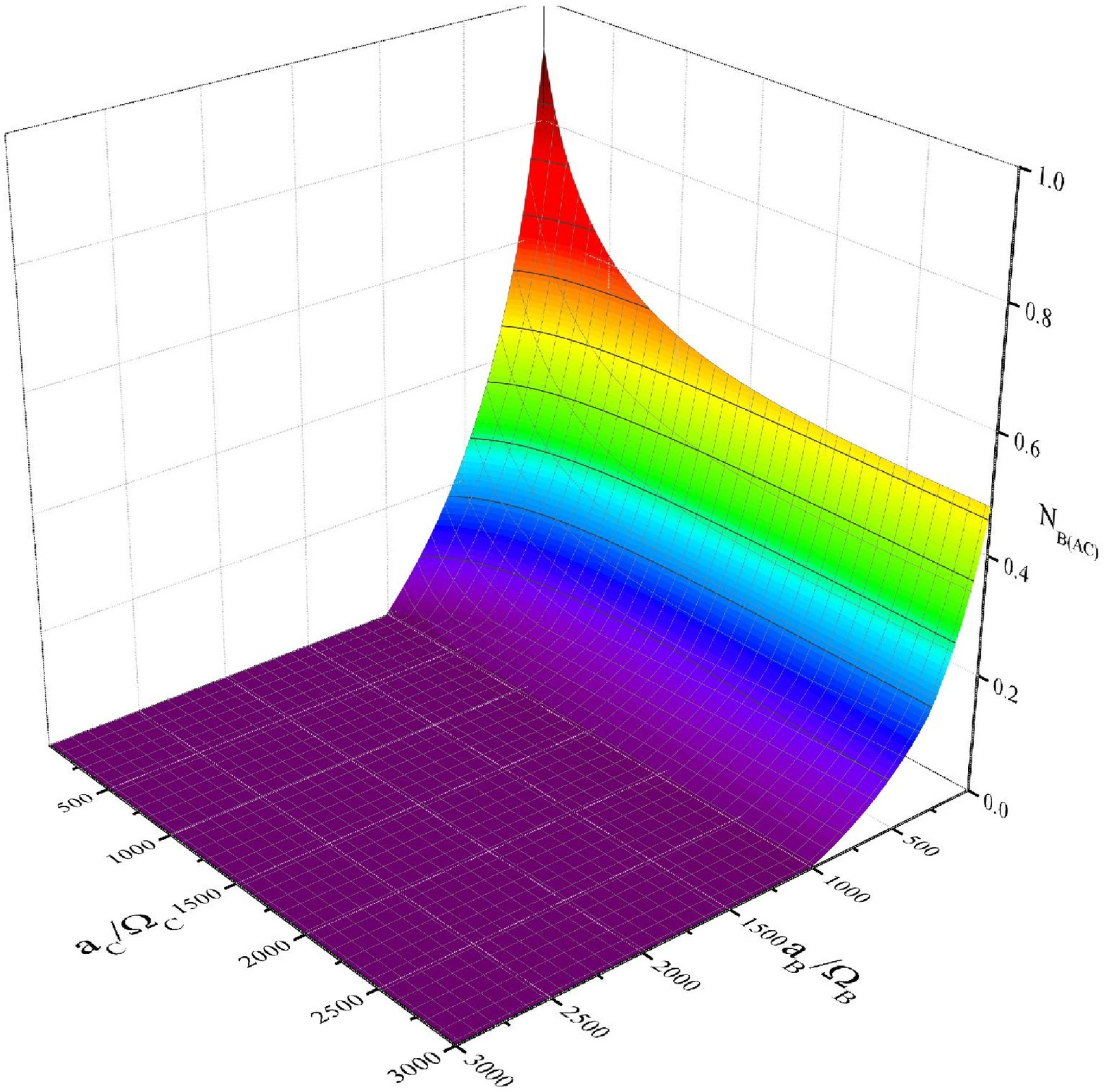}}
\subfigure[]{
\label{w6}
\includegraphics[width=0.35\textwidth]{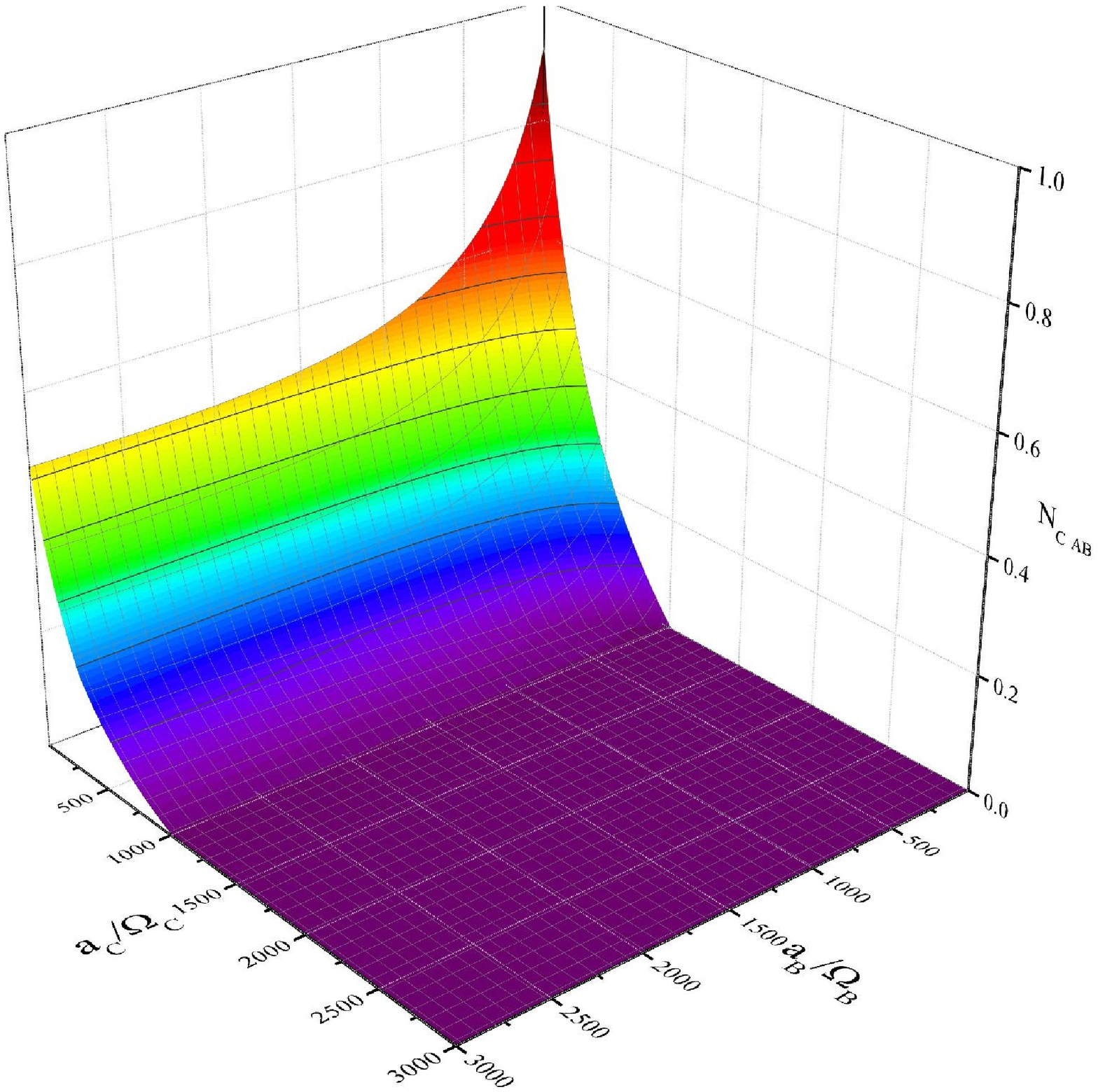}}
\subfigure[]{
\label{w7}
\includegraphics[width=0.35\textwidth]{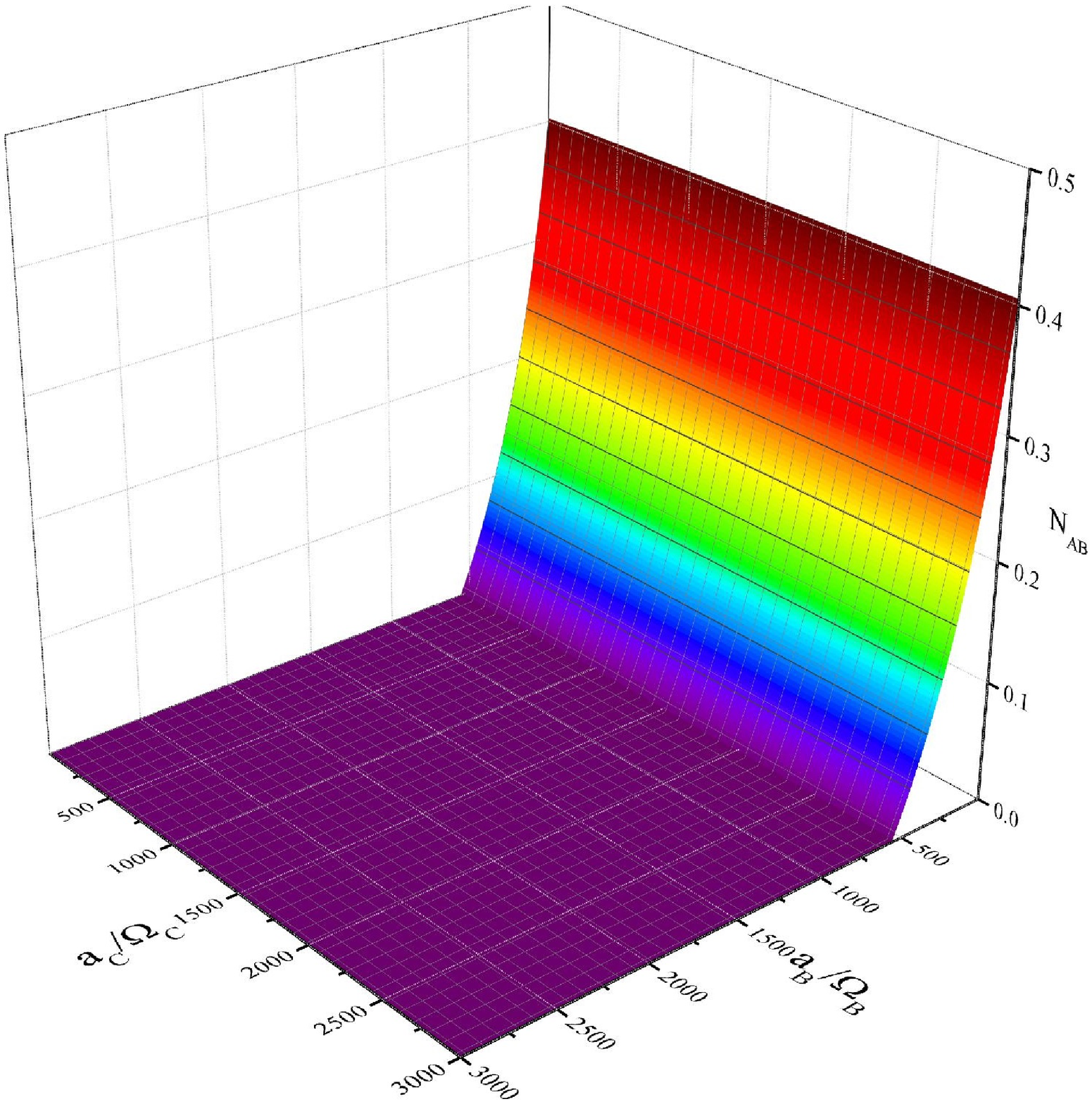}}
\subfigure[]{
\label{w8}
\includegraphics[width=0.35\textwidth]{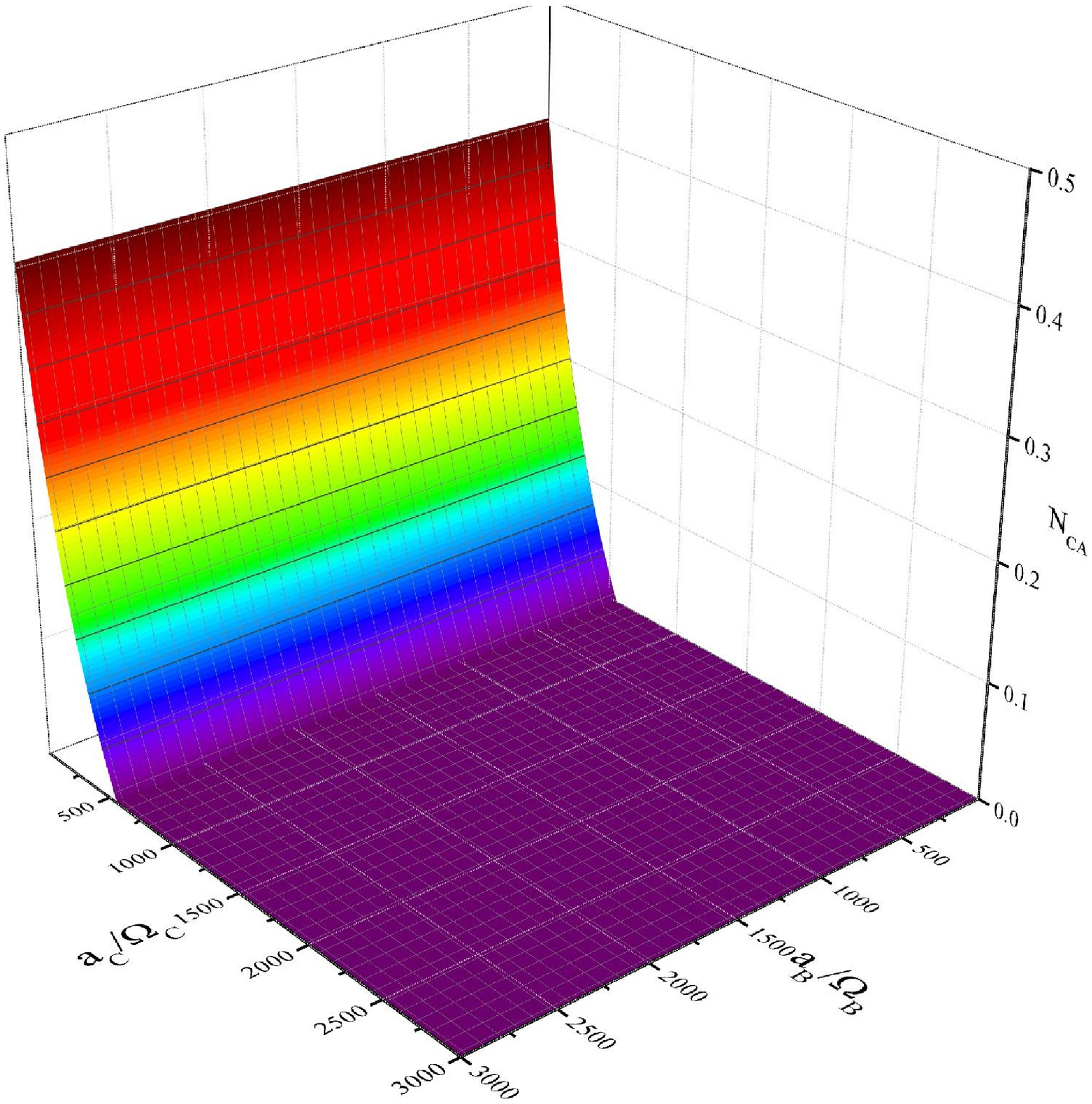}}
\subfigure[]{
\label{w9}
\includegraphics[width=0.35\textwidth]{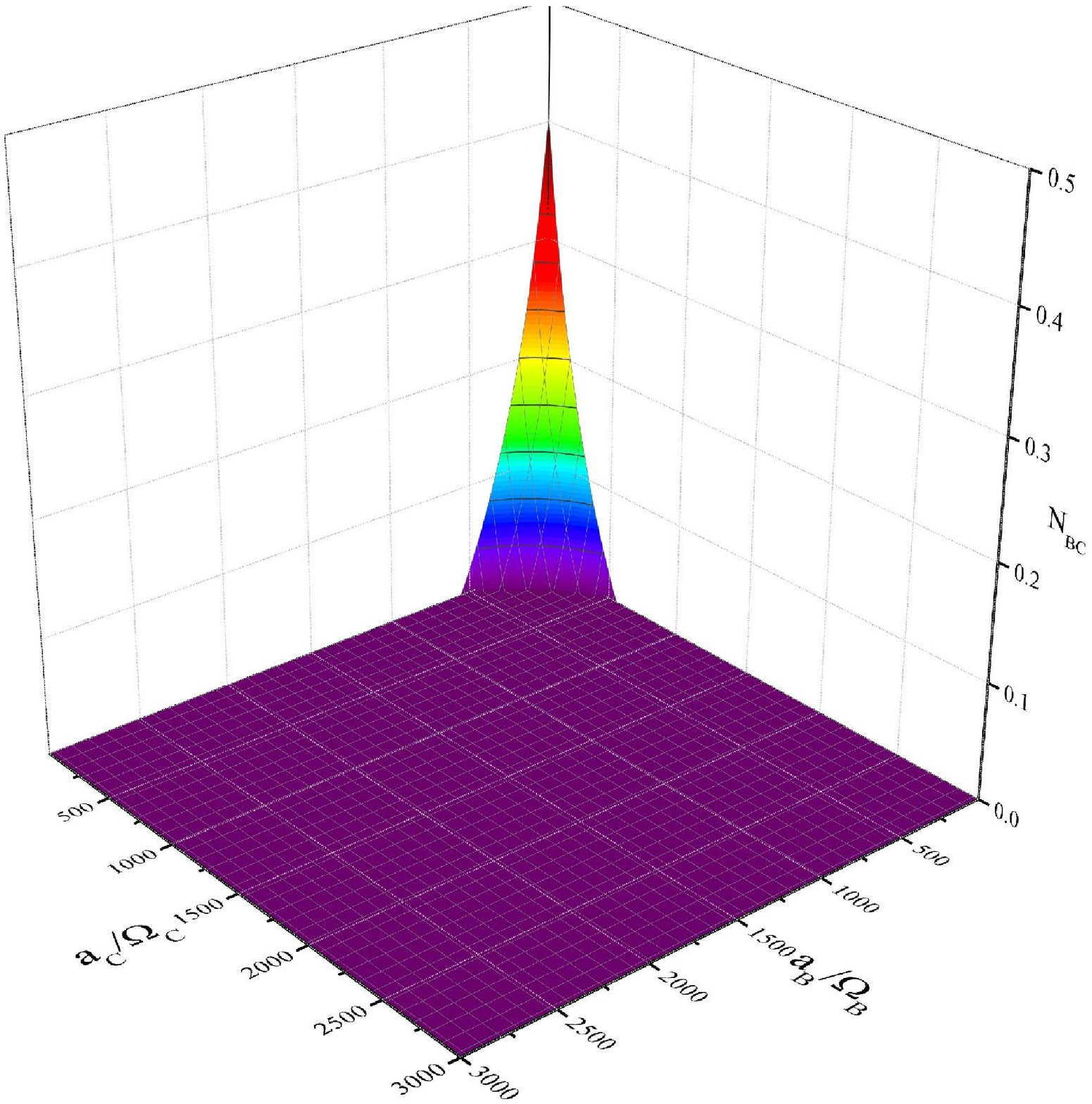}}
\caption{Dependence of the one-tangles (a) ${\cal N}_{A(BC)}$, (b) ${\cal N}_{B(AC)}$, and (c) ${\cal N}_{C(AB)}$ and the two-tangles (d) ${\cal N}_{AB}$, (e) ${\cal N}_{AC}$, and (f) ${\cal N}_{BC}$ on the AFRs of qubits $B$ and $C$. Qubit $A$ moves uniformly. The qubits are in the W state.}
\end{figure}

The one-tangle ${\cal N}_{A(BC)}$ is symmetric with respect to $a_B/\Omega_B$ and $a_C/\Omega_C$, as shown in Fig.~\ref{w4}.
${\cal N}_{B(AC)}$ is shown in Fig.~\ref{w5}, and ${\cal N}_{C(AB)}$ can be obtained from ${\cal N}_{B(AC)}$ by exchanging $B$ and $C$, as shown in Fig.~\ref{w6}. These symmetries are common with the GHZ state.

The two-tangle ${\cal N}_{AB}$ is shown in Fig.~\ref{w7}, and ${\cal N}_{AC}$ can be obtained from ${\cal N}_{AB}$ by replacing $B$ with $C$, as shown in Fig.~\ref{w8}. ${\cal N}_{BC}$ is shown in Fig.~\ref{w9}. All exhibit sudden death.

\begin{figure}

\centering
\subfigure[]{
\label{w10}
\includegraphics[width=0.48\textwidth]{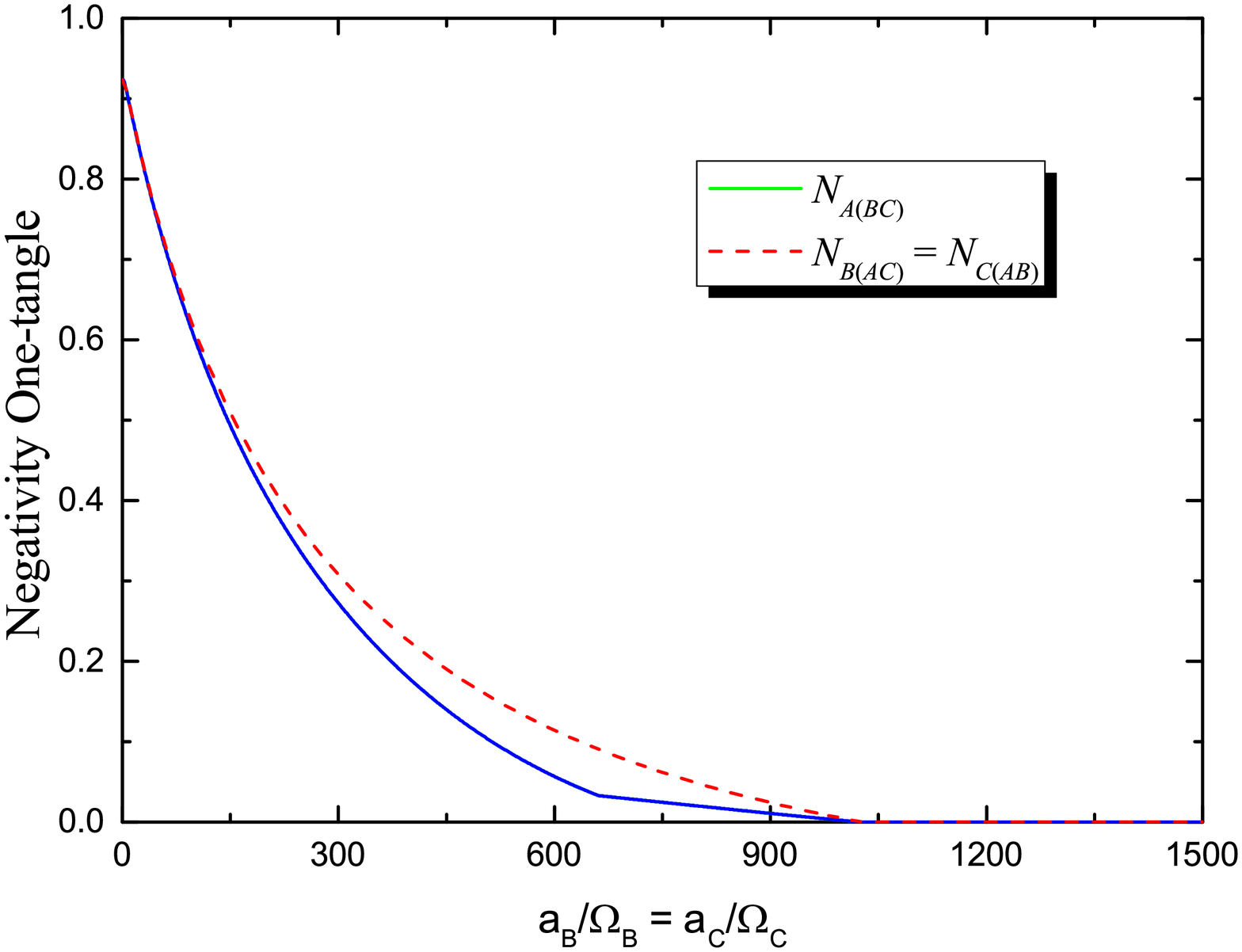}}
\subfigure[]{
\label{w16}
\includegraphics[width=0.48\textwidth]{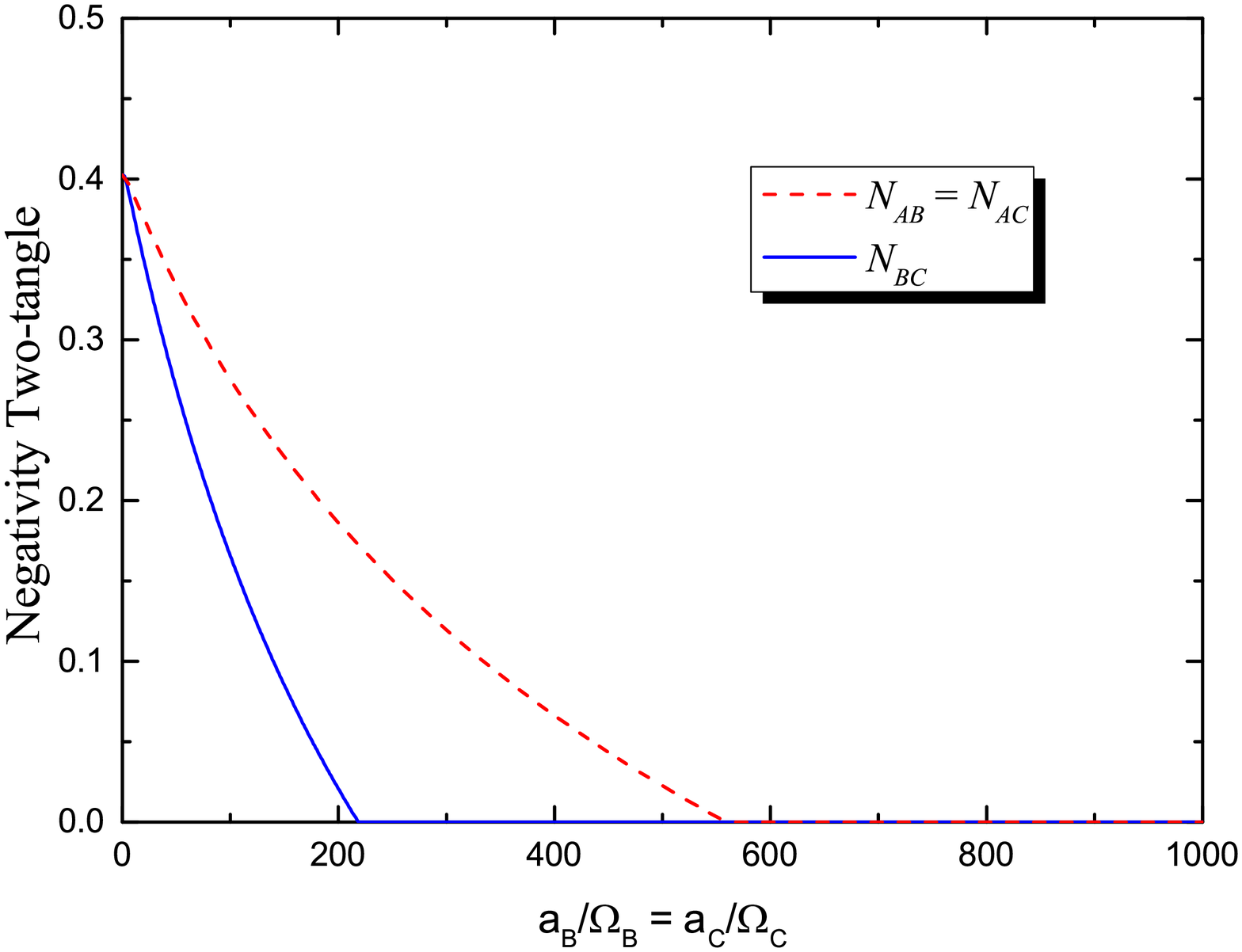}}
\caption{\label{aB=aC} Dependence of (a) the one-tangles and (b) the two-tangles on the AFRs of qubits $B$ and $C$ in the case that they are equal. Qubit $A$ moves uniformly. The qubits are in the W state.}
\end{figure}

\begin{figure}

\centering
\subfigure[]{
\label{w11}
\includegraphics[width=0.48\textwidth]{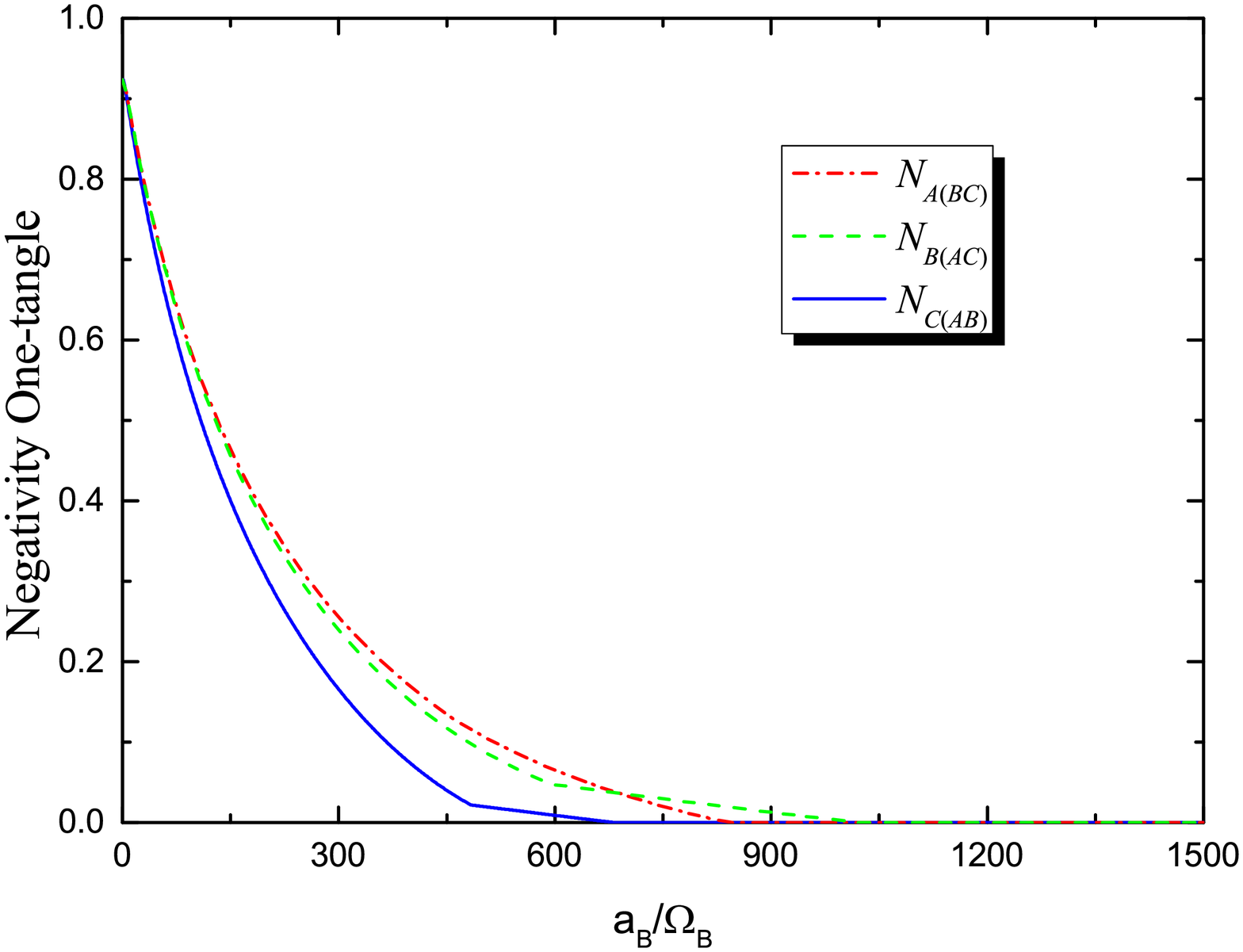}}
\subfigure[]{
\label{w28}
\includegraphics[width=0.48\textwidth]{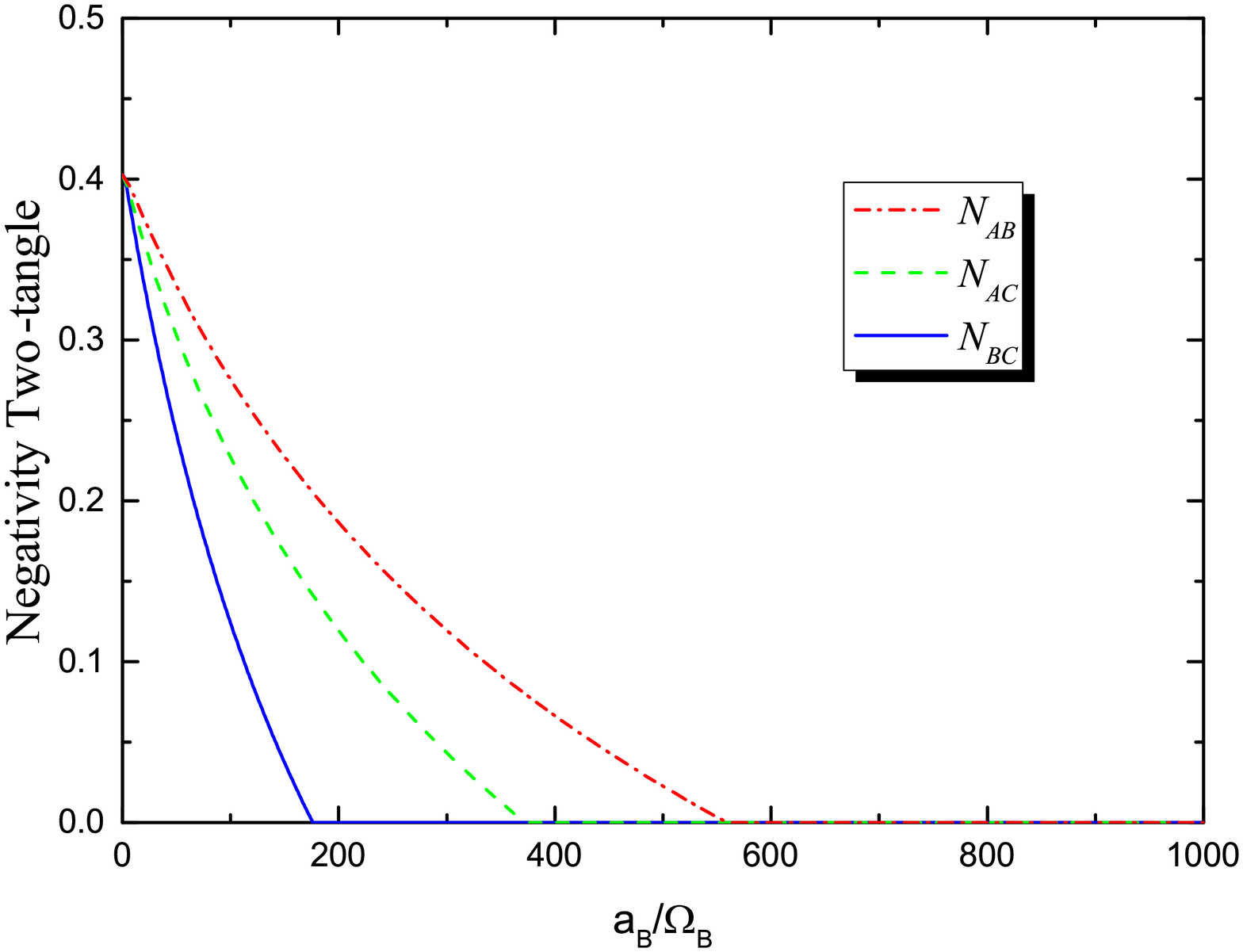}}
\caption{\label{1.5aB=aC} Dependence of (a) the one-tangles and (b) the two-tangles on the AFR of qubit $B$ in the case that $a_C/\Omega_C = 1.5 a_B/\Omega_B$. Qubit $A$ moves uniformly. The qubits are in the W state.}
\end{figure}

\begin{figure}

\centering
\subfigure[]{
\label{w12}
\includegraphics[width=0.48\textwidth]{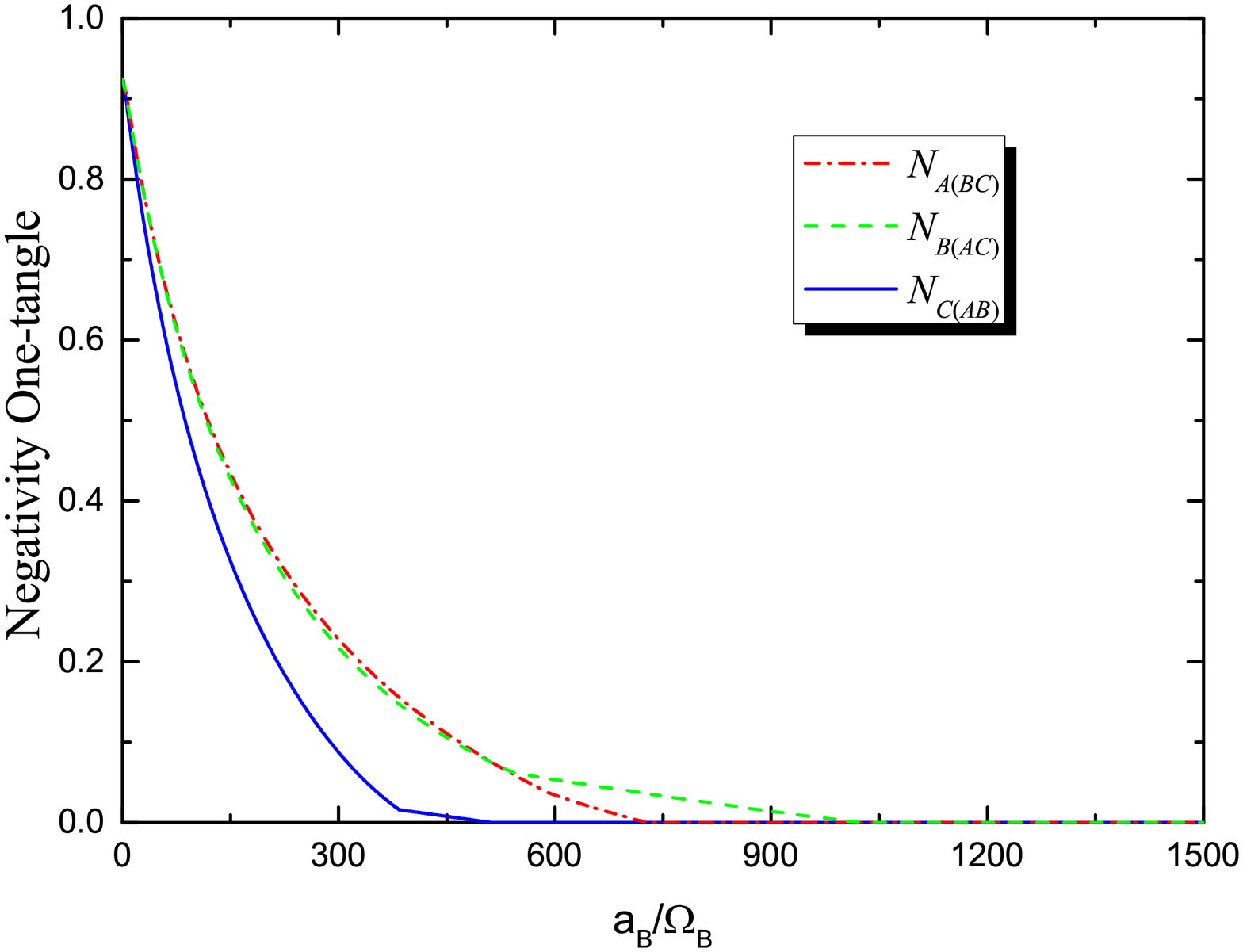}}
\subfigure[]{
\label{w18}
\includegraphics[width=0.48\textwidth]{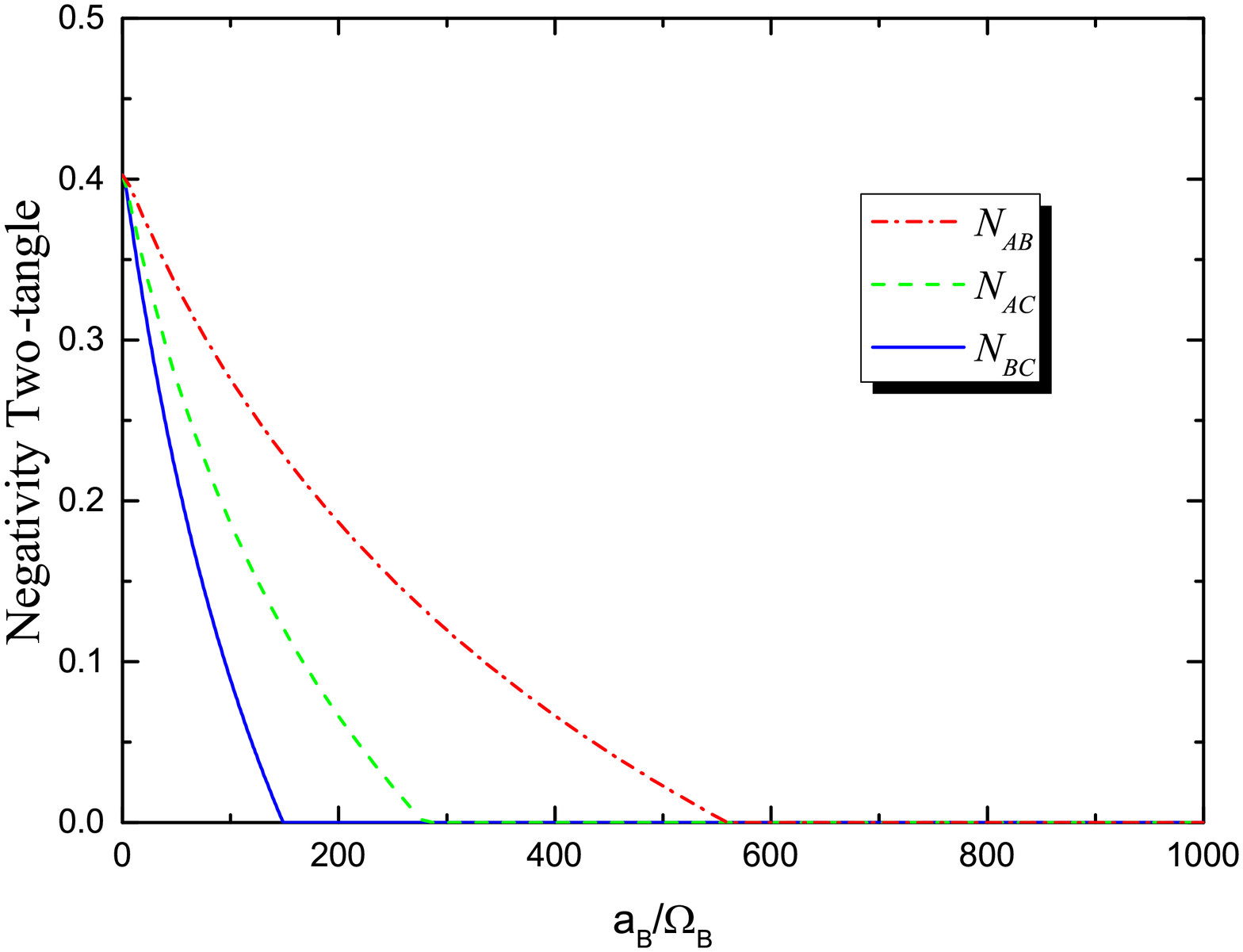}}
\caption{\label{2aB=aC} Dependence of (a) the one-tangles and (b) the two-tangles on the AFR of qubit $B$ in the case that $a_C/\Omega_C = 2 a_B/\Omega_B$. Qubit $A$ moves uniformly. The qubits are in the W state.}
\end{figure}

To see the dependence on $a_B/\Omega_B$ more clearly, we examined the two-dimensional cross sections of the 3D plots with $ a_C/\Omega_C=a_B/\Omega_B$ (Fig.~\ref{aB=aC}),  $a_C/\Omega_C= 1.5a_B/\Omega_B$ (Fig.~\ref{1.5aB=aC}) and $a_C/\Omega_C= 2a_B/\Omega_B$ (Fig.~\ref{2aB=aC}).

In these cases, in which $a_c/\Omega_C \propto a_B/\Omega_B$, the behavior of the one-tangles is similar to that in the GHZ state: each one-tangle suddenly dies at certain values of  $a_B/\Omega_B$. For $0=a_A/\Omega_A <a_B/\Omega_B <a_C/\Omega_C$, ${\cal N}_{A(BC)} > {\cal N}_{B(AC)}>{\cal N}_{C(AB)}$ until the sudden death of the smallest one ${\cal N}_{C(AB)}$, and afterwards ${\cal N}_{B(AC)} > {\cal N}_{C(AB)}$. See Fig.~\ref{w10}, Fig.~\ref{w11}, and Fig.~\ref{w12}. A feature different from GHZ is that in addition to the sudden death, there is also a sudden change.

\begin{figure}

\centering
\subfigure[]{
\label{w13}
\includegraphics[width=0.48\textwidth]{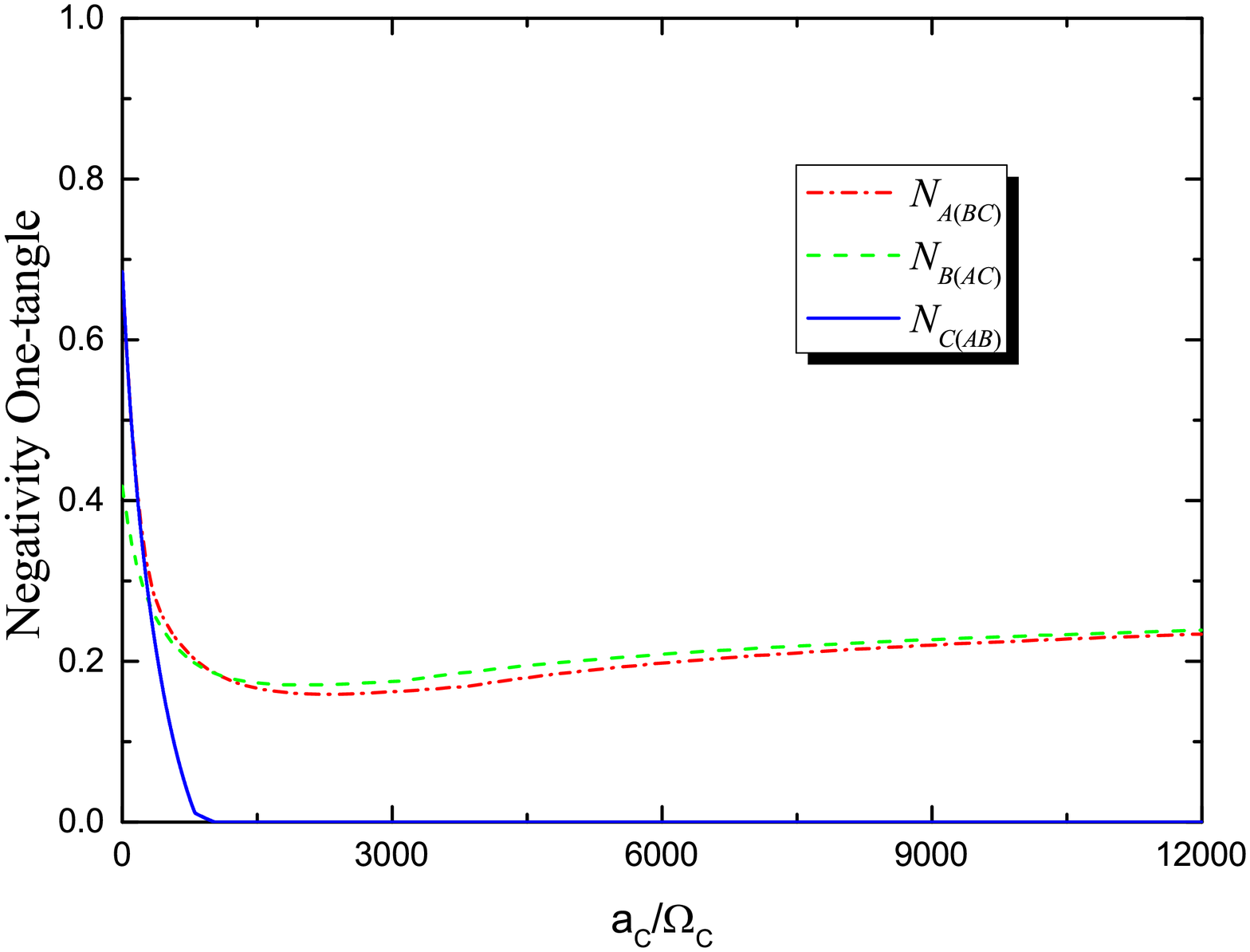}}
\subfigure[]{
\label{w19}
\includegraphics[width=0.48\textwidth]{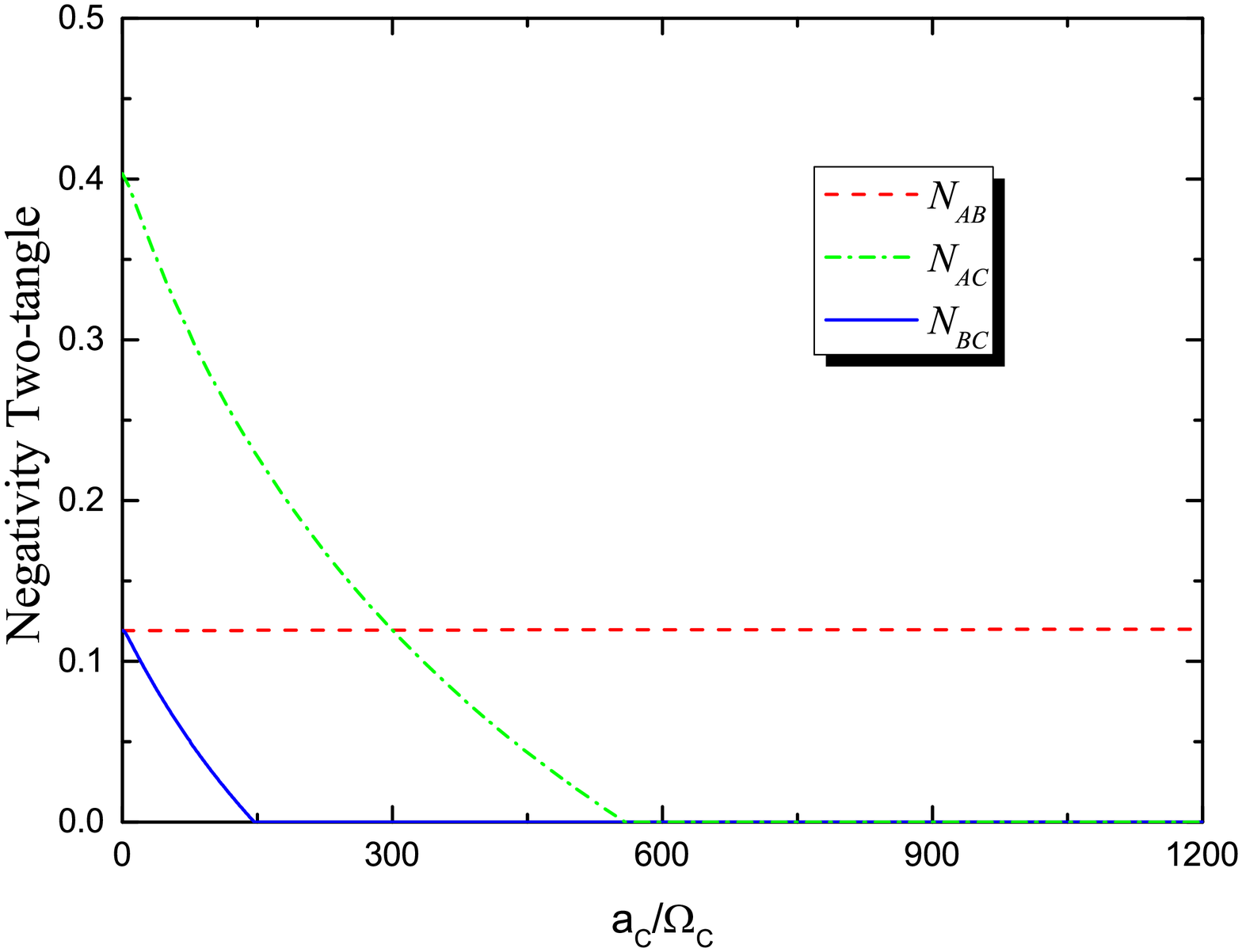}}
\caption{Dependence of (a) the one-tangles and (b) the two-tangles on the AFR of qubit $C$ in
the case that $a_B/\Omega_B = 300$. Qubit $A$ moves uniformly. The qubits are in the W state.}
\end{figure}

\begin{figure}

\centering
\subfigure[]{
\label{w14}
\includegraphics[width=0.48\textwidth]{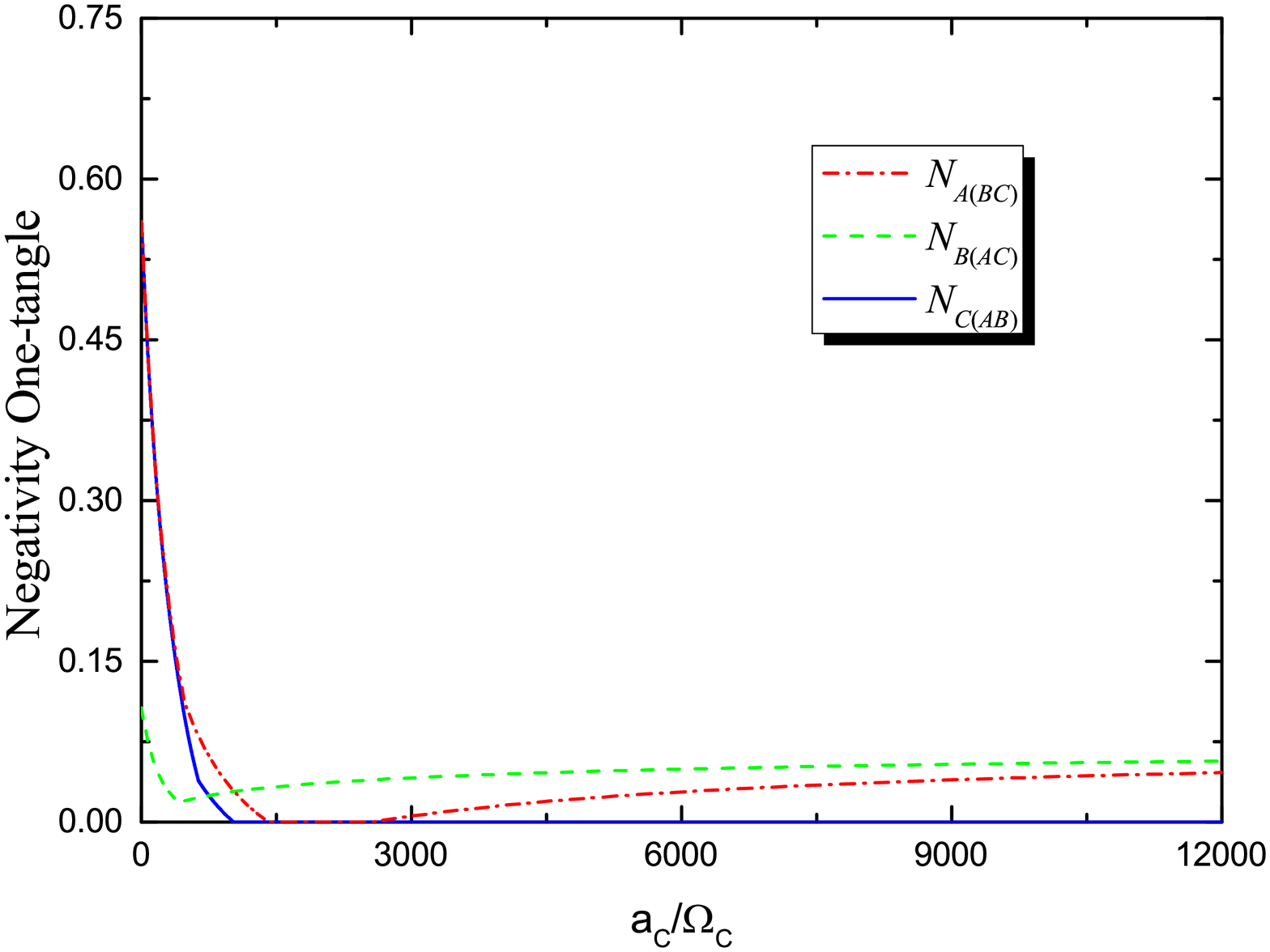}}
\subfigure[]{
\label{w20}
\includegraphics[width=0.48\textwidth]{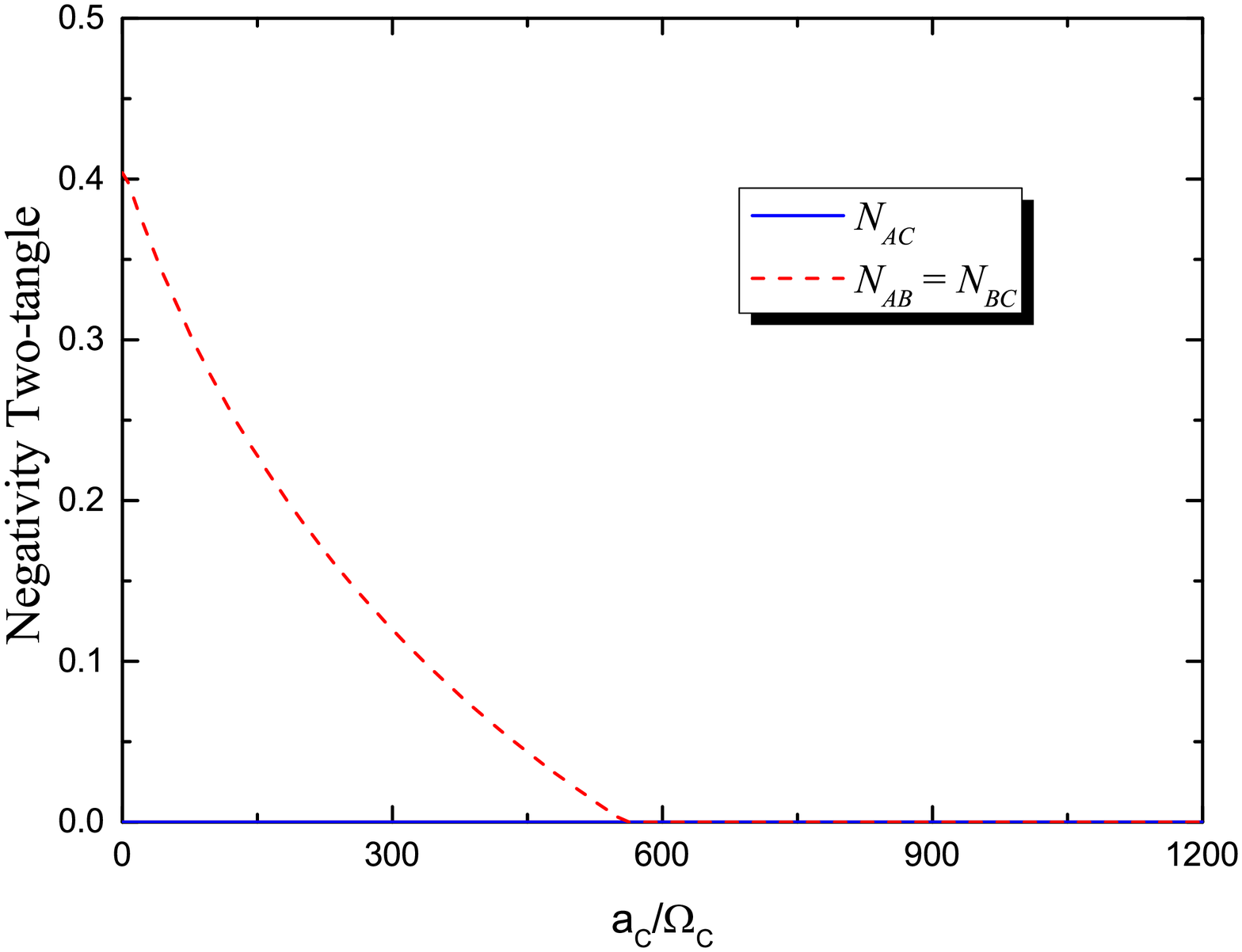}}
\caption{Dependence of (a) the one-tangles and (b) the two-tangles on the AFR of qubit $C$ in
the case that $a_B/\Omega_B = 750$. Qubit $A$ moves uniformly. The qubits are in the W state.}
\end{figure}

We also look at one-tangles for some given values of $a_B/\Omega_B$, as shown in Fig.~\ref{w13} for $a_B/\Omega_B=300$ and in Fig.~\ref{w14} for $a_B/\Omega_B=750$. In both figures, a feature in common with the GHZ state is that when the smallest one-tangle (namely, ${\cal N}_{C(AB)}$) suddenly dies, the other two one-tangles exchange the order of magnitude. Compared with the GHZ state, a new feature is that ${\cal N}_{A(BC)}$ and ${\cal N}_{B(AC)}$ are not monotonic with respect to $a_C/\Omega_C$. As $a_B/\Omega_B$ is larger in Fig.~\ref{w14} than in Fig.~\ref{w13}, the one-tangles decrease faster. In Fig.~\ref{w13}, only ${\cal N}_{C(AB)}$ has sudden death, while the minima of the other two one-tangles are nonzero. In Fig.~\ref{w14}, both  ${\cal N}_{C(AB)}$ and  ${\cal N}_{A(BC)}$ have sudden death, but  ${\cal N}_{A(BC)}$ revives and increases at  larger values of $a_C/\Omega_C$, because $B$ and $C$ constitute one party, and $B$ is fixed to be not large enough. In each of the two figures, ${\cal N}_{B(AC)}$  has a nonzero minimum, as $a_B/\Omega_B$ is now fixed, while $C$ is only one of the two qubits constituting  the other party. Note that $C$ is the qubit on which the acceleration-frequency can always be large enough.

Now we look at the two-tangles. The 2D cross sections at $a_C/\Omega_C=a_B/\Omega_B$,  $a_C/\Omega_C= 1.5a_B/\Omega_B$, and $a_C/\Omega_C= 2a_B/\Omega_B$ are shown in Figs.~\ref{w16}, \ref{w28}, and \ref{w18}, respectively. ${\cal N}_{AB}$ decreases with $a_B/\Omega_B$,  ${\cal N}_{AC}$  decreases with  $a_C/\Omega_C$,  while ${\cal N}_{BC}$ decreases with both. For a given value of $a_B/\Omega_B$, ${\cal N}_{BC}$ is the smallest two-tangle in each  of these cross sections. For the same value of $a_B/\Omega_B$,   ${\cal N}_{AC}$ in Fig.~\ref{w16} is larger than in Fig.~\ref{w28}, which is larger than in Fig.~\ref{w18}, because $a_C/\Omega_C$  in  Fig.~\ref{w16} is smaller than in Fig.~\ref{w28}, which is smaller than in Fig.~\ref{w18}.  Each  two-tangle suddenly dies when the AFR of one or both of the parties are  large enough. This is consistent with the behavior of entanglement of a two-qubit system \cite{dai}.

In Fig.~\ref{w19} and Fig.~\ref{w20} we show the cases of $a_B/\Omega_B=300$ and $a_B/\Omega_B=750$, respectively. In Fig.~\ref{w19}, ${\cal N}_{AB}$, measuring the entanglement between $A$ and $B$,  is a constant independent of $a_C/\Omega_C$.  In Fig.~\ref{w20}, because $a_B/\Omega_B$ is large enough, ${\cal N}_{AB}$ and ${\cal N}_{BC}$ both remain zero, independent of $a_C/\Omega_C$.

\begin{figure}
\centering
\includegraphics[width=0.6\textwidth]{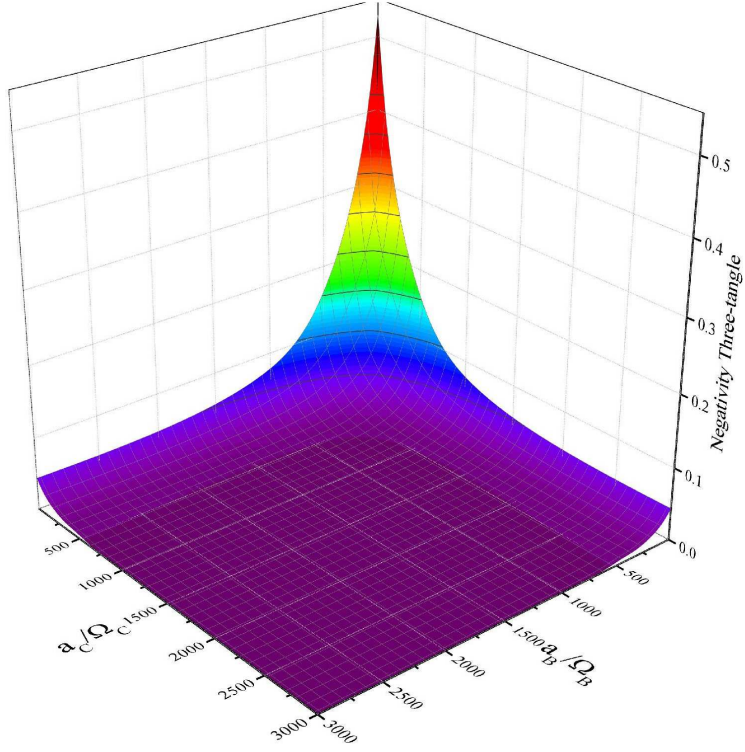}
\caption{\label{w21} Dependence of the three-tangle on the AFR of qubits $B$ and $C$, while qubit $A$ moves uniformly. The qubits are in the W state.}
\end{figure}

\begin{figure}

\centering
\subfigure[]{
\label{w27}
\includegraphics[width=0.48\textwidth]{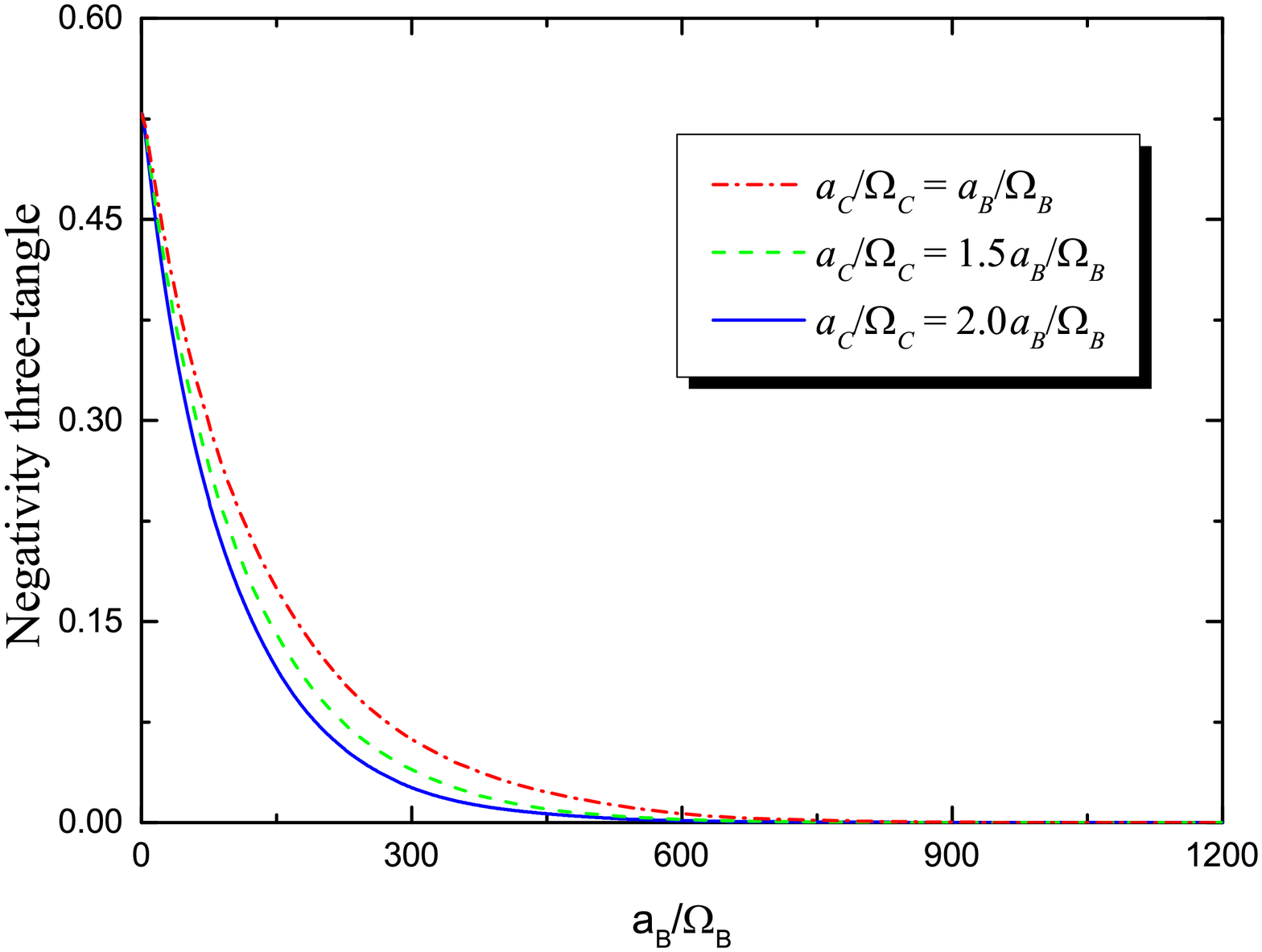}}
\subfigure[]{
\label{w31}
\includegraphics[width=0.48\textwidth]{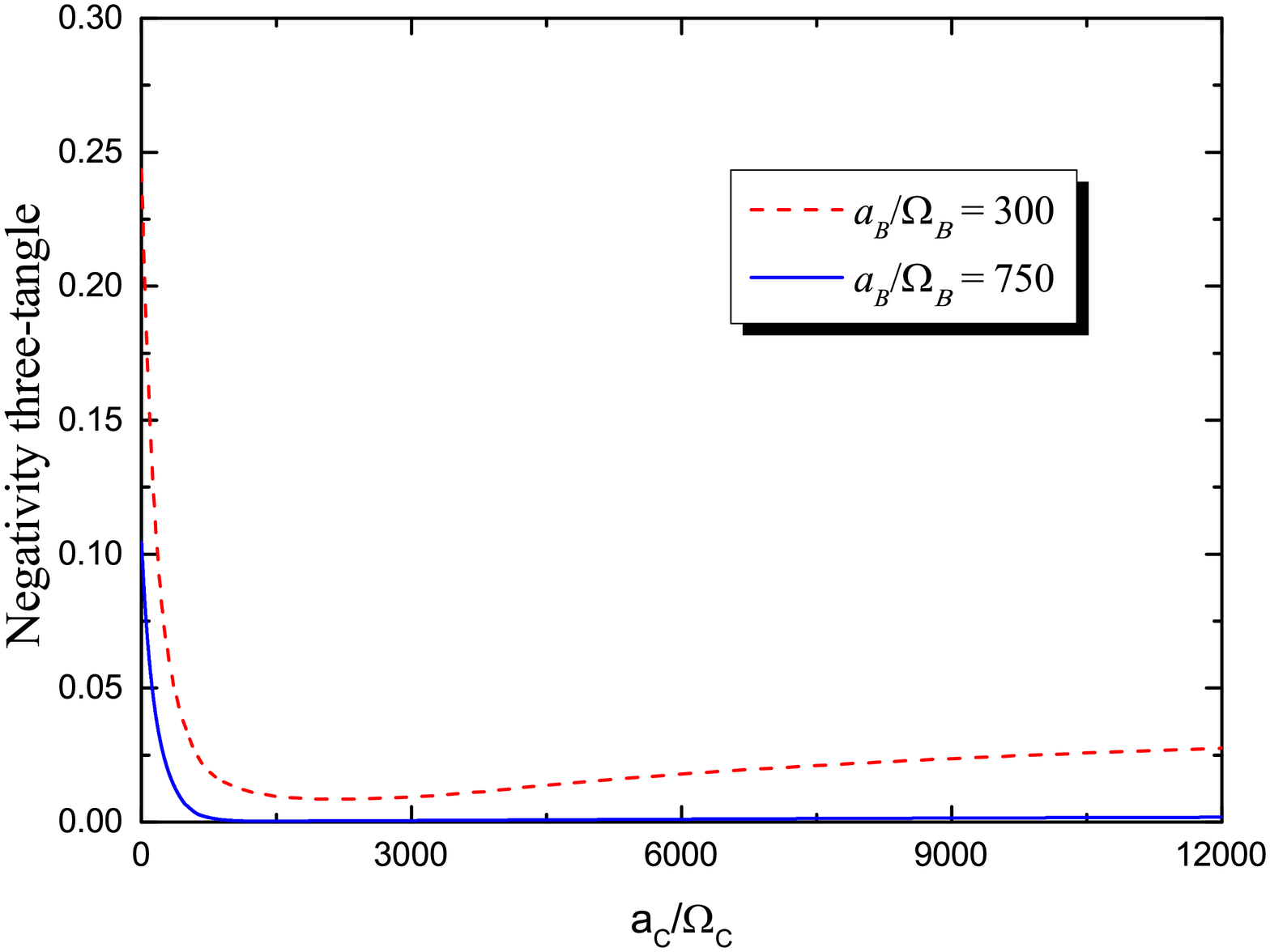}}
\caption{(a) Dependence of the three-tangle on the AFR of qubit $B$ in the case that $a_C/\Omega_C = a_B/\Omega_B$, $1.5a_B/\Omega_B$ and $2a_B/\Omega_B$. (b) Dependence of the three-tangle on the AFR of qubit $C$ in the case that $a_B/\Omega_B = 300$ and $a_B/\Omega_B = 750$. Qubit $A$ moves uniformly. The qubits are in the W state.}
\end{figure}

Finally, let us look at the three-tangle, whose 3D plot is shown in Fig.~\ref{w21}. To clearly see how it is different from the three-tangle of the GHZ state (Fig.~\ref{g12}), we examine some 2D cross sections. In Fig.~\ref{w27}, in which it is set that $a_C/\Omega_C \propto a_B/\Omega_B$, the three-tangle dies when $a_B/\Omega_B$ is large enough, and the larger $a_C/\Omega_C$, the quicker the decrease of the three-tangle with $a_B/\Omega_B$. This feature is similar to that of the GHZ state [Fig.~\ref{g13}]. But on the other hand, when two of the three qubits' accelerations are not large enough---for example, when it is fixed that $a_A/\Omega_A=0$ while $a_B/\Omega_B=300$ or $750$ [Fig.~\ref{w31}]---with the increase of $a_C/\Omega_C$, the three-tangle  decreases to a nonzero minimum, and then increases towards an asymptotic value.    Also refer to Fig.~\ref{w3} for the case  of $a_A/\Omega_A=a_B/\Omega_B=0$. Therefore it is implied that the AFRs of at least two  qubits should be large enough in order that the three-tangle has sudden death.

Recall that in the GHZ state, when the AFRs of two qubits are not  large enough, the three-tangle approaches zero asymptotically while the AFR of the remaining qubit tends to infinity [Figs.~\ref{g2} and ~\ref{g14}]. This is a difference between the two states.

\subsection{$A$, $B$, and $C$ all accelerating}

Now we consider the case that all three qubits accelerate, obtaining

\begin{eqnarray}
{\rho _{ABC}} &=& \eta_A^2\eta_B^2\eta_C^2\sum\limits_{{n_A},{n_B},{n_C}} {\frac{{{e^{ - 2\pi \left( {{n_A}{\Omega _A}/{a_A} + {n_B}{\Omega _B}/{a_B} + {n_C}{\Omega _C}/{a_C}} \right)}}}}{{{Z_{{n_A},{n_B},{n_C}}}}}} \left[ {|001\rangle \langle 010|} \right.\nonumber\\
 &&+ |001\rangle \langle 100| + |100\rangle \langle 001| + |100\rangle \langle 010| + |010\rangle \langle 001| + |010\rangle \langle 100|\nonumber\\
 &&+ n_B|\mu_B|^2|011\rangle \langle 110| + n_A|\mu_A|^2|101\rangle \langle 110| + n_C|\mu_C|^2|011\rangle \langle 101|\nonumber\\
 &&+ n_A|\mu_A|^2|110\rangle \langle 101| + n_C|\mu_C|^2|101\rangle \langle 011| + n_B|\mu _B|^2|110\rangle \langle 011|\nonumber\\
 &&+ \left( {\left( n_A + 1 \right)|\mu_A|^2 + \left( n_B + 1 \right)|\mu_B|^2 + \left( n_C + 1 \right)|\mu_C|^2} \right)|000\rangle \langle 000|\nonumber\\
 &&+ \left( {1 + \left( n_A + 1 \right)n_C|\mu_A|^2|\mu_C|^2 + \left( n_B + 1\right)n_C|\mu_B|^2|\mu_C|^2} \right)|001\rangle \langle 001|\nonumber\\
 &&+ \left( {1 + \left( n_A + 1 \right)n_B|\mu_A|^2|\mu _B|^2 + n_B\left( n_C + 1\right)|\mu_B|^2|\mu_C|^2} \right)|010\rangle \langle 010|\nonumber\\
 &&+ \left( {\left( n_A + 1 \right)n_Bn_C|\mu_A|^2|\mu_B|^2|\mu_C|^2 + n_B|\mu_B|^2 + n_C|\mu_C|^2} \right)|011\rangle \langle 011|\nonumber\\
 &&+ \left( {1 + n_A\left( n_B + 1 \right)|\mu_A|^2|\mu _B|^2 + n_A\left( n_C + 1 \right)|\mu _A|^2|\mu_C|^2} \right)|100\rangle \langle 100|\nonumber\\
 &&+ \left( {n_A\left( n_B + 1 \right)n_C|\mu _A|^2|\mu _B|^2|\mu_C|^2 + n_A|\mu_A|^2 + n_C|\mu_C|^2} \right)|101\rangle \langle 101|\nonumber\\
 &&+ \left( {n_A|\mu_A|^2 + n_B|\mu _B|^2 + n_An_B\left( n_C + 1 \right)|\mu_A|^2|\mu_B|^2|\mu_C|^2} \right)|110\rangle \langle 110|\nonumber\\
 &&+ \left.{\left( {n_An_B|\mu_A|^2|\mu_B|^2 + n_An_C|\mu_A|^2|\mu_C|^2 + n_Bn_C|\mu_B|^2|\mu_C|^2} \right)|111\rangle \langle 111|} \right],\nonumber\\*
 \,
\end{eqnarray}
where
\begin{equation}
\begin{split}
{Z_{{n_A},{n_B},{n_C}}} =& 3 + (3n_An_Bn_C + n_An_B + n_An_C + n_Bn_C)|\mu_A|^2|\mu_B|^2|\mu_C|^2 \\
 &+ (3n_A+1)|\mu_A|^2 + (3n_B n_C+n_B+n_C)|\mu_B|^2|\mu_C|^2 \\
 &+ (3n_B+1)|\mu_B|^2 + (3n_A n_C+n_A+n_C)|\mu_A|^2|\mu_C|^2\\
 &+ (3n_C+1)|\mu_C|^2 + (3n_A n_B+n_A+n_B)|\mu_A|^2|\mu_B|^2.
\end{split}
\end{equation}

\begin{figure}
\subfigure[]{
\label{w40}
\includegraphics[width=0.48\textwidth]{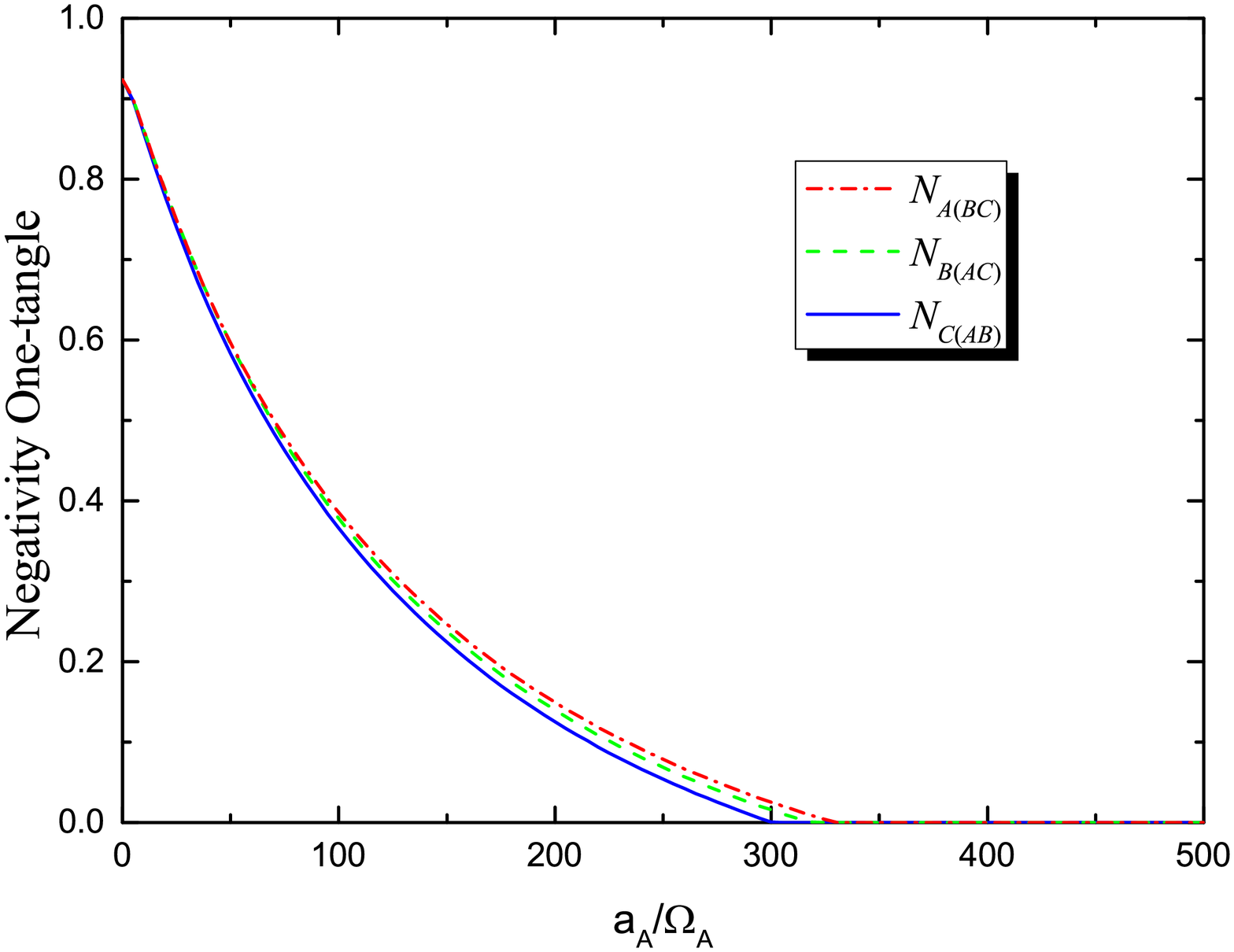}}
\subfigure[]{
\label{w41}
\includegraphics[width=0.48\textwidth]{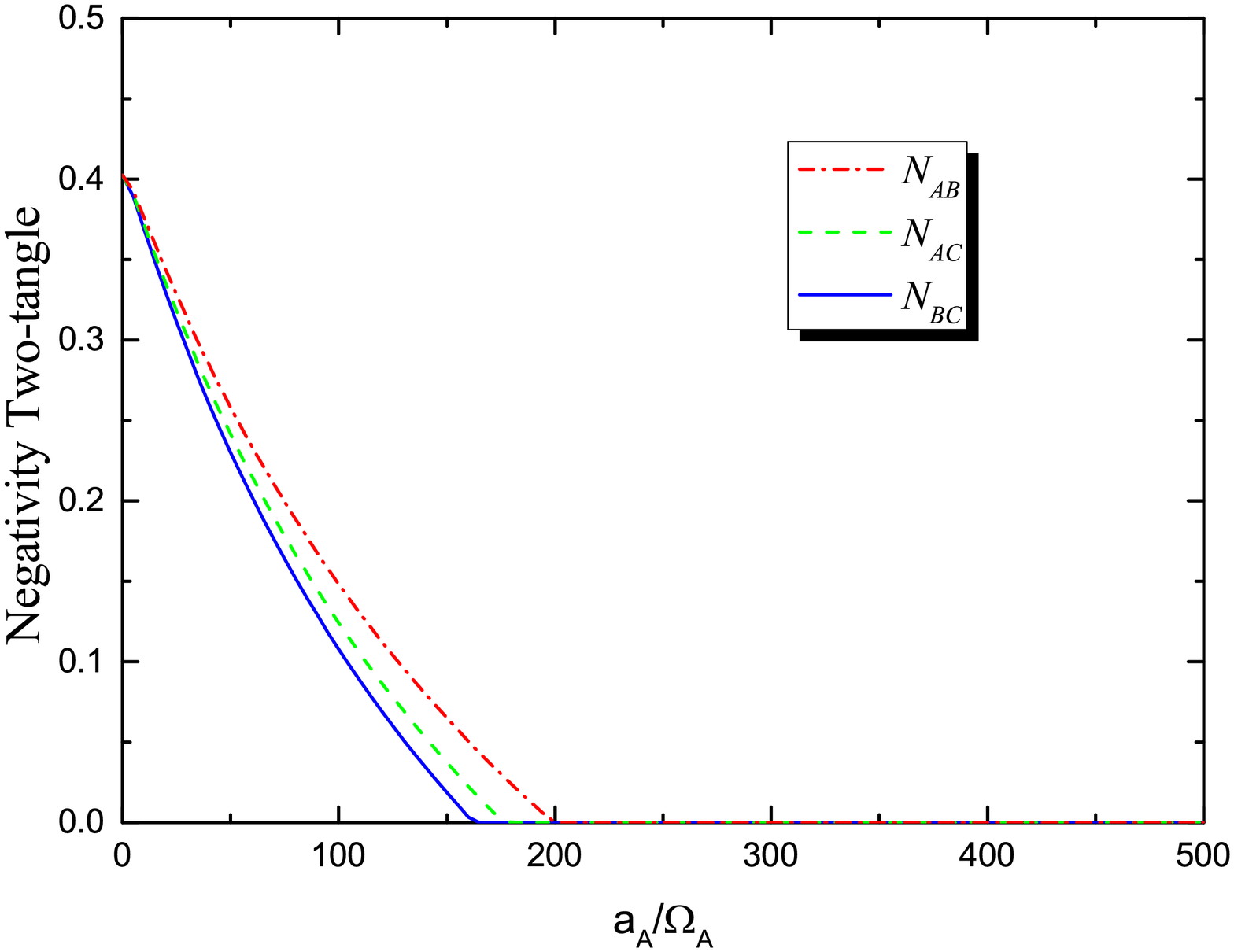}}
\subfigure[]{
\label{w42}
\includegraphics[width=0.48\textwidth]{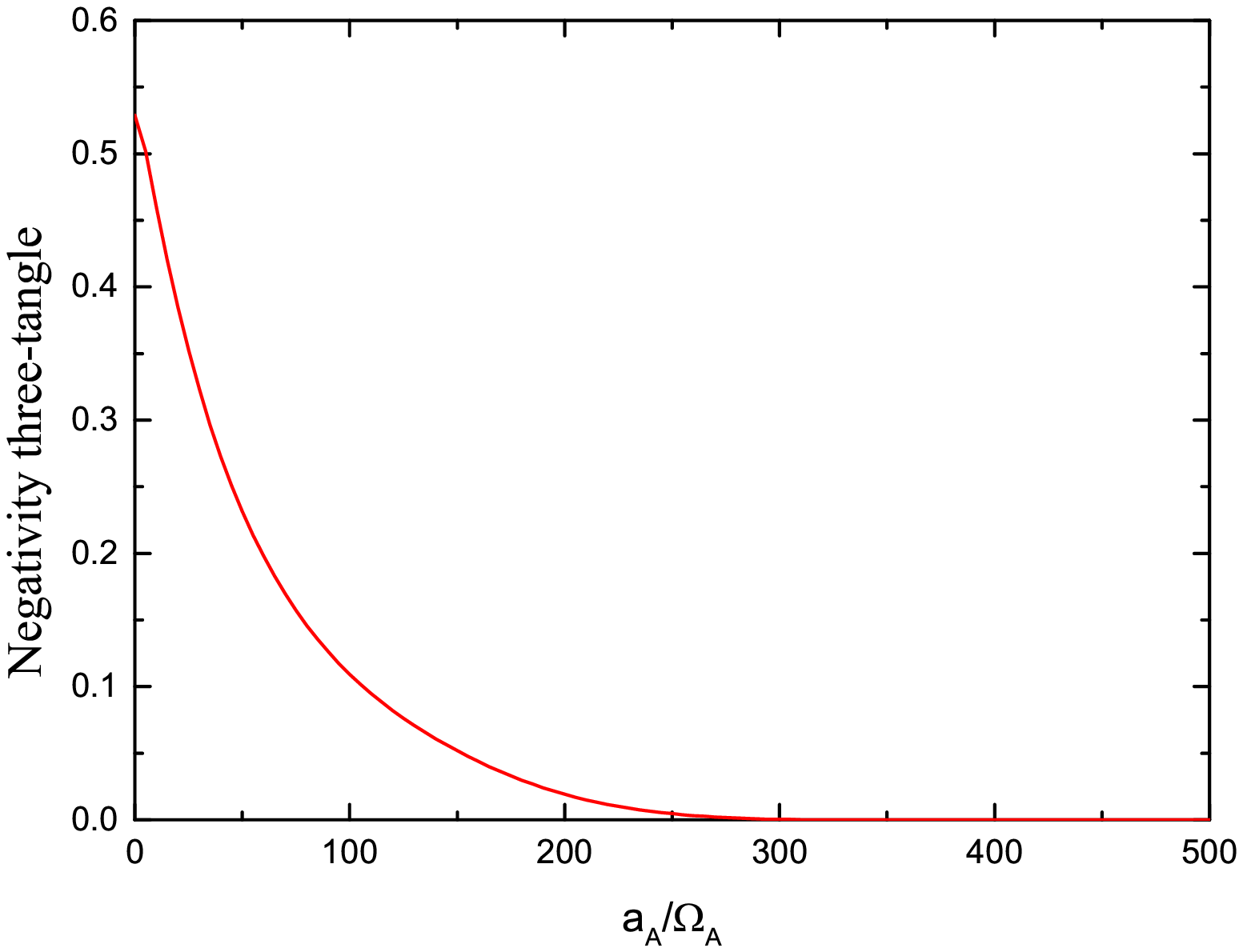}}
\caption{Dependence of (a) the one-tangles, (b) the two-tangles, and (c) the three-tangle on the AFR of qubit $A$ in
the case that $a_B/\Omega_B = 1.2a_A/\Omega_A$, $a_C/\Omega_C = 1.5a_A/\Omega_A$. The qubits are in the W state.}
\end{figure}

As an example, we specify $a_B/\Omega_B = 1.2a_A/\Omega_A$, $a_C/\Omega_C = 1.5a_A/\Omega_A$. One-tangles, two-tangles, and three-tangle are shown in  Figs.~\ref{w40}, \ref{w41}, and  Fig.~\ref{w42}, respectively. The behavior of the one-tangles and three-tangle of the W state is similar to that of the GHZ state as shown in Fig.~\ref{15to16}, but under the same conditions, the values of these tangles are smaller in the W state. Like in the GHZ case, ${\cal N}_{C(AB)}$ suddenly dies first, because of the larger value of $a_C/\Omega_C$. Afterwards,  ${\cal N}_{A(BC)}$ and ${\cal N}_{B(AC)}$ do not intersect because $a_A/\Omega_A$ is too large. The three-tangle suddenly dies when all one-tangles become zero.
Note that when a one-tangle dies, the relevant two   two-tangles have already died because of the relation (\ref{mono}).    Also note that a sudden change occurs in the three-tangle when the one-tangles and two-tangles suddenly die.

The cases that one or more AFRs are not large enough can be inferred from the cases when only one or two qubits accelerate discussed above.

\section{Summary}

In this paper, we have studied how the Unruh effect influences various kinds of entanglement of three qubits, each of which is coupled with an ambient scalar field. The initial states have been considered to be two typical three-qubit entangled states---the GHZ and the W states---which represent two types of tripartite entanglement. We have studied the cases of one qubit accelerating, two qubits accelerating, and three qubits accelerating. The details are summarized in Table~\ref{table}.

A two-tangle measures bipartite entanglement between two qubits, with the other qubit traced out. For the GHZ state as the initial state, the two-tangles always remain zero. For the W state as the initial state, the  two-tangle between two qubits remains a nonzero constant if these two
qubits move uniformly. If at least one of the two qubits accelerates, their two-tangle  suddenly dies when the AFR of one or both of these two qubits are large enough.

When two qubits move uniformly while the other accelerates, the one-tangle between the accelerating qubit and the party consisting of the two uniformly moving qubits  suddenly dies  if the AFR of the accelerating qubit is large enough. However, each of the other two one-tangles---between one uniformly moving qubit and the party consisting of the other uniformly moving qubit and the accelerating qubit---approaches an asymptotic value as the  AFR tends to infinity. For  the GHZ state, the asymptotic value is zero. For the W state, the asymptotic value is nonzero and is larger than the local minima existing  at finite values of the AFR.

\begin{longtable}{|c|l|l|l|}
\hline
\multicolumn{2}{|c|}{}
&\multicolumn{1}{|c|}{\textbf{GHZ}}
&\multicolumn{1}{|c|}{\textbf{W}}\\
\hline
\endhead
\multirow{3}{*}{\tabincell{l}{$a_A=0$,\\$a_B=0$,\\ $a_C\neq 0$.}}    &1-tangles
&\tabincell{l}{${\cal N}_{C(AB)}$ SD at a finite $a_C/\Omega_C$.\\
${\cal N}_{A(BC)}={\cal N}_{B(AC)} \rightarrow 0$ as \\$a_C/\Omega_C\rightarrow \infty$.}
&\tabincell{l}{${\cal N}_{C(AB)}$ SD at a finite $a_C/\Omega_C$. \\${\cal N}_{A(BC)}={\cal N}_{B(AC)}$  is non-\\monotonic, with  nonzero local \\minima, $\rightarrow \frac{2}{3}$ as $a_C/\Omega_C\rightarrow\infty$.}\\
\cline{2-4}
&2-tangles    & $0$.   &\tabincell{l}{${\cal N}_{AB}$ remains constant. \\${\cal N}_{AC}={\cal N}_{BC}$ SD at a finite \\$a_C/\Omega_C$.}\\
\cline{2-4}
&3-tangle  &\tabincell{l}{$\rightarrow 0$ as $a_C/\Omega_C\rightarrow \infty$. There \\is a sudden change where \\${\cal N}_{C(AB)}$ SD.}  &\tabincell{l}{Nonmonotonic, with a nonzero \\local minimum at a finite \\$a_C/\Omega_C$, $\rightarrow (\sqrt{5}-1)/27$ as \\$a_C/\Omega_C\rightarrow \infty$.}\\
\hline
\multirow{3}{*}{\tabincell{l}{$a_A=0$,
\\$a_B\neq 0$,\\ $a_C\neq 0$.}}   &1-tangles
&\tabincell{l}{${\cal N}_{A(BC)}$ SD at finite values of \\both $a_B/\Omega_B$ and $a_C/\Omega_C$.\\
${\cal N}_{\beta (A\gamma)}$ SD at a finite value \\of  $a_\beta/\Omega_\beta$, while weakly varies \\with $a_\gamma/\Omega_\gamma$.     }
&\tabincell{l}{${\cal N}_{A(BC)}$ eventually SD if both\\  $a_B/\Omega_B$ and $a_C/\Omega_C$ are large \\enough. It is nonmonotonic, \\with nonzero local minima, and \\can SD and revive with the \\increase of one AFR while the \\other is not large enough. \\${\cal N}_{\beta (A\gamma)}$ SD when $a_\beta/\Omega_\beta$ is large \\enough. It has nonzero local \\minima and has sudden changes \\if $a_\beta/\Omega_\beta$ is not large enough no \\matter how large is $a_\gamma/\Omega_\gamma$.}\\
\cline{2-4}
&2-tangles   &$0$.   &\tabincell{l}{SD when one or both of the \\two relevant qubits have large \\enough AFRs. }\\
\cline{2-4}
&3-tangle    &\tabincell{l}{SD if $a_B/\Omega_B$ and $a_C/\Omega_C$ are \\both large enough.}
&\tabincell{l}{SD if $a_B/\Omega_B$ and $a_C/\Omega_C$ are \\ both large enough.  It has\\ nonmonotonicity   and local \\ nonzero minima, $\rightarrow$ a nonzero \\asymptotic value when only \\one  AFR  $\rightarrow \infty$ while the other \\nonzero AFR is not large enough.}\\
\hline
\multirow{3}{*}{\tabincell{l}{$a_A\neq 0$,\\$a_B\neq 0$,\\ $a_C\neq 0$.}}
&1-tangles
&\tabincell{l}{${\cal N}_{\alpha(\beta\gamma)}$ SD if  the AFR  of $\alpha$ or \\AFRs of both $\beta$ and $\gamma$ are \\large enough. }
&\tabincell{l}{${\cal N}_{\alpha(\beta\gamma)}$ SD if  the AFR of $\alpha$ or \\AFRs of both $\beta$ and $\gamma$ are \\large enough. There exists \\nonmonotonicity. }\\
\cline{2-4}
&2-tangles   &$0$.   &\tabincell{l}{SD when one or both of the \\two relevant  qubits have \\large enough AFRs.}\\
\cline{2-4}
&3-tangle    &\tabincell{l}{SD after all 1-tangles SD, \\that is, if at least two of \\three qubits have large \\enough AFRs.}
&\tabincell{l}{  SD after all 1-tangles SD, that \\is, if at least two of three qubits \\have large enough AFRs.  }\\
\hline
\caption{\label{table}    Comparison  between the GHZ and the W states of the entanglement behavior  caused by the Unruh fields. SD is the acronym for ``sudden death'' or ``suddenly die''. By nonmonotonicity, it is with respect to one AFR.  }
\end{longtable}

When two qubits accelerate while the other moves uniformly, the one-tangle between the uniformly moving qubit and the party consisting of the two accelerating qubits suddenly dies if the AFRs of both of the two   accelerating qubits are large enough. Each of the other two one-tangles---between one accelerating qubit and the party consisting of the other accelerating qubit and the uniformly moving qubit---suddenly dies when the AFR of the qubit which is by itself one party is large enough. It weakly depends on the AFR of the other accelerating qubit. These features are common in the GHZ and the W states.

When all qubits accelerate,
for both the GHZ and the W states,  the one-tangle ${\cal N}_{\alpha(\beta\gamma)}$ suddenly dies if $a_\alpha/\Omega_\alpha$ is large enough, or both $a_\beta/\Omega_\beta$ and $a_\gamma/\Omega_\gamma$ are large enough.

Therefore, generally speaking, all the one-tangles  eventually suddenly die if at least two of the three qubits have large enough AFRs. When all the one-tangles suddenly die, all the two-tangles must have also died, as dictated by the monogamy relation (\ref{mono}), and consequently the three-tangles also suddenly die. The main difference between the W state and the GHZ state is that for the W state, there exists nonmonotonicity with respect to the AFR of each qubit alone.

It is well known that near the horizon $r=2m$ of a black hole, the Schwarzschild metric can be approximated as the Rindler metric with the acceleration $a = \frac{m}{{{r^2}}}{ ( {1 - \frac{{2m}}{r}}  )^{ - 1/2}}$,  while  the uniform movement   corresponds to free falling into the black hole~\cite{rindler}.  Therefore, the above result can be translated to be near the horizon of a black hole, with $r\approx 2m[1-\frac{1}{(4m a)^2}]^{-1}$ corresponding to the acceleration $a$.

The calculations in this paper imply that near the horizon of a black hole, for three qubits coupled with scalar fields,
all the one-tangles and then all the two-tangles and the three-tangle  eventually  die if at least two of the three qubits are close enough to the horizon $2m$. That is, all kinds of entanglement of the field-coupled qubits are eventually killed by the black hole horizon. Finally, we  conjecture   that for $N$  particles, each of which is coupled with a scalar field,   all kinds of entanglement  suddenly die if $N-1$ particles are close enough to the horizon of a black hole.

\acknowledgments

This work was supported by the National Science Foundation of China (Grant No. 11374060).

\end{document}